\documentclass{article}
\pdfoutput=1

\usepackage{arxiv}
\usepackage[utf8]{inputenc} 
\usepackage[T1]{fontenc}    
\usepackage{hyperref}       
\hypersetup{unicode=true,
            pdfborder={0 0 0},
            breaklinks=true,
            colorlinks=true,
            linkcolor=blue,
            citecolor=blue,
            filecolor=blue,
            urlcolor=blue}
\usepackage{url}            
\usepackage{booktabs}       
\usepackage{amsfonts}       
\usepackage{nicefrac}       
\usepackage{microtype}      
\usepackage{amsmath}
\usepackage{algorithm,algorithmic}
\usepackage{graphicx}
\usepackage{bbm}
\usepackage{caption} 
\usepackage{natbib}
\usepackage{xr}
\usepackage{xcolor}
\newcommand{\beginsupplement}{%
	\setcounter{table}{0}
	\renewcommand{\thetable}{S\arabic{table}}%
	\setcounter{figure}{0}
	\renewcommand{\thefigure}{S\arabic{figure}}%
	\setcounter{section}{0}
	\renewcommand{\thesection}{S\arabic{section}}%
}

\title{$R^*$: A robust MCMC convergence diagnostic with uncertainty using decision tree classifiers}

\author{%
	 Ben Lambert\\
	 MRC Centre for Global Infectious Disease Analysis\\
	 School of Public Health\\
	 Imperial College London\\
	 W2 1PG, United Kingdom\\
	 \texttt{ben.c.lambert@gmail.com} \\
	 \And
	 Aki Vehtari \\
	 Department of Computer Science\\
	 Aalto University\\
	 Finland\\
	 \texttt{aki.vehtari@aalto.fi}
}

\begin{document}

\maketitle

\begin{abstract}
	Markov chain Monte Carlo (MCMC) has transformed Bayesian model inference over the past three decades: mainly because of this, Bayesian inference is now a workhorse of applied scientists. Under general conditions, MCMC sampling converges asymptotically to the posterior distribution, but this provides no guarantees about its performance in finite time. The predominant method for monitoring convergence is to run multiple chains and monitor individual chains' characteristics and compare these to the population as a whole: if within-chain and between-chain summaries are comparable, then this is taken to indicate that the chains have converged to a common stationary distribution. Here, we introduce a new method for diagnosing convergence based on how well a machine learning classifier model can successfully discriminate the individual chains. We call this convergence measure $R^*$. In contrast to the predominant $\widehat{R}$, $R^*$ is a single statistic across all parameters that indicates lack of mixing, although individual variables' importance for this metric can also be determined. Additionally, $R^*$ is not based on any single characteristic of the sampling distribution; instead it uses all the information in the chain, including that given by the joint sampling distribution, which is currently largely overlooked by existing approaches. We recommend calculating $R^*$ using two different machine learning classifiers — gradient-boosted regression trees and random forests — which each work well in models of different dimensions. Because each of these methods outputs a classification probability, as a byproduct, we obtain uncertainty in $R^*$. The method is straightforward to implement and could be a complementary additional check on MCMC convergence for applied analyses.
\end{abstract}

	\section{Introduction}
	Markov chain Monte Carlo (MCMC) is the class of exact-approximate methods that has contributed most to applied Bayesian inference in recent years. In particular, MCMC has made Bayesian inference widely available to a diverse community of practitioners through the many software packages that use it as an internal inference engine: from Gibbs sampling \citep{geman1984stochastic}, which underpins the popular BUGS \citep{lunn2000winbugs} and JAGS \citep{plummer2003jags} libraries, to more recent algorithms: for example, Hamiltonian Monte Carlo (HMC) \citep{neal2011mcmc}, the No U-Turn Sampler (NUTS) \citep{hoffman2014no}, and a dynamic HMC variant \citep{betancourt2017conceptual}, which Stan \citep{carpenter2017stan}, PyMC3 \citep{salvatier2016probabilistic}, Turing \citep{ge2018turing}, TensorFlow Probability  \citep{dillon2017tensorflow} and Pyro \citep{bingham2019pyro} implement. MCMC methods are currently the most effective tools for sampling from many classes of posterior distributions encountered in applied work, and it seems unlikely that this trend will change soon.
	
	Its importance in applied scientists' toolkits means it is essential that MCMC is used properly and with adequate care. A cost of automated inference software is that it is increasingly easy to regard MCMC as oracular: giving uncompromised views onto the posterior. Because of this, software packages (Stan \citep{carpenter2017stan}, for example), go to great lengths to communicate to users any issues with sampling.
	
	The most important determination of whether MCMC has worked is how closely the sampling distribution has converged to the posterior \citep{brooks2011handbook}. MCMC methods are thus created because of an asymptotic property: that given an infinite number of draws, their sampling distribution approaches the posterior (under general conditions). Although the guarantees are asymptotic, MCMC estimates can have negligible bias with only a relatively small number of draws.
	
	The one diagnostic method for determining whether practical convergence has occurred relies on the fact that the posterior distribution is the unique stationary distribution for an MCMC sampler. Therefore, it would appear that, if an MCMC sampling distribution stops changing, then convergence has occurred. Unfortunately, anyone who uses MCMC knows that it is full of false dawns: chains can easily become stuck in areas of parameter space, and observation over short intervals mean the sampling distribution \textit{appears} converged \citep{gelman1992single}. Like furious bees trapped in a room of a house \citep{lambertbees}, MCMC samplers may fail to move due to the narrow gaps that join neighbouring areas. With MCMC, absence of evidence of new areas of high posterior density is, time and again, not evidence of their absence.
	
	To combat this curse of hindsight, running multiple, independent chains, which have been initialised at diverse areas of parameter space is recommended \citep{gelman1992inference}. If the chains appear not to ``mix'' -- a term essentially meaning that it is difficult to resolve an individual chain's path from the mass of paths overlaid on top of one another -- they are yet to converge. This approach makes it less likely that faux-convergence will occur due to chains becoming stuck in an area of parameter space, and running multiple chains is standard practice in applied inference \citep{lambert2018Student}. The predominant approach to quantitatively measuring this mixing is to compare each chain's sampling distribution to that of the population of chains as a whole: specifically, $\widehat{R}$ -- the main convergence statistic used -- compares within-chain variance to that between-chains \citep{gelman1992inference}. If these variances are similar, $\widehat{R}\approx 1$, and chains are deemed to have mixed. Recently, Stan has adopted more advanced variations on the original $\widehat{R}$ formula: for example, splitting individual chains in two to combat poor intra-chain mixing \citep{gelman2013bayesian}; and using ranks of parameter draws rather than the raw values themselves to calculate $\widehat{R}$ \citep{vehtari2019rank}. Additionally, there has been more focus on ensuring that the effective sample size (ESS), a measure of sample quality (see, for example, \cite{lambert2018Student}), is sufficient, and accordingly, new measures of this quantity have been proposed \citep{vehtari2019rank} and adopted \citep{carpenter2017stan}. Collectively, these statistics help alert users of MCMC to issues with sampling (that typically echo issues with the model) meaning that all is not hunky dory.
	
	Here, we introduce $R^*$, a new convergence diagnostic metric. This statistic is built on the intuition that, if chains are mixed, it should not be possible to discern from a draw's value the chain that generated it. Rephrased, it should not be possible to predict which draws come from which chain. In this vein, we use machine learning (ML) classifiers to measure convergence. Specifically, we train classifiers to predict the chain that generated each observation. By evaluating the performance of classifiers on a held-out test set, this provides a new convergence metric. To maximise predictive accuracy, our chosen classifiers naturally exploit differences in the full joint distributions between chains, which means they are sensitive to variations across the joint distribution of target model dimensions unlike most existent convergence diagnostics. Our statistic, unlike its $\widehat{R}$ cousins, is scalar-valued for multivariate distributions: one model provides a single $R^*$, whereas $\widehat{R}$ has separate values for each univariate marginal distribution. However, the ML classifiers we use can straightforwardly be interrogated to estimate which parameters were most important for generating predictive accuracy.
	
	There are, of course, a huge variety of possible ML classifiers. For a method to be useful and widely adopted, however, it needs to satisfy a number of criteria: first, across a range of examples, it should consistently be able to detect poor MCMC convergence; second, the ML classifier should be trainable in a reasonable amount of time; lastly, the methods should be tuned (via their hyperparameters) so as to be standardised, so that $R^*$ computed by one analyst is comparable to that computed by another. Here, we perform comparisons across some of the most popular methods in the ML literature and recommend two tree-based ML classifiers: gradient-boosted regression trees \citep{friedman2001greedy,greenwell2019package} (``GBM'') and random forest classifiers \citep{breiman2001random} (``RF''). Both of these methods performed consistently well across our range of examples, and the models were relatively efficient to train. We recommend calculating $R^*$ using both of these methods: GBMs tend to perform best for low dimensional cases; in moderate dimensional cases upwards, RF dominates. This difference in performance is partly due to hyperparameter tuning: echoing previous results in the literature \cite[Chapters~11\&12]{boehmke2019hands}, our results show that GBMs, in general, are more sensitive to hyperparameter choice than are RFs. For a GBM classifier to perform well in higher dimensions, more rounds of boosting are required, which substantially increases training time; instead, we fix the hyperparameters of GBMs to specific values to ensure practical usefulness. For RF classifiers, empirical studies have demonstrated reliable heuristics for selecting robust hyperparameters \citep{bernard2009influence}, and, in any case, across our set of test cases, these classifiers are relatively insensitive to hyperparameter choice.
	
	For the types of problem we tested, $R^*$ calculation is of a speed comparable to some of the newer $\widehat{R}$ measures calculated (typically $\mathcal{O}$(seconds) to calculate), although for models with 10,000s of parameters and many iterations, the time taken is longer. In addition, since ML classifiers can output predicted class probabilities, we obtain uncertainty measures for $R^*$, which we find provides a useful summary of MCMC convergence. $R^*$ can straightforwardly be incorporated into existing software libraries to provide a complementary convergence metric alongside more established measures.
	
	The structure of this paper is as follows: in \S\ref{sec:method}, we describe in detail the method for calculating $R^*$ and its uncertainty; in \S\ref{sec:results}, we examine the performance of $R^*$ across a range of scenarios. Code for reproducing the analyses is provided at \url{https://github.com/ben18785/ml-mcmc-convergence}.

	\section{Method}\label{sec:method}
\begin{figure}[!htb]
		\centerline{\includegraphics[width=1.0\textwidth]{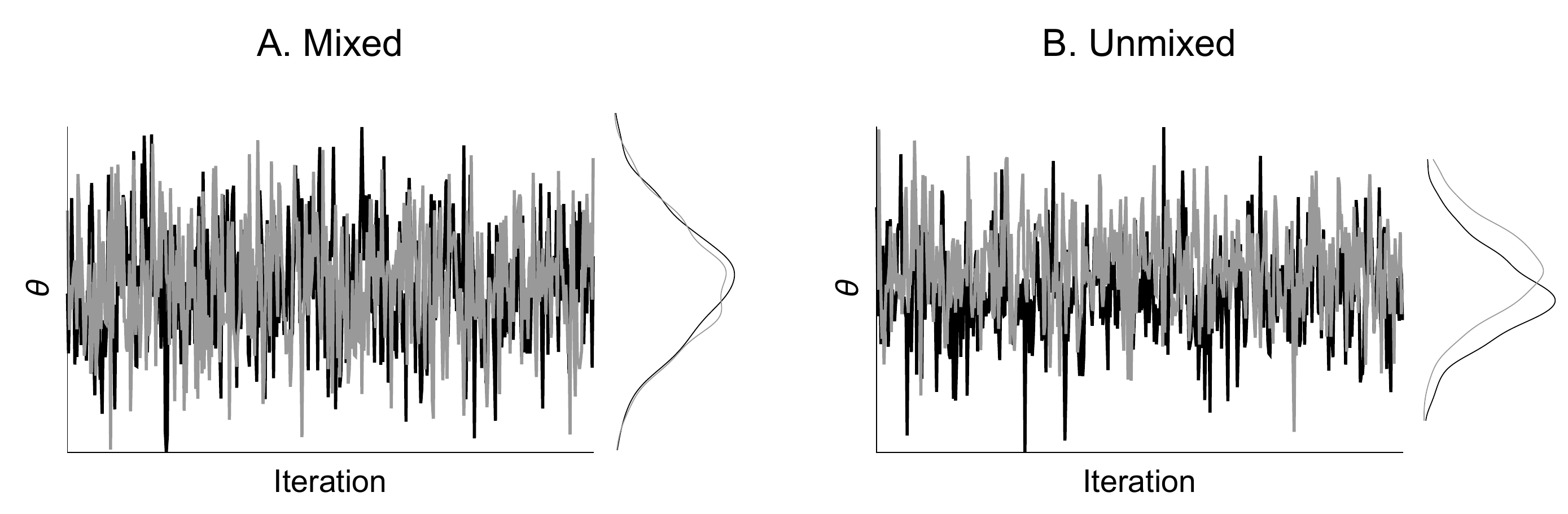}}
		\caption{\textbf{Chain prediction based on the marginal distribution of a single parameter.} A shows the path of two chains that have mixed (with marginal distribution to the right of panel); B shows two chains that have not mixed.}
		\label{fig:marginal}
	\end{figure}

\begin{figure}[h]
		\centerline{\includegraphics[width=1.0\textwidth]{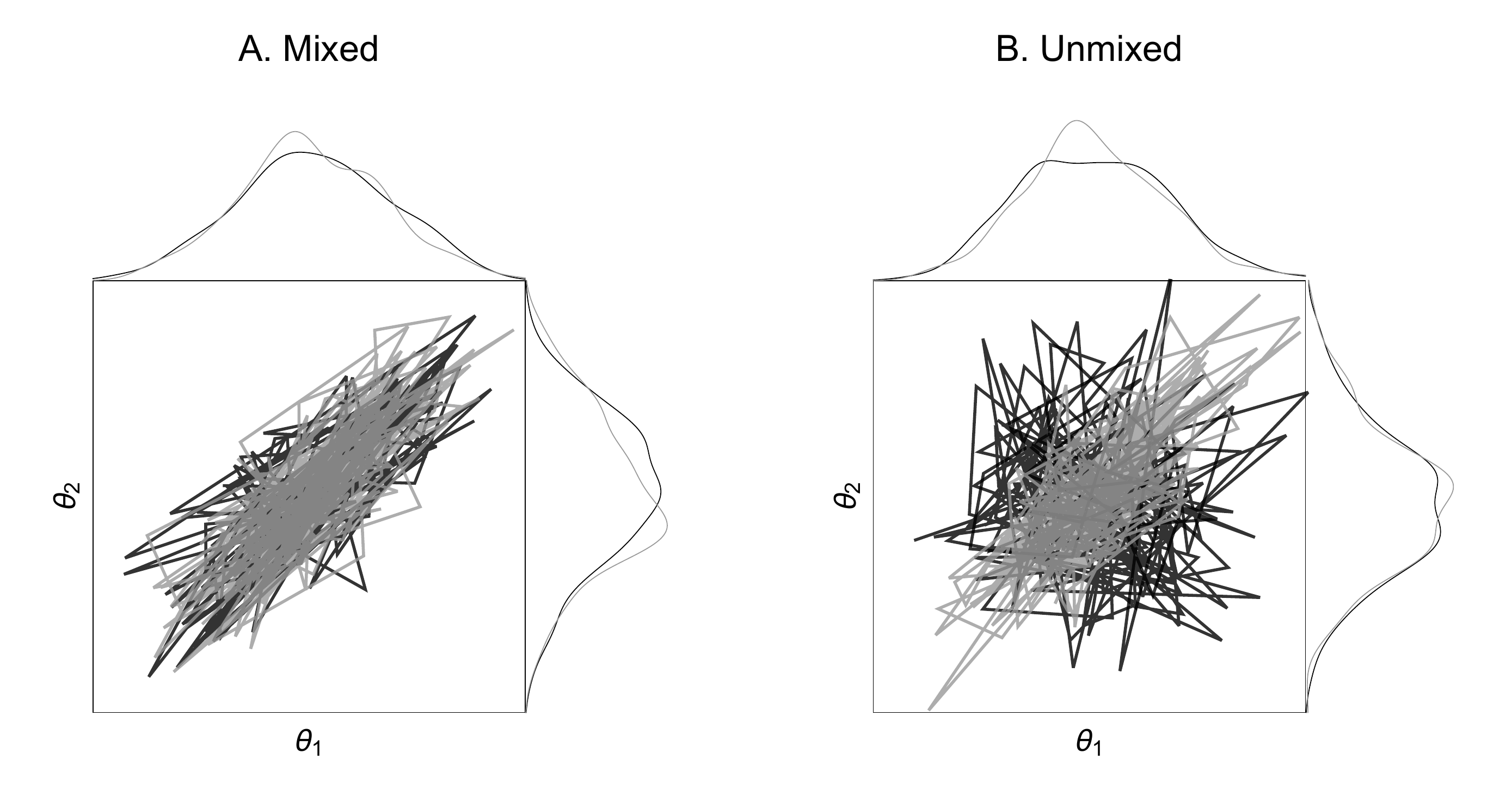}}
		\caption{\textbf{Chain prediction based on the joint distribution of two parameters where each chain's marginals are the same.} A shows the path of two chains that have mixed resulting in similar sampling distributions (to the right and above each panel); B shows two chains that have not mixed.}
		\label{fig:joint}
	\end{figure}

	If Markov chains have not mixed, it is possible to guess (with more accuracy than chance) to which chain a draw belongs from its value. This is possible if there are differences in the sampling distribution for any dimension in the target distribution (Fig. \ref{fig:marginal}): in this case, if the marginal distributions differ between chains, this information can be used to predict which chain a draw belongs to. It  is also possible to predict the chain that generated a given draw if there are differences in the joint distribution of two (or more) dimensions of the target, even if the marginal distributions are the same (Fig. \ref{fig:joint}).
	
	These two cases, whilst simple, illustrate the basis of our approach. To determine if a set of Markov chains has converged to the same distribution, we train a supervised ML model to classify the chain to which each draw belongs. By evaluating its performance on a separate test set, we delineate whether chains have mixed based on whether classification accuracy is above the ``null'' case, where accuracy is $1/{N}$, and $N$ is the number of chains. By taking the ratio of classification accuracy to this null accuracy, we obtain a statistic that is interpretable in a similar way to $\widehat{R}$ \citep{vehtari2019rank}. In a nod to this established statistic, we call our statistic $R^*$, and, by design, $R^*\approx 1$ signifies convergence. Algorithm \ref{alg:R_star} gives a recipe for calculating $R^*$.
	
	\begin{algorithm}[tb]
		\caption{$R^*$ calculation}
		\label{alg:R_star}
		\begin{algorithmic}
			\STATE Given chain-wise draws from the target, $\{X^{\{1\}},X^{\{2\}},...,X^{\{N\}}\}$ and a test set length, $S_\text{test}$:
			\FOR{$m=1$ to $N$}
			\STATE Create train and test sets by random-sampling (w/o replacement), $X^{\{m\}}\rightarrow\{X^{\{m\}}_\text{train},X^{\{m\}}_\text{test}\}$
			\ENDFOR
			\STATE Stack $X_\text{train} = (X^{\{1\}}_\text{train},X^{\{2\}}_\text{train},...,X^{\{N\}}_\text{train})^T$
			\STATE Stack $X_\text{test} = (X^{\{1\}}_\text{test},X^{\{2\}}_\text{test},...,X^{\{N\}}_\text{test})^T$
			\STATE Train ML model to classify chain id from any draw, $x$: $\text{ML}(x|X_\text{train}) \rightarrow c$
			\FOR{$s=1$ to $S_\text{test}$}
			\STATE Obtain test draw, $x^{\{s\}}=X_\text{test}(s)\in \mathbb{R}^K$
			\STATE Predict chain id, $c^{\{s\}} = \text{ML}(x^{\{s\}}|X_\text{train})$
			\STATE Compare with actual id, $c^s$: $a^{\{s\}}=\mathbbm{1}(c^{\{s\}}=c^s)$
			\ENDFOR
			\STATE Calculate predictive accuracy, $\bar{a} = \frac{1}{S_\text{test}} \sum_{s=1}^{S_\text{test}} a^{\{s\}}$
			\STATE Calculate ratio to null model accuracy, $R^* = \bar{a} / (1 / N) = N \bar{a}$
			\RETURN $R^*$
		\end{algorithmic}
	\end{algorithm}
	
	To identify promising candidate ML methods, we run a series of experiments using a number of popular classifiers (see \S\ref{sec:ml_model}). Two methods performed consistently well across our examples: gradient-boosted regression trees (also known as a type of gradient-boosted machine or GBM, introduced in \cite{friedman2001greedy}) and random forest classifiers \citep{breiman2001random} (``RF''). Both of these approaches are based on decision trees: GBMs use an iterative approach known as ``boosting'', where each subsequent decision tree aims to predict the residuals from the previous one; RFs use an approach known as ``bagging'' which trains decision trees on many bootstrapped copies of the training data, yielding a collection of independently trained trees whose individual decisions, when aggregated, yield a class. For the implementations of these classifiers, we use those available in \textbf{\textsf{R}}'s ``Caret'' package \citep{kuhn2008building}, which, in turn, uses the ``gbm'' \citep{greenwell2019package} and ``randomForest'' \citep{liaw2002classification} packages.
	
	The data for each chain has dimensions: $X\in \mathbb{R}^{S}\times \mathbb{R}^{K}$, where $S$ is the number of draws taken (here assumed the same for each chain, but this is not a binding constraint), and $K$ is the number of parameters. We split each chain's draws into randomly divided training and testing sets: here, we use 70\% of draws for training and 30\% for testing. Both approaches typically took $\mathcal{O}$(seconds) on a desktop computer to execute training then prediction on a testing set for most models we consider in \S\ref{sec:results}.
	
	Since different posteriors present different challenges to MCMC samplers, the nature of classification boundaries is problem-specific. Because of this, there is no unique optimal classifier across all problems, and the performance of the algorithms we investigate depends on their hyperparameters (see \S\ref{sec:hyperparameters}). GBM models have a number of hyperparameters and have previously been demonstrated to be hard to tune \cite[Chapter~12]{boehmke2019hands}. To ensure practical run time for this method, we suggest running it using a default hyperparameter set: an interaction depth of 3, a shrinkage parameter of 0.1, 10 observations being the minimum required for each node, and that 50 trees be grown. This choice of hyperparameters was chosen to present a balance between training cost and classification accuracy, producing an $R^*$  that was a stringent measure of convergence — in general, more stringent than $\widehat{R}$. RFs are generally less sensitive to variation in their single hyperparameter: $m_{\text{try}}$, the size of the subset of all features over which to search for an optimal split \cite[Chapter~11]{boehmke2019hands}. Our experiments replicate these results (see \S\ref{sec:hyperparameters}), and we suggest running RFs using the heuristic $m_{\text{try}}=\sqrt{K}$, which has previously been shown to be a choice resulting in robust classification performance \citep{bernard2009influence,boehmke2019hands}. Unless otherwise stated, in the examples explored in \S\ref{sec:results}, the hyperparameters for these two classifiers were set at these defaults.
	
	From a classifier fit, predicted chain probabilities can also be obtained, which we leverage to produce an uncertainty distribution for $R^*$. Algorithm \ref{alg:R_star_uncertainty} gives a recipe for generating draws from this distribution, which we now elaborate on in words. For each draw, $s$, in our testing set, classifiers output a simplex of chain probabilities: $\boldsymbol{p}^{\{s\}}=(p_1^{\{s\}},p_2^{\{s\}},...,p_N^{\{s\}})$, which forms a categorical distribution that can be sampled from to yield a unique chain prediction, $c^{\{s\}}$. By comparing this classification to the true classification, $c^s$, we obtain a binary measure, $a^{\{s\}}=\mathbbm{1}(c^{\{s\}}=c^s)$, of whether this prediction was correct. We repeat this process for each draw in the testing set, generating $\boldsymbol{a}=(a^{\{1\}},a^{\{2\}},...,a^{\{S_\text{test}\}})$, whose average yields a single $R^{*\{i\}}=N \bar{a}$ estimate for iteration $i$. We then iterate this process, for $i=1,2,...,I$, producing a set of $(R^{*\{1\}},R^{*\{2\}},...,R^{*\{I\}})$, which collectively represent a distribution for $R^*$.

	\begin{algorithm}[tb]
		\caption{Procedure to generate $I$ draws of $R^*$}
		\label{alg:R_star_uncertainty}
		\begin{algorithmic}
			\STATE Given test data $X_\text{test}$, number of chains $N$, number of iterations $I$, and fitted
			\STATE model, $\text{ML}(x|X_\text{train})\rightarrow(p_1,p_2,...,p_N)$:
			\FOR{$i=1$ to $I$}
			\FOR{$s=1$ to $S_\text{test}$}
			\STATE Obtain test draw, $x^{\{s\}}=X_\text{test}(s)\in \mathbb{R}^K$
			\STATE Predict chain id probabilities, $(p_1^{\{s\}},p_2^{\{s\}},...,p_N^{\{s\}})= \text{ML}(x^{\{s\}}|X_\text{train})$
			\STATE Draw a chain id, $c^{\{s\}} \sim \text{categorical}(p_1^{\{s\}},p_2^{\{s\}},...,p_N^{\{s\}})$
			\STATE Compare with actual id, $c^s$: $a^{\{s\}}=\mathbbm{1}(c^{\{s\}}=c^s)$
			\ENDFOR
			\STATE Calculate predictive accuracy, $\bar{a} = \frac{1}{S_\text{test}} \sum_{s=1}^{S_\text{test}} a^{\{s\}}$
			\STATE Calculate ratio to null model accuracy, $R^{*\{i\}} = \bar{a} / (1 / N) = N\bar{a}$
			\ENDFOR
			\RETURN $(R^{*\{1\}},R^{*\{2\}},...,R^{*\{I\}})$
		\end{algorithmic}
	\end{algorithm}

	\section{Results}\label{sec:results}
	To illustrate the versatility of $R^*$, we use a range of examples that demonstrate how this statistic fares across a range of scenarios. Table \ref{tab:results} summarises the examples and provides a rationale for their inclusion. The experiments not detailed in the main text are briefly described in \S\ref{sec:further_experiments} and more fully in the relevant sections given in Table \ref{tab:results}. In all cases where $\widehat{R}$ was calculated, unless otherwise stated, we followed the approach in \cite{vehtari2019rank}, by calculating it as the maximum of rank-normalised split-$\widehat{R}$ and rank-normalised folded split-$\widehat{R}$: for simplicity, we refer to this as rank-normalised split-$\widehat{R}$.
	
	\begin{table}[]
		\footnotesize
		\begin{tabular}{l|l|l}
			\textbf{Example }&  \textbf{Relevance} & \textbf{Section} \\
			\midrule
			Autoregressive & \textbf{Examining $R^*$ and sensitivities to its calculation} & \ref{sec:heterogeneity} \\
			& Detecting heterogeneous chain variance using $R^*$ & \ref{sec:heterogeneity_performance} \\
			& Stochasticity in $R^*$ & \ref{sec:heterogeneity_stochasticity}\\
			& Generating $R^*$ uncertainty measure & \ref{sec:heterogeneity_uncertainty}\\
			& Sensitivity of $R^*$ to number of chains & \ref{sec:heterogeneity_numchains}\\
			\midrule
			Multivariate normals & \textbf{Detecting convergence in joint distributions} &  \ref{sec:multivariate_normal}\\
			& Unconverged joint distribution in bivariate normal & \ref{sec:multivariate_normal_bivariate}\\
			& High correlations between dims in 250D normal  & \ref{sec:multivariate_normal_250}\\
			& Measuring contributions of variables to poor convergence & \ref{sec:multivariate_normal_varimportance}\\
			\midrule
			Cauchy & \textbf{Detecting convergence for long-tailed distributions} & \ref{sec:cauchy}\\
			& Comparing $R^*$ and existing measures to & \ref{sec:cauchy_objective}\\
			& objective convergence &\\
			\midrule
			Eight schools model & \textbf{Hierarchical Bayesian model slow convergence} & \ref{sec:eight_shools}\\
			\midrule 
			Wide multivariate normal & \textbf{Detecting convergence when \# draws $\sim$ \# dims} & \ref{sec:wide}\\
			\midrule
			Non-stationary marginals & \textbf{Detecting time-varying sampling distributions} & \ref{sec:non-stationary}\\
			& Trends in mean across all chains & \ref{sec:non-stationary_chains}\\
			& Trends in mean in a single dimension & \ref{sec:non-stationary_single}\\
			& Trends in covariance & \ref{sec:non-stationary_covariance}\\
			& Sensitivity of $R^*$ to chain persistence & \ref{sec:non-stationary_persistence}\\
			\midrule
			Ovarian and prostate models & \textbf{Bayesian models with many parameters} & \ref{sec:prostate}\\
			&  \textbf{and multimodal posteriors} &\\
			\midrule		
			Discrete Markov model & \textbf{Evaluating $R^*$ on discrete parameter models} & \ref{sec:discrete}\\
			& Small state-space & \ref{sec:discrete_small}\\
			& Large state-space & \ref{sec:discrete_large}\\
			\midrule
			Various & \textbf{Sensitivity of $R^*$ to ML model} & \ref{sec:ML_sensitivity}\\
			& Comparing different ML classifiers & \ref{sec:ml_model}\\
			& Sensitivity of $R^*$ to GBM and RF hyperparameters & \ref{sec:hyperparameters}\\
			\midrule
			Multivariate normal \& & \textbf{Comparing GBM and RF classifiers} & \ref{sec:comparison_gbm_rf}\\
			Student-t dists. & &\\
			& Detecting convergence in joint distributions & \ref{sec:joint_distribution}\\
			& Tail convergence & \ref{sec:tail_fatness}\\
		\end{tabular}\caption{\textbf{Summarising the example problems and reasons for their inclusion.}}\label{tab:results}
	\end{table}
	
	\subsection{Heterogeneity in chain variance: autoregressive example}\label{sec:heterogeneity}
	In this section, we use a simple example to illustrate how $R^*$ works. Specifically, we show how $R^*$ can detect heterogeneous variance across Markov chains. This example aims to illustrate the basic mechanics behind how $R^*$ works, so we use only the GBM classifier here. This section also investigates the sensitivity of this measure: across different training and testing sets (\S\ref{sec:heterogeneity_stochasticity}) and different draws from the ML model-predicted probability simplex (\S\ref{sec:heterogeneity_uncertainty}); and to differing numbers of chains (\S\ref{sec:heterogeneity_numchains}). The experiments we use to study these issues are all of similar form to the following data generating process: four Markov chains are generated, where each samples from an autoregressive order 1 (AR(1)) process of the form,
	\begin{equation}\label{eq:ar1}
	X_t = \rho X_{t-1} + \epsilon_t,
	\end{equation}
	where $\epsilon_t\stackrel{i.i.d.}{\sim}\text{normal}(0, \sigma)$, $\rho=0.3$ and $t=1,2,...,2000$. Three of the chains share the same $\sigma=1$, whereas the other chain has $\sigma=1/3$, so that it has $1/3$ of the (unconditional) standard deviation of the others.
	
	\subsubsection{Performance of $R^*$}\label{sec:heterogeneity_performance}
	To illustrate the consistency of $R^*$, we performed 1000 replicates where, in each case, we generated four $\{X_t\}$ series as described (i.e. where one chain has a lower variance). We then fit a GBM to a labelled training set. The fitted model is then used to classify draws in an independent test set according to the chain which generated them. For each replicate, we then calculated $R^*$ as described in Algorithm \ref{alg:R_star}.
	
\begin{figure}[!htb]
		\centerline{\includegraphics[width=1.0\textwidth]{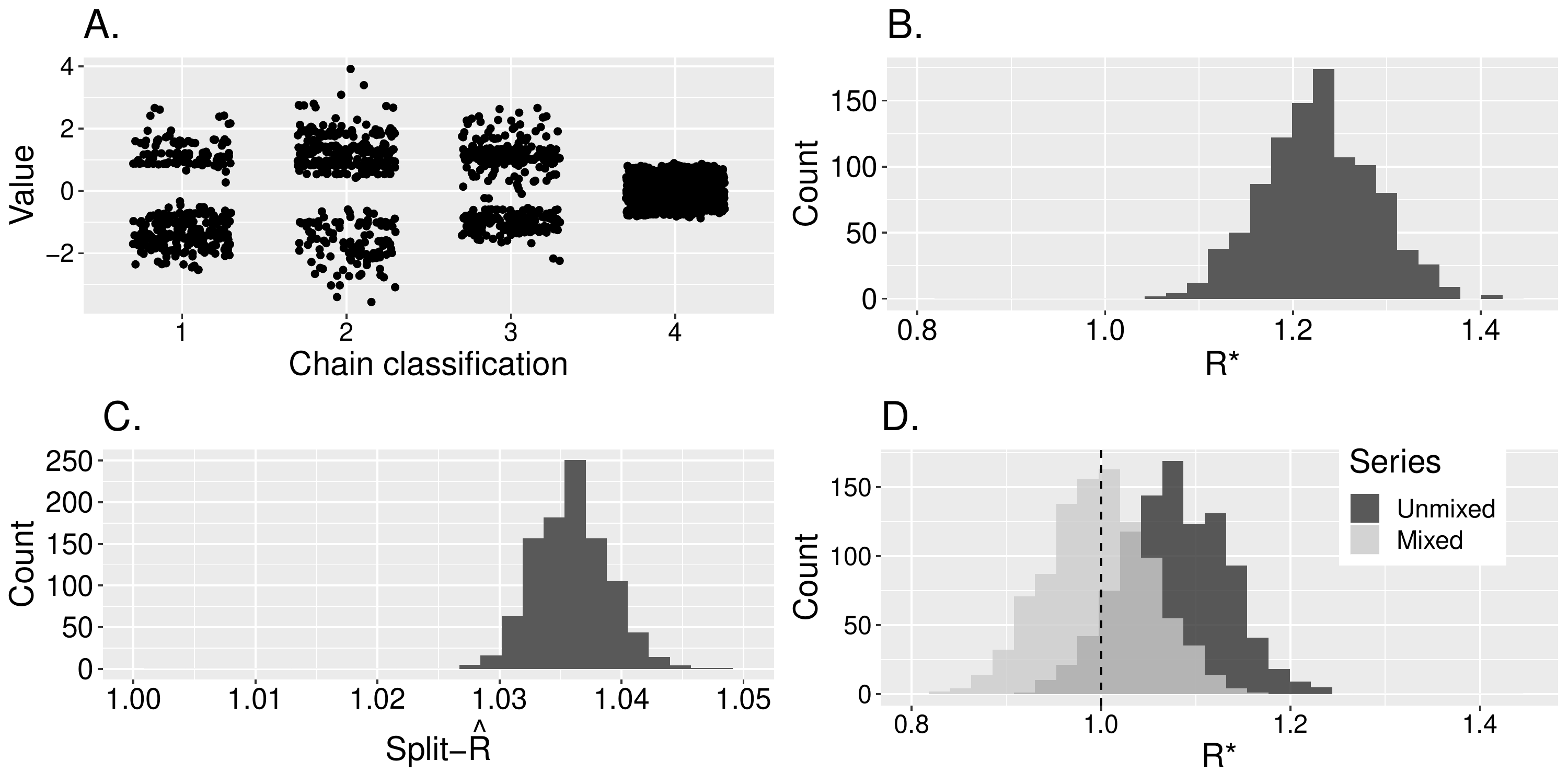}}
		\caption{\textbf{Autoregressive example.} A shows how the GBM's classifications vary according to the draw's value for an example model fit; B shows $R^*$ values generated by Algorithm \ref{alg:R_star} across 1000 replicate datasets; C shows corresponding rank-normalised split-$\widehat{R}$ values for each of the 1000 replicates; and D shows 1000 $R^*$ samples as generated by Algorithm \ref{alg:R_star_uncertainty} for two series: the ``unmixed'' dataset being the same as used for figures A-C; the ``mixed'' where all chains have the same distribution as described in \S\ref{sec:heterogeneity}. Note that, in D, only a single series of each series type is used to generate distribution. All examples used a GBM for classification using the default hyperparameter values given in \S\ref{sec:method}.}
		\label{fig:ar1}
	\end{figure}

	In Fig. \ref{fig:ar1}A, we show how a GBM fitted to one such replicate dataset classifies observations according to a draw's value. Unsurprisingly, since the fourth chain has a smaller variance, observations close to zero are likely to be classified as being generated by this chain.
	
	In Fig. \ref{fig:ar1}B, we show that $R^*>1$ for all replicates, indicating that the chains had not converged in all cases. In Fig. \ref{fig:ar1}C, we show rank-normalised split-$\widehat{R}$ for each replicate;  as for $R^*$, this metric indicates the chains had not converged in all replicates because $\widehat{R}>1.01$.
	
	\subsubsection{Stochasticity in $R^*$}\label{sec:heterogeneity_stochasticity}
	Unlike $\widehat{R}$, $R^*$ is a stochastic convergence measure due to randomness in creating training and testing sets (essentially a form of sampling variation) and randomness in the methods used to train the ML model. This means that even if the same sample is used, $R^*$ will return a different value each time it is calculated if the pseudorandom seed is not fixed. To probe the extent of this randomness, we generated data using the same process as in \S\ref{sec:heterogeneity} but now using varying sample sizes, including samples consisting of 500, 1000, 2000, 4000 and 8000 draws. For each dataset, we computed $R^*$ on it 100 times, allowing the pseudorandom seed to vary between calculations. We stress that, for each sample size, we used the same dataset (so there were 5 datasets created in total -- one for each sample size), so stochasticity comes from $R^*$ calculation, not that from the data generating process.
	
\begin{figure}[!htb]
		\centerline{\includegraphics[width=0.6\textwidth]{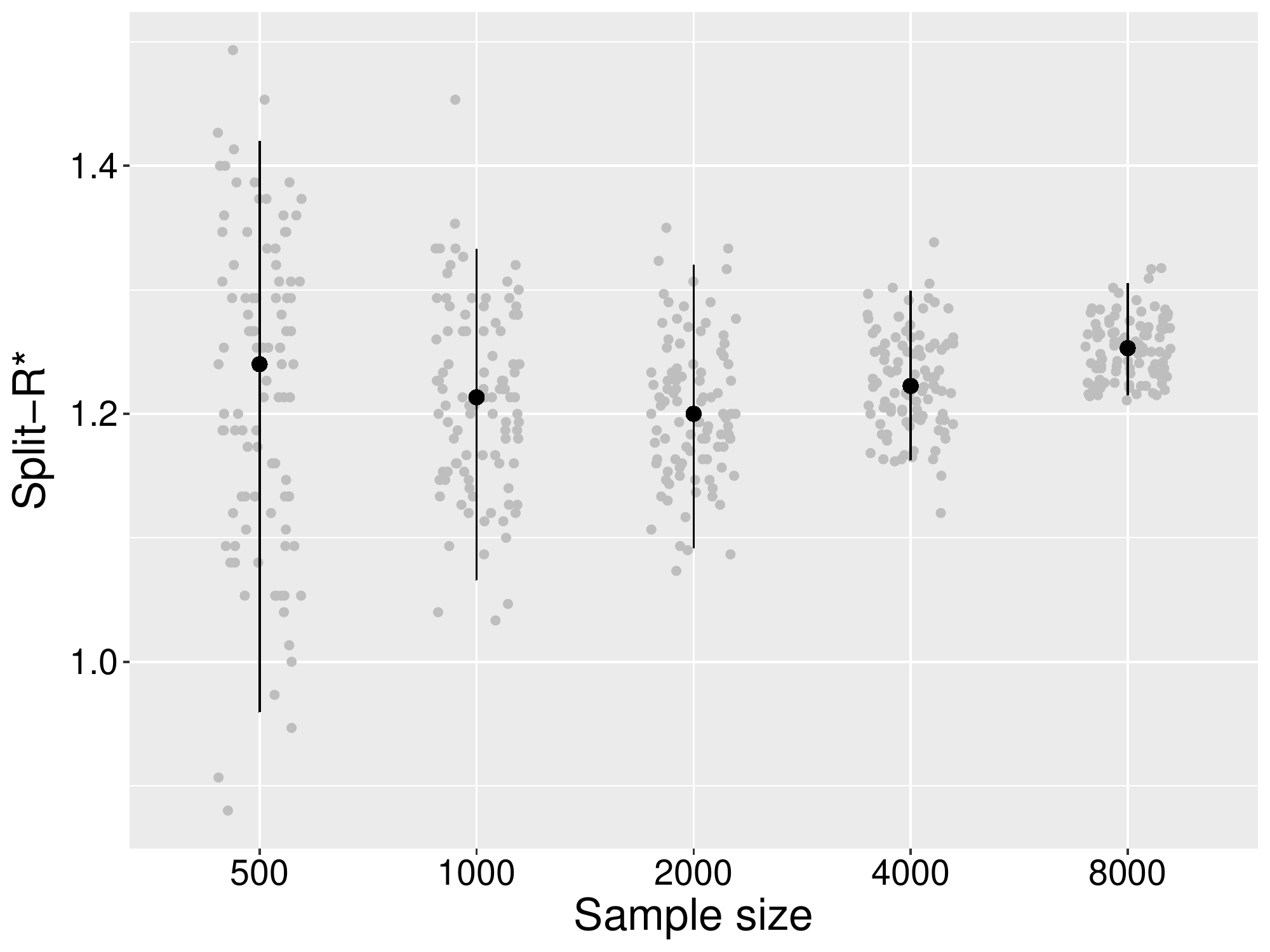}}
		\caption{\textbf{Autoregressive example: $R^*$ stochasticity.} The horizontal axis shows sample size; the vertical axis shows the value of $R^*$ calculated as per Algorithm \ref{alg:R_star} applied to chains split into two halves. Grey points show the value of $R^*$ for each replicate (jitter was added to point positions). Black points show the median $R^*$ value; upper and lower whiskers show 2.5\% and 97.5\% quantiles. For each sample size, a single dataset was created and used for all $R^*$ calculations.}
		\label{fig:ar1_samplesize}
	\end{figure}

	In Fig. \ref{fig:ar1_samplesize}, we show the results of this study. In this figure, the horizontal axis shows the sample size, and the vertical axis, the value of $R^*$ in each repetition. This shows that as the number of samples increased, variation in $R^*$ declined. At a sample size of 500, there were four cases where $R^*<1$; in larger samples, there were none. Intuitively, the reduction in sampling variation when composing training and test sets from larger samples results in lower variability in ML model predictions. We also expect that larger sample sizes should lead to higher $R^*$ values with lower variance, since more training data leads to ML models with lower generalisation error. We may start to see this here, since the median $R^*=1.25$ at a sample size of 8000 was greater than for the smaller samples. 
	
	If randomness in $R^*$ calculation leads to different conclusions about convergence being drawn, this would be problematic. One potential remedy for this is to repeatedly calculate $R^*$ on a given sample, much as we have done here, and consider the distribution of $R^*$ values computed. The computational cost of doing this may, of course, be unreasonable. Instead, in \S\ref{sec:heterogeneity_uncertainty}, we consider an alternative approach based on bootstrapping a single ML model's predictions.
	
	\subsubsection{Uncertainty distribution for $R^*$}\label{sec:heterogeneity_uncertainty}
	GBMs return a probability simplex for each draw, indicating the probability that the draw was generated by a given chain. We can use this simplex to generate a measure of uncertainty in $R^*$ as detailed in Algorithm \ref{alg:R_star_uncertainty}. We demonstrate this idea using two datasets: one generated as described in \S\ref{sec:heterogeneity}, where one chain (out of four) has a lower variance than the others (we call this the ``unmixed'' data); and another, where all chains sample from the same distribution (we call this the ``mixed'' data). In Fig. \ref{fig:ar1}D, we show the $R^*$ distributions in each case. For the unmixed data, the distribution has its bulk of mass away from 1, indicating lack of convergence. For the mixed data, the distribution is centred on 1, indicating convergence. In the mixed case, there are many draws where $R^*<1$: these indicate that, in that particular draw from the probability simplex, chain classification is actually worse than selecting a chain identification uniformly at random. Much like how it is possible for $\hat{R}<1$, this is a sample property, driven by the sampling distribution of the categorical distribution defined by the probability simplex.
	
	It is worth emphasising that the uncertainty distribution obtained by Algorithm \ref{alg:R_star_uncertainty} differs from that obtained from repeatedly calculating $R^*$ via Algorithm \ref{alg:R_star} as was done in \S\ref{sec:heterogeneity_stochasticity}. In Algorithm \ref{alg:R_star_uncertainty}, variation in $R^*$ comes from sampling from the probability simplex: if predicted chain probabilities are close to uniform, there will be greater uncertainty in $R^*$. Repeatedly calculating $R^*$ by applying Algorithm \ref{alg:R_star} to the same dataset yields a distribution whose width derives from sampling variation when forming training and testing sets and the stochasticity in training ML models. Collectively, these differences mean that the two measures of uncertainty will differ.
	
	There is an additional difference, though, in the central points of each distribution: the distribution obtained by Algorithm \ref{alg:R_star_uncertainty} will, in general, have a lower mean than that obtained by repeated application of Algorithm \ref{alg:R_star}. To see this, note that the darker-shaded $R^*$ distribution in Figure \ref{fig:ar1}D was generated via Algorithm \ref{alg:R_star_uncertainty} and has a mean around 1.07; the distribution shown in Figure \ref{fig:ar1}B was generated by repeatedly recomputing $R^*$ using Algorithm \ref{alg:R_star} and has a mean closer to 1.22. This difference in mean is expected since predictive performance when assigning chain identities stochastically when sampling from the categorical distribution of the probability simplex (as is done in Algorithm \ref{alg:R_star_uncertainty}) will generally result in worse prediction than when assigning each chain identity using whichever chain has the highest class probability (as is done in Algorithm \ref{alg:R_star}). Of course, we would prefer it if the uncertainty distribution generated by Algorithm \ref{alg:R_star_uncertainty} had a mean closer to the one obtained by repeated application of Algorithm \ref{alg:R_star}. Nonetheless, in practice, we have found that the mean of the $R^*$ distribution generated by Algorithm \ref{alg:R_star_uncertainty} provides a useful cheaper diagnostic.
	
	\subsubsection{Sensitivity to number of chains}\label{sec:heterogeneity_numchains}
	We have so far focused on the sensitivity of $R^*$ to chain heterogeneity with a fixed number of chains: four. Since classification may become a harder problem when there are more categories, we now demonstrate how $R^*$ (as calculated by Algorithm \ref{alg:R_star}) performs across various number of chains. For comparison, we also illustrate how the performance of rank-normalised split-$\widehat{R}$ varies with number of chains. To do so, we consider an autoregressive example similar to that described in \S\ref{sec:heterogeneity}: where all chains bar one have $\sigma=1$, and the remaining chain has $\sigma=1/3$. We consider cases with 2, 4, 8, 16, and 32 chains. The other hyperparameters of the data generating process remain the same as in \S\ref{sec:heterogeneity}.
	
\begin{figure}[!htb]
		\centerline{\includegraphics[width=0.8\textwidth]{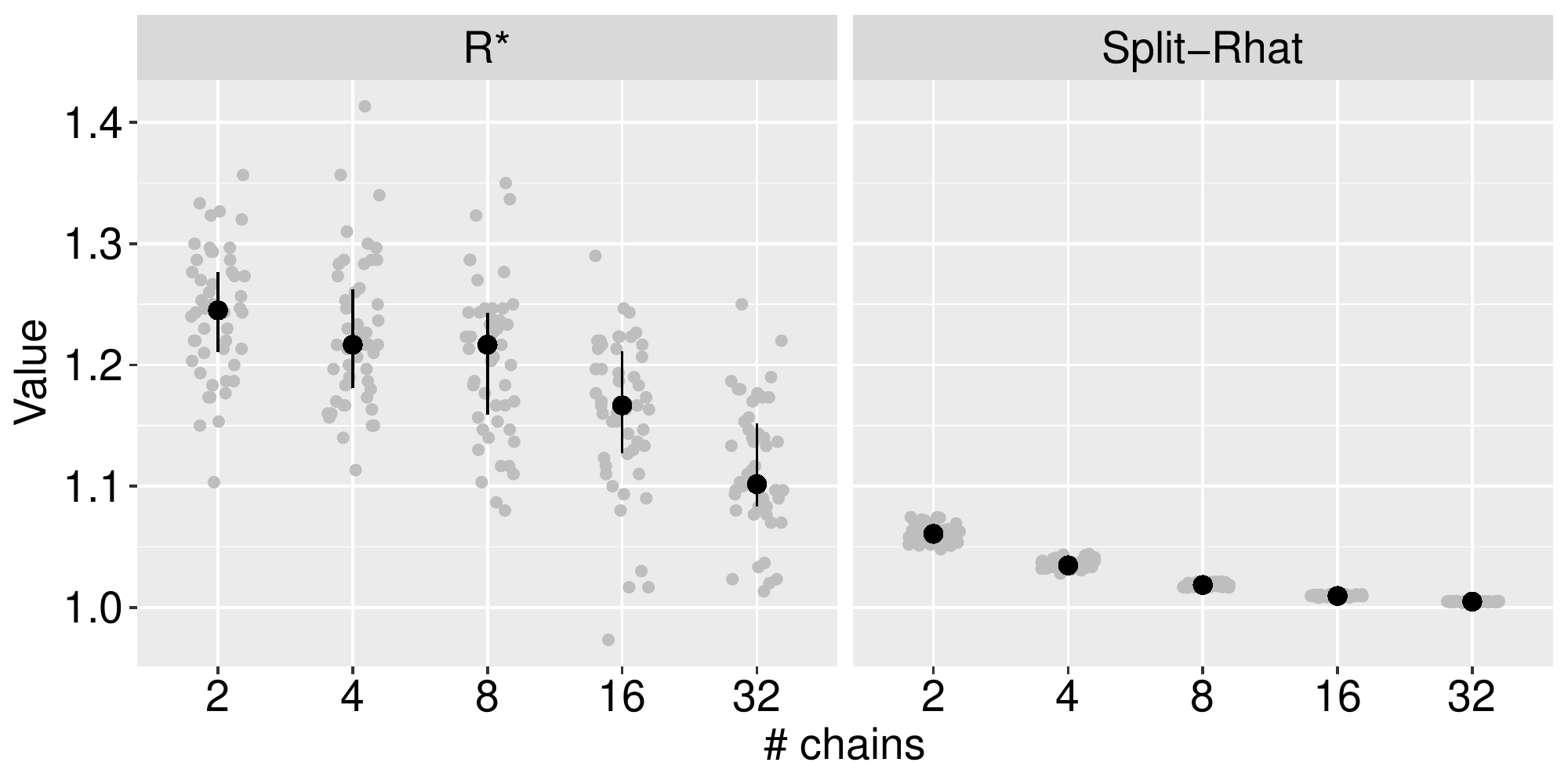}}
		\caption{\textbf{Autoregressive example: sensitivity to number of chains.} The horizontal axis shows the number of chains used in the data generating process described in \S\ref{sec:heterogeneity_numchains}. The vertical axis shows the value of $R^*$ as calculated by Algorithm \ref{alg:R_star} (left panel) on chains split into two halves, and rank-normalised split-$\widehat{R}$ (right panel). Grey points indicate the values of both convergence measures calculated for each replicate; horizontal jitter has been added to points. The point-ranges shown indicate the 25\%, 50\% and 75\% quantiles across 50 replicates at each number of chains.}
		\label{fig:ar1_numchains}
	\end{figure}

	In Fig. \ref{fig:ar1_numchains}, we show the results of these simulations with 50 replicates at each number of chains. On the horizontal axis, we show the number of chains, and on the vertical axis, the value of each of the two convergence measures ($R^*$ in the left panel; $\widehat{R}$ in the right panel). In general, both measures decline with chain count. For $R^*$, this may be because it is harder to classify chains when there are more of them. For $\widehat{R}$, this is because between-chain variance becomes relatively lower to that within them when there are more chains and only one of them differs in its marginal distribution. Across the replicates we ran, median $\widehat{R}<1.01$ for 16 or more chains; a minimum of $R^*=1.10$ was obtained for 32 chains.
	
	The decline of both of these measures when more chains are used hints that perhaps a moving threshold for diagnosing convergence may be pertinent to avoid neglecting those minority of chains with differing information. Here, however, we do not make suggestions on what such guidelines could be and leave this for later work.
	
	\color{black}
	\subsection{Diagnosing convergence in joint distributions: multivariate normal models}\label{sec:multivariate_normal}
	In this section, we illustrate how $R^*$ can diagnose convergence issues in the joint target distribution.
	
	\subsubsection{Bivariate model}\label{sec:multivariate_normal_bivariate}
	First, we consider a bivariate normal density. In all four chains, we use independent sampling to generate 2000 draws from bivariate normal densities with means of zero; in three of these chains, the covariance matrix is an identity matrix; in one chain, the covariance matrix also has unit diagonal terms but has off-diagonal terms of 0.9, indicating strong covariance between the two dimensions. By construction, all chains target the same marginal distribution in each dimension, but the fourth chain has a different joint distribution.
	
	First, we use the code provided in \cite{vehtari2019rank} to calculate rank-normalised $\widehat{R}$ and two different ESS measures that aim to capture how well certain regions of the posterior have been explored: these are known as bulk-ESS and tail-ESS. In all cases, the various quantities were calculated based on chains split into halves. For both dimensions, the two ESS measures were above 7000, and $\widehat{R}<1.001$, indicating no issues with convergence.
	
\begin{figure}[!htb]
		\centerline{\includegraphics[width=0.6\textwidth]{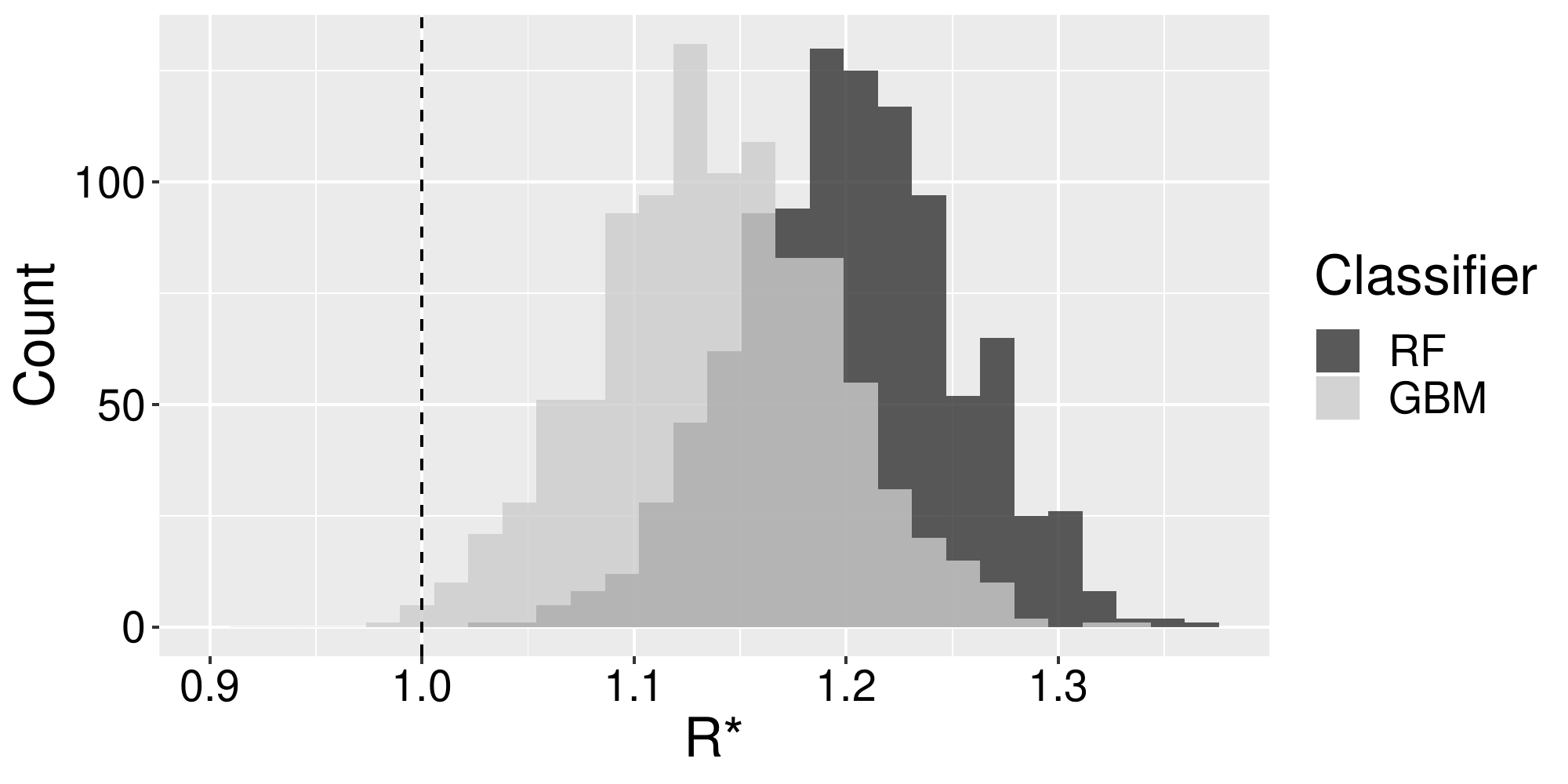}}
		\caption{\textbf{Bivariate normal example.} The distribution for $R^*$ across 1000 draws as calculated using Algorithm \ref{alg:R_star_uncertainty} for both the GBM and RF classifiers.}
		\label{fig:bivariate}
	\end{figure}

	Next, we estimate the $R^*$ distribution using Algorithm \ref{alg:R_star_uncertainty} using both GBM and RF classifiers. These distributions are shown in Fig. \ref{fig:bivariate}. The mean of the GBM-$R^*$ distribution is 1.14, and $>$99\% of $R^*$ draws are above 1; the mean of the RF-$R^*$ distribution was 1.27 and all draws were above 1. Collectively, these measures indicate that the sampling distribution has not converged. By taking account of all the information in the chains, $R^*$ is able to probe issues in joint distribution convergence which are missed by measures that consider only marginals.
	
	\subsubsection{250-dimensional model}\label{sec:multivariate_normal_250}
	We next consider a more challenging problem -- a 250-dimensional multivariate normal target where its precision matrix, $\boldsymbol{A}\in\mathbb{R}^{250}\times\mathbb{R}^{250}$, is generated from a Wishart distribution \citep{hoffman2014no}. We assume that the Wishart distribution's degrees of freedom is 250, resulting in a distribution with high correlations between dimensions. We use Stan's NUTS algorithm \citep{betancourt2017conceptual} to sample from this target distribution and run the algorithm for two different iteration counts (each time across 4 chains): 400 and 10,000 (the latter thinned by a factor of 5). First, we used Stan to sample from the ``centered'' parameterisation of this model, which is of the form,
	\begin{equation}\label{eq:mvt_normal_250}
	\boldsymbol{x}\sim \mathcal{N}(\boldsymbol{0},\boldsymbol{A}^{-1}),
	\end{equation}
	where $\boldsymbol{x}\in\mathbb{R}^{250}$. For each set of draws, we used Algorithm \ref{alg:R_star_uncertainty} with a GBM classifier to generate an uncertainty distribution for $R^*$, which is shown in Fig. \ref{fig:mvt}A (the equivalent plot for a RF classifier is similar and shown in Fig. \ref{fig:mvt_gbm_vs_rf}). From the plot for the 400 iteration case, it is clear that convergence has not yet occurred since $R^*>1$ across the bulk of this distribution. Even in the 10,000 iteration case, the $R^*$ distribution remains stubbornly shifted a little rightwards of $R^*=1$ (its mean is 1.06): in this case, $\widehat{R}<1.01$ for all parameters (Fig. \ref{fig:mvt}B), although 54\% had bulk-$\text{ESS}<400$ and 13\% of parameters had tail-$\text{ESS}<400$ indicating issues with convergence \citep{vehtari2019rank}.
	
	Rather than run the MCMC sampler for more iterations, we move to a ``non-centered'' parameterisation, which introduces auxillary variables $\boldsymbol{z}\in\mathbb{R}^{250}$ that don't affect $p(\boldsymbol{x})$ but facilitate sampling from it. This model has the form,
	\begin{align}
	\boldsymbol{A}^{-1} = \boldsymbol{L}\boldsymbol{L}^T,\qquad
	\boldsymbol{x} = \boldsymbol{L} \boldsymbol{z},\qquad
	z_j\sim \text{normal}(0, 1), \text{ for } j = 1,2,...,250.
	\end{align}
	where $\boldsymbol{L}$ is the Cholesky decomposition of the covariance matrix, $\boldsymbol{A}^{-1}$. Fig. \ref{fig:mvt}A shows the $R^*$ distribution resultant from 10,000 NUTS iterations in this case: now the distribution has mean $R^*=1.00$. Fig. \ref{fig:mvt}B shows the $\widehat{R}$ values for each $x$ parameter in this model, and, echoing the result for $R^*$, $\widehat{R}<1.01$ in all cases; further, bulk- and tail-$\text{ESS}>400$ for all parameters.
	
	\subsubsection{Variable importance}\label{sec:multivariate_normal_varimportance}
\begin{figure}[!htb]
		\centerline{\includegraphics[width=1\textwidth]{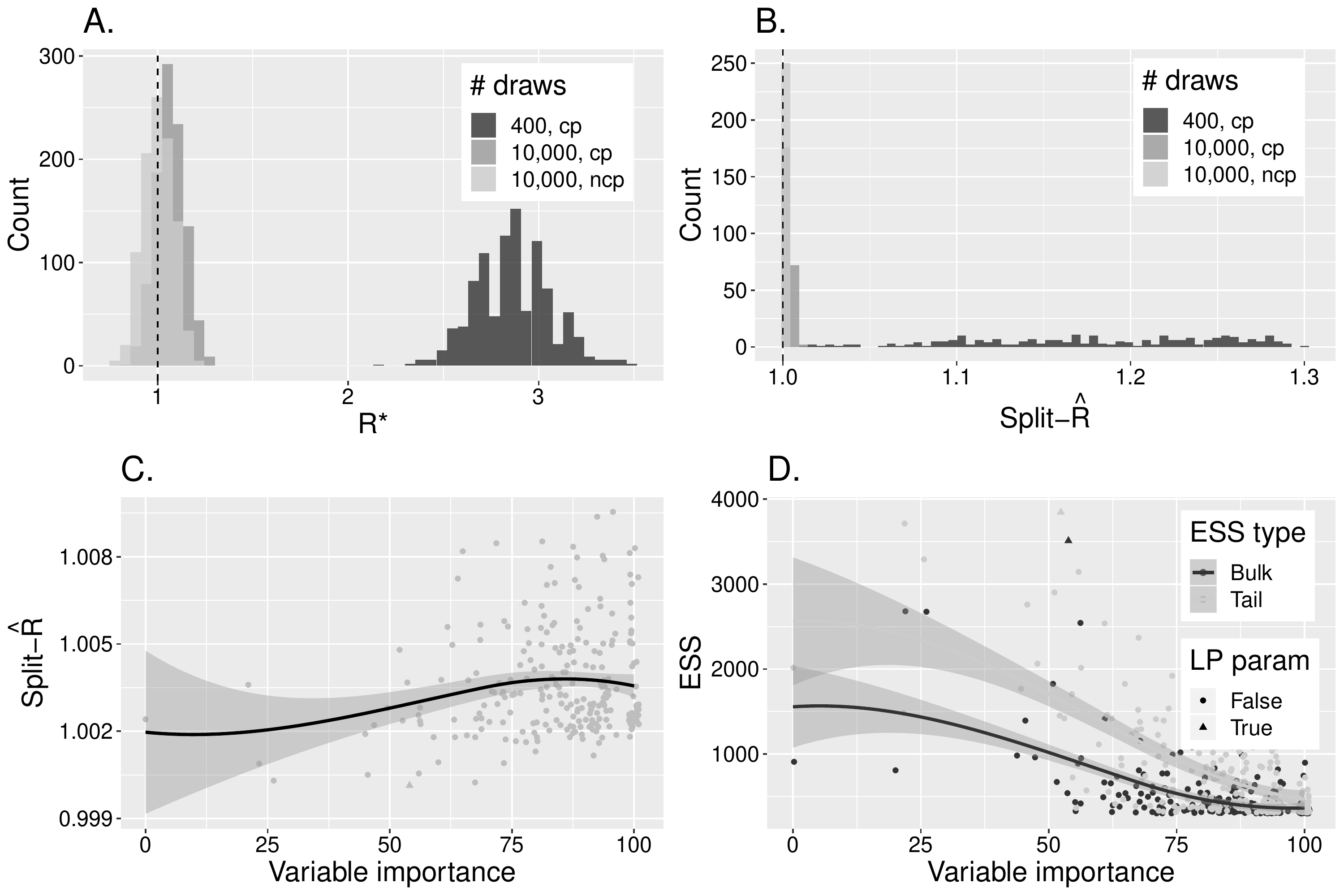}}
		\caption{\textbf{Multivariate normal example with 250 dimensions.} A shows $R^*$ distributions obtained for two MCMC samples (of differing numbers of draws: 400 and 10,000) from the centered parameterisation (``cp'') and one from the non-centered version (``ncp''; with 10,000 draws); B shows the rank-normalised split-$\widehat{R}$ values for all parameters from the same MCMC runs as in A; C shows variable importance versus $\widehat{R}$ for each parameter; and D shows variable importance versus bulk- and tail-ESS as calculated by \cite{vehtari2019rank}. In A, 1000 $R^*$ draws by Algorithm \ref{alg:R_star_uncertainty} are shown for each MCMC run. In plots C and D, horizontal jitter was added to the points and a loess fit line with standard errors overlaid.}
		\label{fig:mvt}
	\end{figure}

	In GBMs, it is possible to calculate variable importance (see, for example, \cite{friedman2001greedy} and \cite{greenwell2019package}), which allows us to determine which variables were most informative for predictions. We now compare these with the more established metrics $\widehat{R}$ and ESS. For a GBM fitted to the centered model of eq. \eqref{eq:mvt_normal_250} with 10,000 MCMC iterations (thinning by a factor of 5) for each chain, we plot in Fig. \ref{fig:mvt}C variable importance (here high values mean a variable is more important) versus $\widehat{R}$ for all dimensions of the target distribution (including Stan's $lp$ quantity, shown as a triangle). In this plot, there is a positive association between GBM's variable importance and $\widehat{R}$ (Spearman's rank correlation: $\rho=0.17, S=2185680, p<0.01$). In Fig.  \ref{fig:mvt}D, we plot variable importance versus two measures: bulk-ESS and tail-ESS, which both exhibited a strong non-linear negative association (Spearman's rank correlation: bulk-ESS: $\rho=-0.57, S=4142470, p<0.01$; tail-ESS: $\rho=-0.56, S=4113709, p<0.01$). Since none of these plots form perfect ``lines'' along which all the plotted points fall, this illustrates that variable importance provides information complementary to $\widehat{R}$ and ESS.
	
	\subsection{Infinite variance: Cauchy example}\label{sec:cauchy}
	We next explore how $R^*$ can be used to determine convergence for distributions with infinite variance. Like \cite{vehtari2019rank}, we first use Stan to sample from independent standard Cauchy distributions for each element of a 50-dimensional vector $x$,
	\begin{equation}
	x_j\sim \text{Cauchy}(0, 1),\; \text{for } j=1,...,50.
	\end{equation}
	We call this parameterisation the ``nominal'' version of this model.
	
	In addition, we also use Stan to sample from an ``alternative'' parameterisation of the Cauchy, based on a scale mixture of Gaussians \citep{vehtari2019rank},
	\begin{align}
	a_j \sim  \text{normal}(0,1), \qquad
	b_j \sim  \text{Gamma}(0.5, 0.5), \qquad
	x_j =  a_j/\sqrt{b_j}.
	\end{align}
	The distribution of the $x$ vector is the same under both parameterisations, although the thin-tailed $(a,b)$ vectors define a higher dimensional posterior that improves sampling efficiency.
	
\begin{figure}[!htb]
		\centerline{\includegraphics[width=1.0\textwidth]{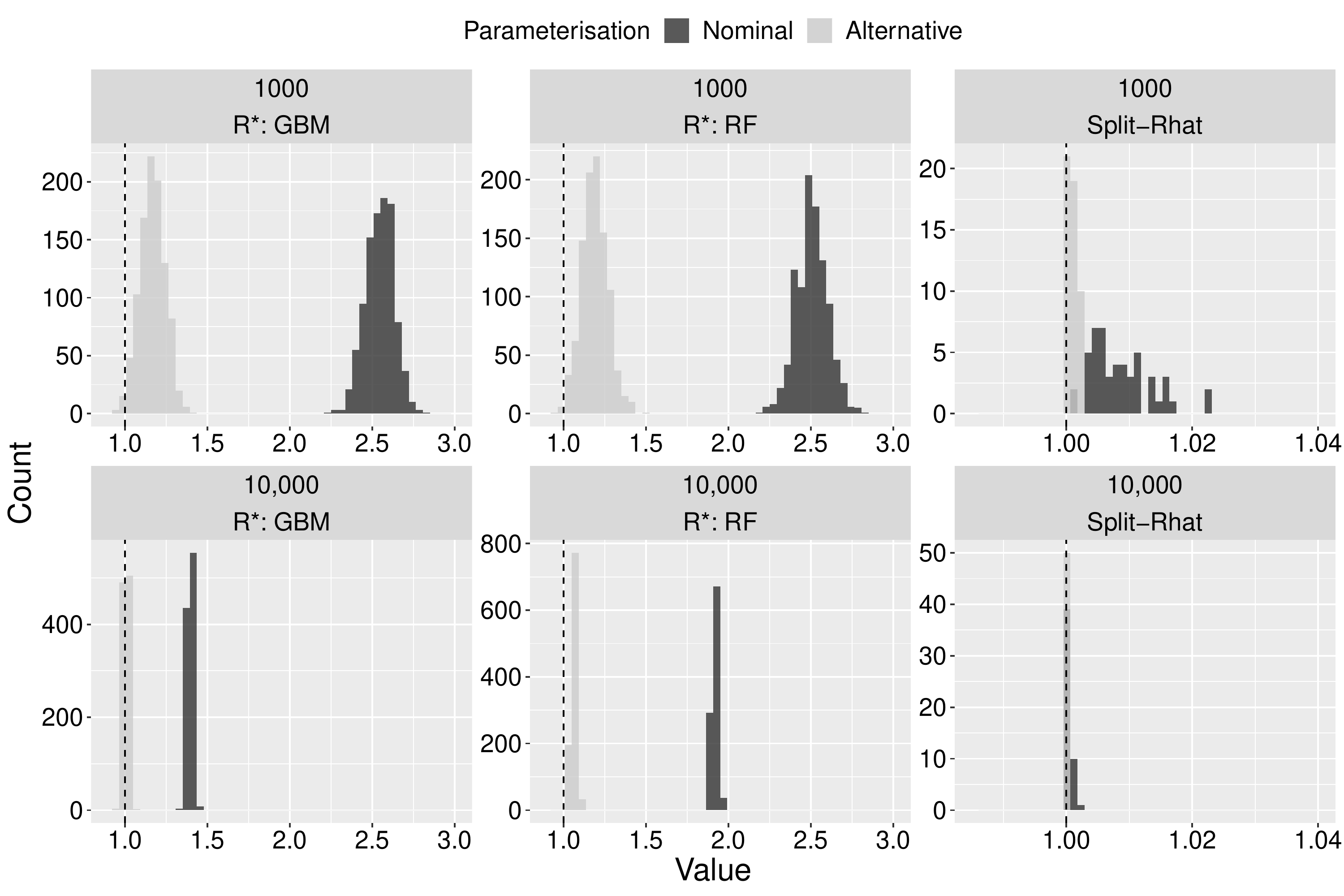}}
		\caption{\textbf{Cauchy example.} Rows show convergence results for MCMC runs with 1000 (top) and 10,000 (bottom; obtained by thinning iterations by a factor of 3) post-warm-up iterations (each with half iterations discarded as warm-up) for each of 4 chains. Columns show the $R^*$ distributions from GBM (left) and RF (middle) and rank-normalised split-$\widehat{R}$ values across all parameters (right). Shadings indicate different model paramerisations as indicated in legend.}
		\label{fig:cauchy}
	\end{figure}

	In the top-left and top-middle panel of Fig. \ref{fig:cauchy}, we show the $R^*$ distribution for GBM and RF classifiers under both parameterisations. As shown in \cite{vehtari2019rank}, the nominal parameterisation results in poor sampling efficiency due to its long tails, meaning that, after 1000 MCMC post-warm-up iterations (with 1000 warm-up iterations discarded) across each of 4 chains, draws still contain information about chain identity, and, accordingly, the $R^*$ distribution is shifted rightwards from $R^*=1$. The alternative parameterisation fares better, and the $R^*$ distribution is nearer $R^*=1$, yet its mean remains above this value. In the top-right panel of Fig. \ref{fig:cauchy}, we show the rank-normalised split-$\widehat{R}$ values across each of the 50 parameters for the same MCMC runs. The nominal parameterisation has some parameters with $\widehat{R}>1.01$, indicative non-convergence, whereas the alternative has $\widehat{R}<1.01$ for all parameters.
	
	Since the $R^*$ distribution indicated non-convergence for both parameterisations, we ran each model for sixty-times as long, although thinned by a factor of 3, resulting in 10,000 post-warm-up iterations across each of 4 chains. In the bottom row of Fig. \ref{fig:cauchy}, we show the results for these longer runs. In these, the alternative parameterisation now has an $R^*$ distribution centred on $R^*=1$ for the GBM classifier although the RF classifier $R^*$ distribution remains slightly rightwards of this target indicating that convergence is nearer but more iterations are likely still required. Despite the added iterations, the $R^*$ distribution from the nominal model remains stubbornly away from 1. The $\widehat{R}$ values are all below 1.01, indicating convergence in both cases.
	
	\subsubsection{Measuring convergence objectively}\label{sec:cauchy_objective}
	To illustrate that $R^*$ provides a reliable metric for capturing convergence, we now calculate a quantitative measure that captures how closely a sampling distribution matches the target. One measure of distributional ``closeness'' is the KL-divergence, which, in this case, could be used to measure the divergence from target to sampling distribution: if the target distribution is known, fitting a kernel density estimator (KDE) to samples allows an approximate (typically univariate) measure of KL-divergence to be calculated for each dimension. The trouble is, for distributions like the Cauchy with fat tails, fitting a KDE to the samples provides a noisy measure of the sampling distribution in the tails. This means that approximate KL-divergence is unreliable for these types of model. We decided not to use the Kolmogorov-Smirnov (KS) test, since it is most sensitive to differences between distributions around the median, whereas, here, we are interested in behaviour in the tails. Additionally, we found that the Anderson-Darling and Cram\'er-Von Mises tests \citep{faraway2019goftest}, which do not suffer the same shortcomings as the KS, behaved equally erratically and provided measures that were hard to intuit. The Wasserstein distance was also trialled but had great uncertainty due to the long-tails of the Cauchy. Instead, we chose a measure of distributional discrepancy based around similarity between target quantiles and sample-estimated equivalents. Specifically, we calculate the $R^2$ for the linear regression of actual quantile values on sample-estimated quantiles, where, if $R^2\sim 1$, the sampling distribution recapitulates well the target quantities. In our example, we consider all percentiles: 0.1\%, 0.2\%,...,99.8\%, 99.9\% and calculate the mean $R^2$ across all 50 dimensions.
	
\begin{figure}[!htb]
		\centerline{\includegraphics[width=1.0\textwidth]{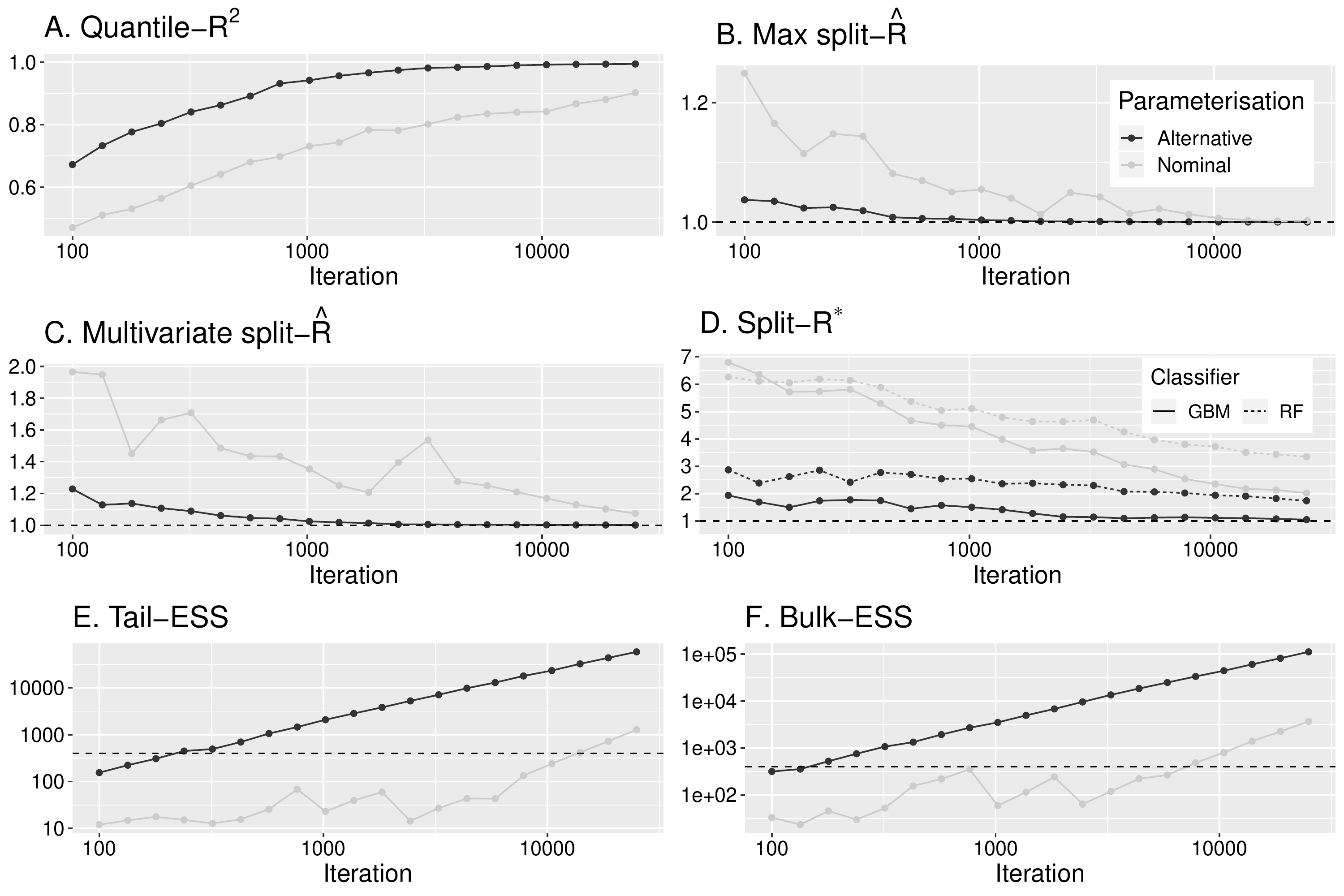}}
		\caption{\textbf{Measuring convergence for the Cauchy model.} A shows a measure of convergence, the mean quantile $R^2$, that requires knowing the target distribution; B shows the maximum value of split-$\widehat{R}$ across each of the 50 dimensions of the target; C shows the multivariate split-$\widehat{R}$ value; D shows the value of split-$R^*$ as calculated by Algorithm \ref{alg:R_star} for both the GBM and RF classifiers; and E and F show tail- and bulk-ESS. Dashed lines indicate recommended thresholds for each convergence statistic.}
		\label{fig:cauchy_convergence}
	\end{figure}

	In Fig. \ref{fig:cauchy_convergence}A, we plot this \textit{quantile-$R^2$} as a function of MCMC sample size for both parameterisations of the Cauchy model. This shows that after c.10,000 iterations, the alternative parameterisation approaches $R^2\approx 1$; at the same number of iterations, the nominal parameterisation still provides a poor measure of tail quantiles. Next, in Figs. \ref{fig:cauchy_convergence}B\&C, we plot two measures of $\widehat{R}$, each calculated from splitting the 4 original chains into two equal halves. The first of these measures is the rank-normalised $\widehat{R}$ \citep{vehtari2019rank}, which provides a separate measurement for each target dimension; in Fig. \ref{fig:cauchy_convergence}B, we show how the maximum of this measurement across all 50 dimensions changes with sample size. After c.500 iterations, the alternative parameterisation achieves $\widehat{R}<1.01$ for all target dimensions, and, after c.10,000 iterations, the nominal model achieves the same maximum $\widehat{R}$ value: in both cases, these suggest convergence. The second measure is multivariate $\widehat{R}$ \citep{brooks1998general}, which, like $R^*$, yields a single measurement across all dimensions; Fig. \ref{fig:cauchy_convergence}C shows how this metric changes with sample size for both Cauchy model parameterisations. After c.1800 iterations, multivariate $\widehat{R}<1.01$ for the alternative parameterisation, whilst after 25,000 iterations multivariate $\widehat{R}>1.07$ indicating more draws are needed. In Fig. \ref{fig:cauchy_convergence}D, we plot $R^*$ against iteration for both models and for both GBM and RF classifiers: these indicate that, after 25,000 iterations, for the alternative model, $R^*\approx 1.05$ for the GBM classifier and $R^*\approx 1.74$ for the RF classifier; for the nominal model, $R^*>2$ for the GBM classifier $R^*>3$ for the RF classifier: all these $R^*$ values suggest lack of convergence. Finally, in Figs. \ref{fig:cauchy_convergence}E\&F, we plot the minimum across all the dimensions of tail- and bulk-ESS calculated as described in \cite{vehtari2019rank}. After c.180 iterations, the alternative parameterisation surpassed a tail-ESS of 400; after c.18,700, the nominal parameterisation did the same. Both models were quicker to pass 400 bulk-ESSs.
	
	Comparing our measure of convergence that requires knowing the actual target distribution (\textit{quantile-$R^2$}; in Fig. \ref{fig:cauchy_convergence}A), with the various heuristic measures, all show a similar pattern: as sample size increases, the various statistics tend towards convergence. The rate at which these converge differs though, and $R^*$ (Fig. \ref{fig:cauchy_convergence}D) appears at least, qualitatively, most similar to \textit{quantile-$R^2$}.
	
	\subsection{Hierarchical model: Eight schools model}\label{sec:eight_shools}
	We now examine a classic example used to highlight difficulties in performing inference for hierarchical models: referred to as the ``Eight schools'' model (see Section 5.5 in \cite{gelman2013bayesian}), which aimed to determine the effects of coaching on SAT scores in eight schools. 
	
	The model can be parameterised in two ways, as described in \cite{vehtari2019rank} (and introduced in \cite{van2001art}). The simplest way is referred to as the ``centered'' parameterisation and exactly mirrors the underlying statistical model,
	\begin{align*}
	\theta_j &\sim \text{normal}(\mu, \tau), \\
	y_j &\sim \text{normal}(\theta_j, \sigma_j).
	\end{align*}
	The ``non-centered'' parameterisation (first introduced in \cite{van2001art}) recodes this model in a way that does not affect the joint distribution of $(\theta, \mu, \tau, \sigma)$ but makes it easier to sample from it, by introducing auxillary variables, $\tilde \theta_j$. This can be written as,
	\begin{align*}
	\tilde{\theta}_j &\sim \text{normal}(0, 1), \\
	\theta_j &= \mu + \tau \tilde{\theta}_j,\\
	y_j &\sim \text{normal}(\theta_j, \sigma_j).
	\end{align*}
	In both cases, $\theta_j$ are the treatment effects in the eight schools, and $(\mu, \tau)$ represent the population mean and standard deviation of the distribution of these effects. In the centered parameterization, the $\theta_j$ are parameters, whereas in the non-centered parameterization, the $\tilde{\theta}_j$ are parameters and $\theta_j$ is a derived quantity.
	
	We first used Stan \citep{carpenter2017stan} to sample from the centered model using 4 chains. Like \cite{vehtari2019rank}, we used settings that reduce the chance of divergent iterations for the dynamic HMC algorithm \citep{betancourt2017conceptual} (called using the ``NUTS'' option in Stan), meaning that the resultant sampling distribution is likely to be biased. We also used the same algorithm settings to sample from the non-centered model.
	
\begin{figure}[!htb]
		\centerline{\includegraphics[width=1.0\textwidth]{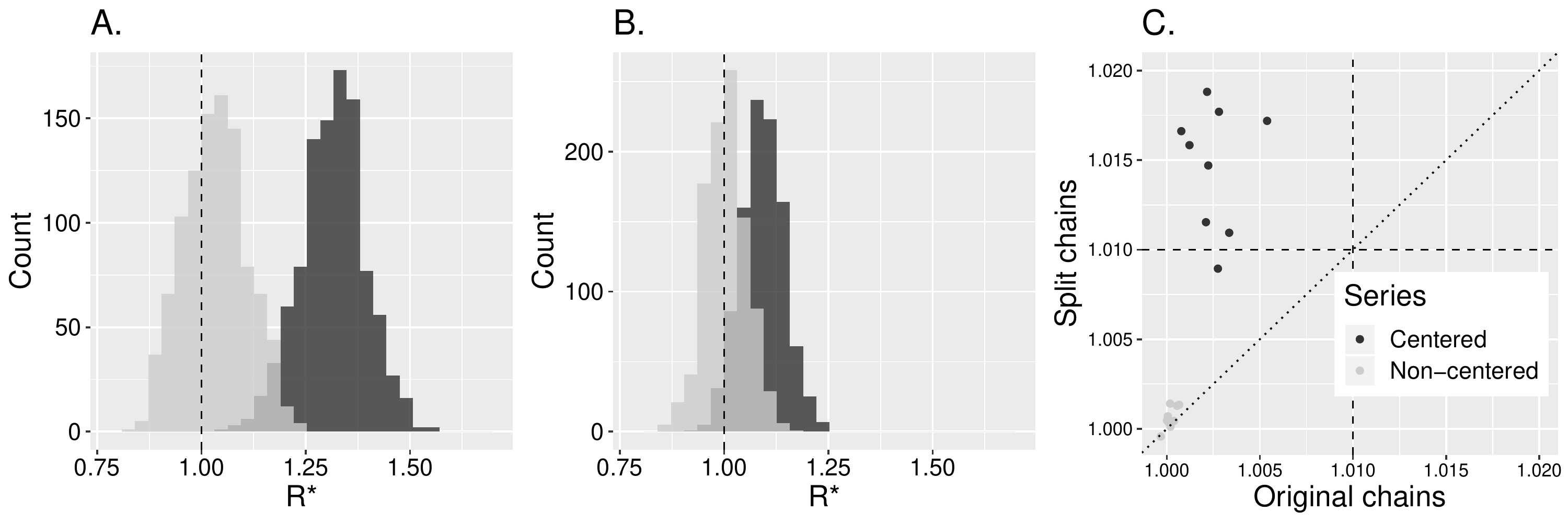}}
		\caption{\textbf{Eight schools example: $R^*$ distributions.} A shows draws from the $R^*$ distribution when splitting chains in two (resulting in 8 chains); B shows the same but using the 4 original chains; C shows rank-normalised $\widehat{R}$ for original 4 chains versus those for the 8 chains case for all ten parameters defined by the centered model -- in this case, we plot horizontal and vertical dashed lines to illustrate the $\widehat{R}=1.01$ cutoff and a $y=x$ line. The legend inset in panel C provides a key for all panels. The MCMC samples comprised 2000 draws in all cases, with 1000 used as post-warm-up iterations. In panels A and B, the plots show 1000 $R^*$ draws using Algorithm \ref{alg:R_star_uncertainty} using a GBM classifier for each parameterisation.}
		\label{fig:eight_schools}
	\end{figure}

	To see how $R^*$ performed on this example, we first split each of the (post-warm-up) chains in two, as is done by default in Stan \citep{carpenter2017stan} and in \cite{vehtari2019rank}, resulting in 500 iterations across 8 chains. Following the same approach as in Algorithm \ref{alg:R_star_uncertainty}, we generated $R^*$ distributions for both the centered and non-centered models using a GBM classifier. The resultant distributions for $R^*$ are shown in Fig.\ref{fig:eight_schools}A. In this plot, the centered model is close to convergence, whereas the non-centered is not.
	
	In addition, to illustrate the power of $R^*$, we also repeat the analysis but, this time, do not split the chains in two. The results are shown in Fig.\ref{fig:eight_schools}B. In this case, because the unsplit chains do not mix with themselves, it is harder to accurately predict the chain that generated each draw, meaning that the centered model $R^*$ values are shifted leftwards. Despite this, however, the centered model distribution for $R^*$ still does not strongly overlap with $R^*=1$, indicating that the model has not converged, contrasting with the non-centered model which appears near convergence.
	
	Fig. \ref{fig:eight_schools_rf} shows the equivalent of Fig. \ref{fig:eight_schools} except using a RF classifier. The results are similar, although the $R^*$ distributions are shifted slightly rightwards: indicating, for example, that more than 2000 draws from the non-centered model may be required for convergence.
	
	It is recommended that $\widehat{R}$, like $R^*$, be calculated using split chains. In Fig. \ref{fig:eight_schools}C, we plot $\widehat{R}$ values obtained when using the original 4 chains (horizontal axis) versus those when using the split chains (vertical axis) for the ten parameters in this model; we do this for both the centered and non-centered models. These show that the values of $\widehat{R}$ for the centered model using the unsplit chains were below 1.01; when using the chains split into two halves, $\widehat{R}>1.01$ for all but a single parameter. All parameters for the non-centered models were below 1.01, indicating convergence.
	
	\subsubsection{Understanding chain classification}
\begin{figure}[!htb]
		\centerline{\includegraphics[width=1.0\textwidth]{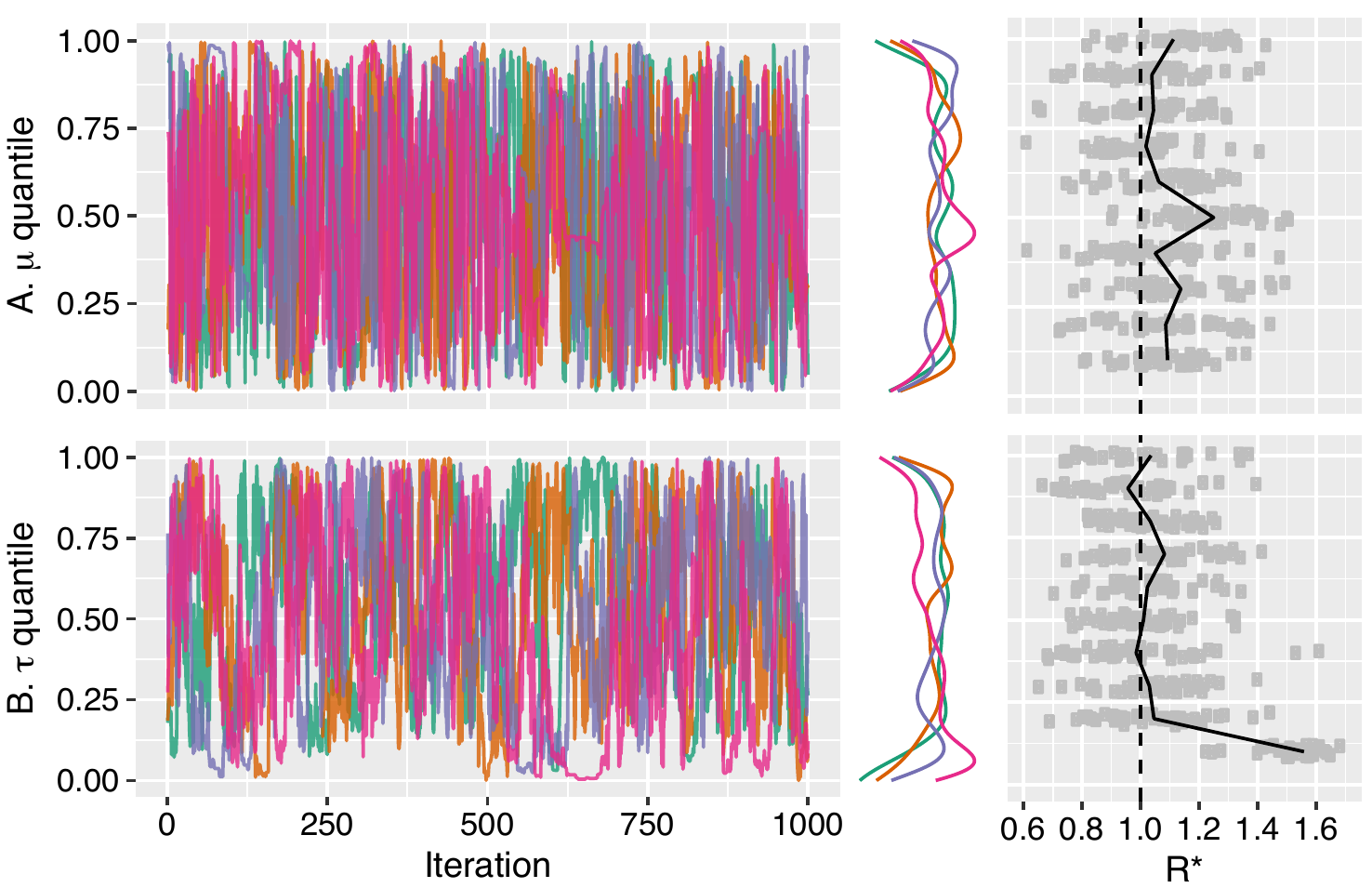}}
		\caption{\textbf{Eight schools example: quantile $R^*$ plots.} Row A shows plots for $\mu$; Row B for $\tau$. In each row, we show the path of the four individual chains above and 40 $R^*$ draws obtained using Algorithm \ref{alg:R_star_uncertainty} for each parameter quantile below. To the right of each trace plot, we show the marginal distribution of each chain estimated via kernel density estimation using Gaussian kernels. Note that, in the right-hand plots, jitter has been added to the data points.}
		\label{fig:eight_schools_r_star_quantiles}
	\end{figure}

	To probe the predictive power of the ML classifier, we investigated how predictive accuracy varies across parameter space. After fitting the GBM model, we group MCMC draws in the test set into deciles and draw from the $R^*$ distribution for each decile. In Fig. \ref{fig:eight_schools_r_star_quantiles}, we show the results of this exercise for (A) $\mu$ and (B) $\tau$. In the left-hand column of this figure, we show the path of four MCMC chains (here we did not split chains when calculating $R^*$ to simplify visualisations) across the quantiles of each parameter space. To the right of each trace plot, we show the marginal distributions for each chain. In the right-hand column, we show 40 $R^*$ draws for each decile, which were generated according to Algorithm \ref{alg:R_star_uncertainty} using a GBM fit to all draws. In essence, the left-hand panels explain the variation in $R^*$ in the right-hand panels: if chains become stuck in regions of parameter space, this causes differences between the marginal distributions of the chains; these differences, in turn, allow a ML model to predict the generative chain in those same sticky regions. For example, for $\mu$, the purple chain became stuck around the middle quantile, forcing a difference in its marginal distribution in that region, which resulted in $R^*>1$ for the corresponding decile. Similarly, for $\tau$, the purple chain became stuck in the lowest quantiles, elevating its marginal distribution there and resulting in improved predictive accuracy.
	
	Fig. \ref{fig:eight_schools_r_star_quantiles} also indicates a potential limitation of $R^*$: namely, that as chains are progressively thinned, those regions where chains behave most idiosyncratically can be missed, resulting in a reduction in classification accuracy and falsely concluding that convergence has occurred.
	
	\subsection{Further experiments}\label{sec:further_experiments}
	Alongside the examples included in the main text, there are a number of supplementary text examples, which we briefly outline here.
	
	In \S\ref{sec:wide}, we illustrate how $R^*$ can provide a reasonable measure of convergence when the number of dimensions of a distribution is comparable to the number of draws. Specifically, this was to test that classification didn't become prone to overfitting in this limit. To test this hypothesis, we investigated two scenarios using a multivariate normal target: one with a 250-multivariate normal with high correlation between dimensions using 250 post-warm-up iterations; and another normal with 10,000 independent dimensions using up to 500 post-warm-up iterations. In both cases, sampling was done using Stan's NUTS algorithm. In both cases, $R^*$ and rank-normalised split-$\widehat{R}$ reached similar conclusions about convergence: namely, that more iterations were needed in all experiments considered. Overall, these experiments show that $R^*$ is a conservative convergence measure that will tend to diagnose unconvergence when there are insufficient draws.
	
	In \S\ref{sec:non-stationary}, we illustrate the importance of splitting chains before calculating $R^*$ to ensure poor within-chain convergence is diagnosed. We illustrate this via four examples: (a) sampling from a univariate normal and adding a linear trend over sampling time, to ensure that the sampling distributions were non-stationary; (b), similar to (a) but across a range of target distribution dimensions where only a single dimension had a non-stationary mean; (c), a bivariate normal with a non-stationary covariance; and (d), an autocorrelated sampling distribution with a univariate normal target with a range of different autocorrelations. The results of (a) echoed those presented in \cite{vehtari2019rank} for $\widehat{R}$ and showed that $R^*$ is insensitive to sampling non-convergence if it occurs within chains; splitting chains into two halves alleviates this issue. The results of (b) show that $R^*$ calculated on split chains is able to diagnose non-stationarity in mean in a single dimension in a way that did not diminish as the numbers of dimensions considered increased. Example (c) showed that split-$R^*$ opposed to split-$\widehat{R}$ is able to diagnose non-stationary covariance between dimensions of a target distribution. In (d), we show that $R^*$ is able to differentiate between distributions with non-stationary target distributions and stationary ones. It also shows that $R^*$ still functions reasonably at higher levels of chain persistence: yielding a conservative convergence measure when there are insufficient draws.
	
	In \S\ref{sec:prostate}, we show that $R^*$ performs well for two Bayesian logistic regression problems with highly multimodal posteriors. Each of these models have 1000s of parameters, and we found that it was slow to compute both $\widehat{R}$ and $R^*$ for them. That said, the computational time for calculating $\widehat{R}$ was considerably less than was needed for $R^*$.
	
	In \S\ref{sec:discrete}, we evaluate $R^*$ on univariate discrete examples: one with four states (in \S\ref{sec:discrete_small}); another, with a larger state-space consisting of 20 states (in \S\ref{sec:discrete_large}). In these examples, we use a discrete Markov model to generate draws from a given target. The small and larger state-space cases, show that $R^*$ behaves as expected: given a sufficient sample size, it is able to detect differences in the transition probability matrix between chains that result in differences in the target distribution.
	
	In \S\ref{sec:ML_sensitivity}, we investigate how two decisions about classifiers — which classifier to use (in \S\ref{sec:ml_model}) and what hyperparameters to use for it (in \S\ref{sec:hyperparameters}) — affect calculation of $R^*$. In \S\ref{sec:ml_model}, we test a range of popular classifiers: GBMs, RFs, k-nearest-neighbour models, support vector machines and generalised linear models across examples. This indicated that GBMs and RFs consistently had the highest classification accuracy across the examples: in higher dimensional problems, RFs tended to best GBMs. In \S\ref{sec:hyperparameters}, we show how these two best classifiers — GBMs and RFs — depend on their hyperparameters. Across the examples we test, GBMs are more sensitive to hyperparameter choices than are RFs. Additionally, GBMs require typically more rounds of boosting in higher dimensional problems, leading to more extensive algorithm runtime. Because of this, we suggest fixing GBM's hyperparameters to values that result in reasonable performance and reasonable runtime. For RFs, we suggest using a heuristic for choosing its hyperparameters that was derived in a previous empirical evaluation of RFs \citep{bernard2009influence}.
	
	In \S\ref{sec:comparison_gbm_rf}, we use a slew of examples to compare $R^*$ calculated using GBM and RF classifiers using our suggested default hyperparameter sets (given in \S\ref{sec:method}). In \S\ref{sec:joint_distribution}, we compare how both methods are able to diagnose lack of convergence in a joint distribution (using a multivariate normal target). In \S\ref{sec:tail_fatness}, we compare the ability of both approaches to diagnose differences in the tails of the marginal distributions between chains (using Student-t targets). In both examples, draws are generated by independent sampling meaning that an optimal $R^*$ can be calculated based on the Bayes optimal classifier (see, for example, \citep{devroye2013probabilistic} and \S\ref{sec:comparison_gbm_rf} for more information). Collectively, the results of \S\ref{sec:joint_distribution} and \S\ref{sec:tail_fatness} suggest that both GBM and RF classifiers can detect differences in sampling distributions, should they exist, between chains. The classification rates achieved by these two approaches exhibited similar trends to the optimal classifier, albeit with lower predictive accuracy. In higher dimensions, it is likely that the difference between optimal classification rates and those from the GBM or RF will increase: particularly, when searching for between-chain differences in tail fatness. These results also suggest a different region of optimality for each classifier: GBMs tend to perform best for low dimensional targets and RFs for moderate-high ones. This is a function of the different rules used to set the hyperparameters of each classifier (see \S\ref{sec:method} and \S\ref{sec:hyperparameters}): for GBMs to perform well in higher dimensions, they need more rounds of tree boosting which substantially increases training time. Because of this, we fixed the hyperparameters of the GBM to values that yield reasonable computation time. RFs are less sensitive to hyperparameter variation (\S\ref{sec:hyperparameters}), and useful heuristics for adapting these with target dimensions are known, which we follow here, as described in \S\ref{sec:method}. Despite this dynamic choice of hyperparameters for RFs, its runtime remained reasonable over the range of examples we test in this paper.
	
	\section{Discussion}
	If an MCMC sampler has converged on the target distribution, the chains must be well-``mixed'', that is, given a draw, it should be impossible to discern which chain generated it. Based on this observation, we used supervised machine learning (ML) classifiers to quantify the information about the generative chain identity contained in draws. By taking the ratio of model predictive accuracy obtained on an independent test set to the accuracy of a null model (which predicts a chain's identity uniformly at random), this defines our $R^*$ statistic. By extracting classifier-predicted chain probabilities from each prediction in the test set, we can additionally generate an uncertainty distribution for $R^*$. Across a range of previously published examples, $R^*$ was shown to be predictive of whether chains had converged.
	
	The predominant methods for diagnosing MCMC convergence rely heavily on looking for between-chain differences in the marginal distributions along each dimension of the target. $R^*$ naturally includes this information in building a model capable of predicting the chain that generated each draw. It also naturally includes information about the joint distribution across all dimensions of the target. Since converged chains should have similar joint distributions (implying similar marginals), any measure of convergence should account for both of these aspects. Indeed, in \S\ref{sec:multivariate_normal}, we show that more established measures may indicate convergence whereas $R^*$ shows otherwise. This indicates the complementarity of $R^*$ to existing measures.

	Different target distributions present different challenges to sampling. Because of this, there is not a unique optimal ML classifier across all cases: this is just a manifestation of the no free lunch theorem \citep{wolpert1997no}. Across a range of examples we tested (see \S\ref{sec:ml_model}), GBM and RF classifiers performed consistently well, and we then used these across all other illustrative examples. It is possible — indeed, likely — that another ML model may exist or be invented that consistently outperforms both these classifiers. We do not see this as a problem: across the examples we considered, $R^*$ calculated using both classifiers tended to provide a measure of convergence as or more stringent than existing diagnostics. In that sense, it is a step in the right direction. If a better classifier is found, the same apparatus we develop here can be used, and this will present a harsher test of convergence.
	
	A different question is, ``When is such a measure of convergence of practical use?''. In our examples, $R^*$ is able to diagnose poor convergence in the tails of marginal distributions: likely of practical relevance for many applications that require tail quantiles. $R^*$ is also able to diagnose lack of convergence in the joint distribution — even if the marginals appear converged. Indeed, it is less clear how the joint distribution can be unconverged when the marginals appear so, but we found examples where this appeared true. A consequence of poor convergence of the joint distribution would be for prediction and, by corollary, model comparison. Further work examining these consequences further is needed, and $R^*$ can help to identify and monitor fruitful candidate systems.

	In \S\ref{sec:prostate}, we fit Bayesian models with many 1000s of parameters then used $R^*$ to diagnose convergence, finding that $R^*$ was considerably more expensive to calculate than $\widehat{R}$. The time complexity of training RFs is thought to be $\mathcal{O}(m_{\text{try}} n_{\text{tree}} N \text{log} N)$ \cite[chapter~5]{louppe2014understanding}, and given the similarities with GBMs (which also builds many decision trees), it is likely to be similar. If so, this suggests that larger statistical models (usually needing more MCMC iterations) may currently be beyond the reach of $R^*$. That said, it is possible to reduce the runtime for $R^*$ using thinned draws (although this risks losing chain idiosyncracies) and using a subset of dimensions (although this risks losing problematic dimensions). Indeed, in \S\ref{sec:multivariate_normal_250}, \S\ref{sec:cauchy} and \S\ref{sec:prostate}, we use these strategies and, nonetheless, find that $R^*$ provides a stringent measure of convergence.
	
	Many implementations of $\widehat{R}$ suggest splitting chains in two before calculating it. In a number of examples, we trial this before calculating $R^*$ and find that this approach leads to more accurate chain prediction. We recommend that this practice be adopted whenever $R^*$ is calculated to ensure that this measure is maximised. Additionally, our non-parametric calculation method for $R^*$ makes it possible to include any covariates which may be useful features for prediction, such as an ``iteration block'' indicator variable taking values $1, 2, ..., K$ in each of $K$ blocks of contiguous iterations. If each chain is thoroughly mixed with itself, including this additional information shouldn't change $R^*$; by contrast, if the chains are random walk-like, this information should boost $R^*$.
	
	MCMC enables inference across a wide range of models encountered across the social, biological and physical sciences. Its ease of implementation, however, masks important underlying fragilities in the method. Namely, that unless the chains have converged to a truly stationary distribution, the draws generated are not faithful depictions of the posterior. In this paper, we introduce a new metric, $R^*$, that is especially good at diagnosing poor convergence in the joint sampling distribution -- an area that has received insufficient attention thus far. $R^*$ can straightforwardly be introduced into existing MCMC libraries and could provide a measure of convergence complementary to existing metrics.
	
	\section{Contributions}
	BL conceived of the original idea for $R^*$, carried out all analyses and wrote the original draft of the paper. AV reviewed the paper and made a host of recommendations: including suggesting a large range of new case studies on which to trial the method and also suggesting new visualisations for diagnosing how predictive performance depends on parameter quantiles; these collectively widened the scope of the original paper and substantially improved its quality. BL and AV reviewed and edited the draft of the paper.
	
	\section{Acknowledgements}
	The authors would like to thank the anonymous reviewers for comments on previous drafts of the paper that lead to significant improvements. We would also like to thank Paul B{\"u}rkner and Jonah Gabry, with whom useful discussions were had during the preparation of this manuscript.
	
	\bibliographystyle{chicago}
	\bibliography{bibliography} 

\begin{thebibliography}{}

\bibitem[\protect\citeauthoryear{Bernard, Heutte, and Adam}{Bernard
  et~al.}{2009}]{bernard2009influence}
Bernard, S., L.~Heutte, and S.~Adam (2009).
\newblock Influence of hyperparameters on random forest accuracy.
\newblock In {\em International Workshop on Multiple Classifier Systems}, pp.\
  171--180. Springer.

\bibitem[\protect\citeauthoryear{Betancourt}{Betancourt}{2017}]{betancourt2017conceptual}
Betancourt, M. (2017).
\newblock A conceptual introduction to {H}amiltonian {M}onte {C}arlo.
\newblock {\em arXiv preprint arXiv:1701.02434\/}.

\bibitem[\protect\citeauthoryear{Bingham, Chen, Jankowiak, Obermeyer, Pradhan,
  Karaletsos, Singh, Szerlip, Horsfall, and Goodman}{Bingham
  et~al.}{2019}]{bingham2019pyro}
Bingham, E., J.~Chen, M.~Jankowiak, F.~Obermeyer, N.~Pradhan, T.~Karaletsos,
  R.~Singh, P.~Szerlip, P.~Horsfall, and N.~Goodman (2019).
\newblock Pyro: Deep universal probabilistic programming.
\newblock {\em The Journal of Machine Learning Research\/}~{\em 20\/}(1),
  973--978.

\bibitem[\protect\citeauthoryear{Boehmke and Greenwell}{Boehmke and
  Greenwell}{2019}]{boehmke2019hands}
Boehmke, B. and B.~Greenwell (2019).
\newblock {\em Hands-on machine learning with R}.
\newblock CRC Press.

\bibitem[\protect\citeauthoryear{Breiman}{Breiman}{2001}]{breiman2001random}
Breiman, L. (2001).
\newblock Random forests.
\newblock {\em Machine Learning\/}~{\em 45\/}(1), 5--32.

\bibitem[\protect\citeauthoryear{Brooks, Gelman, Jones, and Meng}{Brooks
  et~al.}{2011}]{brooks2011handbook}
Brooks, S., A.~Gelman, G.~Jones, and X.~Meng (2011).
\newblock {\em Handbook of {M}arkov chain {M}onte {C}arlo}.
\newblock CRC press.

\bibitem[\protect\citeauthoryear{Brooks and Gelman}{Brooks and
  Gelman}{1998}]{brooks1998general}
Brooks, S.~P. and A.~Gelman (1998).
\newblock General methods for monitoring convergence of iterative simulations.
\newblock {\em Journal of Computational and Graphical Statistics\/}~{\em
  7\/}(4), 434--455.

\bibitem[\protect\citeauthoryear{Carpenter, Gelman, Hoffman, Lee, Goodrich,
  Betancourt, Brubaker, Guo, Li, and Riddell}{Carpenter
  et~al.}{2017}]{carpenter2017stan}
Carpenter, B., A.~Gelman, M.~Hoffman, D.~Lee, B.~Goodrich, M.~Betancourt,
  M.~Brubaker, J.~Guo, P.~Li, and A.~Riddell (2017).
\newblock Stan: A probabilistic programming language.
\newblock {\em Journal of Statistical Software\/}~{\em 76\/}(1).

\bibitem[\protect\citeauthoryear{Devroye, Gy{\"o}rfi, and Lugosi}{Devroye
  et~al.}{2013}]{devroye2013probabilistic}
Devroye, L., L.~Gy{\"o}rfi, and G.~Lugosi (2013).
\newblock {\em A probabilistic theory of pattern recognition}, Volume~31.
\newblock Springer Science \& Business Media.

\bibitem[\protect\citeauthoryear{Dillon, Langmore, Tran, Brevdo, Vasudevan,
  Moore, Patton, Alemi, Hoffman, and Saurous}{Dillon
  et~al.}{2017}]{dillon2017tensorflow}
Dillon, J., I.~Langmore, D.~Tran, E.~Brevdo, S.~Vasudevan, D.~Moore, B.~Patton,
  A.~Alemi, M.~Hoffman, and R.~Saurous (2017).
\newblock Tensorflow distributions.
\newblock {\em arXiv preprint arXiv:1711.10604\/}.

\bibitem[\protect\citeauthoryear{Faraway, Marsaglia, Marsaglia, and
  Baddeley}{Faraway et~al.}{2019}]{faraway2019goftest}
Faraway, J., G.~Marsaglia, J.~Marsaglia, and A.~Baddeley (2019).
\newblock {\em goftest: Classical Goodness-of-Fit Tests for Univariate
  Distributions}.
\newblock R package version 1.2-2.

\bibitem[\protect\citeauthoryear{Friedman}{Friedman}{2001}]{friedman2001greedy}
Friedman, J. (2001).
\newblock Greedy function approximation: a gradient boosting machine.
\newblock {\em Annals of Statistics\/}, 1189--1232.

\bibitem[\protect\citeauthoryear{Ge, Xu, and Ghahramani}{Ge
  et~al.}{2018}]{ge2018turing}
Ge, H., K.~Xu, and Z.~Ghahramani (2018).
\newblock Turing: A language for flexible probabilistic inference.

\bibitem[\protect\citeauthoryear{Gelman and Rubin}{Gelman and
  Rubin}{1992a}]{gelman1992inference}
Gelman, A. and D.~Rubin (1992a).
\newblock Inference from iterative simulation using multiple sequences.
\newblock {\em Statistical Science\/}~{\em 7\/}(4), 457--472.

\bibitem[\protect\citeauthoryear{Gelman and Rubin}{Gelman and
  Rubin}{1992b}]{gelman1992single}
Gelman, A. and D.~Rubin (1992b).
\newblock A single series from the {G}ibbs sampler provides a false sense of
  security.
\newblock {\em Bayesian Statistics\/}~{\em 4}, 625--631.

\bibitem[\protect\citeauthoryear{Gelman, Stern, Carlin, Dunson, Vehtari, and
  Rubin}{Gelman et~al.}{2013}]{gelman2013bayesian}
Gelman, A., H.~Stern, J.~Carlin, D.~Dunson, A.~Vehtari, and D.~Rubin (2013).
\newblock {\em Bayesian data analysis}.
\newblock Chapman and Hall/CRC.

\bibitem[\protect\citeauthoryear{Geman and Geman}{Geman and
  Geman}{1984}]{geman1984stochastic}
Geman, S. and D.~Geman (1984).
\newblock Stochastic relaxation, {G}ibbs distributions, and the {B}ayesian
  restoration of images.
\newblock {\em IEEE Transactions on Pattern Analysis and Machine
  intelligence\/}~(6), 721--741.

\bibitem[\protect\citeauthoryear{Greenwell, Boehmke, Cunningham, Developers,
  and Greenwell}{Greenwell et~al.}{2019}]{greenwell2019package}
Greenwell, B., B.~Boehmke, J.~Cunningham, G.~Developers, and M.~B. Greenwell
  (2019).
\newblock Package ‘gbm’.

\bibitem[\protect\citeauthoryear{Hern{\'a}ndez-Lobato, Hern{\'a}ndez-Lobato,
  and Su{\'a}rez}{Hern{\'a}ndez-Lobato et~al.}{2010}]{hernandez2010expectation}
Hern{\'a}ndez-Lobato, D., J.~Hern{\'a}ndez-Lobato, and A.~Su{\'a}rez (2010).
\newblock Expectation propagation for microarray data classification.
\newblock {\em Pattern Recognition Letters\/}~{\em 31\/}(12), 1618--1626.

\bibitem[\protect\citeauthoryear{Hoffman and Gelman}{Hoffman and
  Gelman}{2014}]{hoffman2014no}
Hoffman, M. and A.~Gelman (2014).
\newblock The {N}o-{U}-turn {S}ampler: adaptively setting path lengths in
  {H}amiltonian {M}onte {C}arlo.
\newblock {\em Journal of Machine Learning Research\/}~{\em 15\/}(1),
  1593--1623.

\bibitem[\protect\citeauthoryear{Karatzoglou, Smola, Hornik, and
  Zeileis}{Karatzoglou et~al.}{2004}]{karatzoglou2004kernlab}
Karatzoglou, A., A.~Smola, K.~Hornik, and A.~Zeileis (2004).
\newblock kernlab-an s4 package for kernel methods in r.
\newblock {\em Journal of Statistical Software\/}~{\em 11\/}(9), 1--20.

\bibitem[\protect\citeauthoryear{Kuhn et~al.}{Kuhn
  et~al.}{2008}]{kuhn2008building}
Kuhn, M. et~al. (2008).
\newblock Building predictive models in {R} using the {C}aret package.
\newblock {\em Journal of Statistical Software\/}~{\em 28\/}(5), 1--26.

\bibitem[\protect\citeauthoryear{Lambert}{Lambert}{2018a}]{lambert2018Student}
Lambert, B. (2018a).
\newblock {\em A Student's Guide to Bayesian Statistics}.
\newblock Sage Publications Ltd.

\bibitem[\protect\citeauthoryear{Lambert}{Lambert}{2018b}]{lambertbees}
Lambert, B. (2018b).
\newblock You{T}ube video: Bob’s bees: the importance of using multiple bees
  (chains) to judge mcmc convergence.

\bibitem[\protect\citeauthoryear{Lewandowski, Kurowicka, and Joe}{Lewandowski
  et~al.}{2009}]{lewandowski2009generating}
Lewandowski, D., D.~Kurowicka, and H.~Joe (2009).
\newblock Generating random correlation matrices based on vines and extended
  onion method.
\newblock {\em Journal of Multivariate Analysis\/}~{\em 100\/}(9), 1989--2001.

\bibitem[\protect\citeauthoryear{Liaw and Wiener}{Liaw and
  Wiener}{2002}]{liaw2002classification}
Liaw, A. and M.~Wiener (2002).
\newblock Classification and regression by randomforest.
\newblock {\em R news\/}~{\em 2\/}(3), 18--22.

\bibitem[\protect\citeauthoryear{Louppe}{Louppe}{2014}]{louppe2014understanding}
Louppe, G. (2014).
\newblock Understanding random forests: From theory to practice.
\newblock {\em arXiv preprint arXiv:1407.7502\/}.

\bibitem[\protect\citeauthoryear{Lunn, Thomas, Best, and Spiegelhalter}{Lunn
  et~al.}{2000}]{lunn2000winbugs}
Lunn, D., A.~Thomas, N.~Best, and D.~Spiegelhalter (2000).
\newblock Winbugs-a {B}ayesian modelling framework: concepts, structure, and
  extensibility.
\newblock {\em Statistics and Computing\/}~{\em 10\/}(4), 325--337.

\bibitem[\protect\citeauthoryear{Neal et~al.}{Neal et~al.}{2011}]{neal2011mcmc}
Neal, R. et~al. (2011).
\newblock {MCMC} using {H}amiltonian dynamics.
\newblock {\em Handbook of {M}arkov chain {M}onte {C}arlo\/}~{\em 2\/}(11), 2.

\bibitem[\protect\citeauthoryear{Paananen, Piironen, Bürkner, and
  Vehtari}{Paananen et~al.}{2019}]{paananen2019implicitly}
Paananen, T., J.~Piironen, P.~Bürkner, and A.~Vehtari (2019).
\newblock Implicitly adaptive importance sampling.
\newblock {\em arXiv\/}.

\bibitem[\protect\citeauthoryear{Piironen and Vehtari}{Piironen and
  Vehtari}{2017}]{piironen2017sparsity}
Piironen, J. and A.~Vehtari (2017).
\newblock Sparsity information and regularization in the horseshoe and other
  shrinkage priors.
\newblock {\em Electronic Journal of Statistics\/}~{\em 11\/}(2), 5018--5051.

\bibitem[\protect\citeauthoryear{Plummer et~al.}{Plummer
  et~al.}{2003}]{plummer2003jags}
Plummer, M. et~al. (2003).
\newblock Jags: A program for analysis of {B}ayesian graphical models using
  {G}ibbs sampling.
\newblock In {\em Proceedings of the 3rd international workshop on Distributed
  Statistical Computing}, Volume 124. Vienna, Austria.

\bibitem[\protect\citeauthoryear{Ripley, Venables, and Ripley}{Ripley
  et~al.}{2016}]{ripley2016package}
Ripley, B., W.~Venables, and B.~Ripley (2016).
\newblock Package ‘nnet’.
\newblock {\em R package version\/}~{\em 7}, 3--12.

\bibitem[\protect\citeauthoryear{Salvatier, Wiecki, and Fonnesbeck}{Salvatier
  et~al.}{2016}]{salvatier2016probabilistic}
Salvatier, J., T.~Wiecki, and C.~Fonnesbeck (2016).
\newblock Probabilistic programming in {P}ython using {P}y{MC}3.
\newblock {\em PeerJ Computer Science\/}~{\em 2}, e55.

\bibitem[\protect\citeauthoryear{Schummer, Ng, Bumgarner, Nelson, Schummer,
  Bednarski, Hassell, Baldwin, Karlan, and Hood}{Schummer
  et~al.}{1999}]{schummer1999comparative}
Schummer, M., W.~Ng, R.~Bumgarner, P.~Nelson, B.~Schummer, D.~Bednarski,
  L.~Hassell, R.~Baldwin, B.~Karlan, and L.~Hood (1999).
\newblock Comparative hybridization of an array of 21,500 ovarian c{DNA}s for
  the discovery of genes overexpressed in ovarian carcinomas.
\newblock {\em Gene\/}~{\em 238\/}(2), 375--385.

\bibitem[\protect\citeauthoryear{Singh, Febbo, Ross, Jackson, Manola, Ladd,
  Tamayo, Renshaw, D'Amico, and Richie}{Singh et~al.}{2002}]{singh2002gene}
Singh, D., P.~Febbo, K.~Ross, D.~Jackson, J.~Manola, C.~Ladd, P.~Tamayo,
  A.~Renshaw, A.~D'Amico, and J.~Richie (2002).
\newblock Gene expression correlates of clinical prostate cancer behavior.
\newblock {\em Cancer Cell\/}~{\em 1\/}(2), 203--209.

\bibitem[\protect\citeauthoryear{Van~Dyk and Meng}{Van~Dyk and
  Meng}{2001}]{van2001art}
Van~Dyk, D. and X.~Meng (2001).
\newblock The art of data augmentation.
\newblock {\em Journal of Computational and Graphical Statistics\/}~{\em
  10\/}(1), 1--50.

\bibitem[\protect\citeauthoryear{Vehtari, Gelman, Simpson, Carpenter, and
  B{\"u}rkner}{Vehtari et~al.}{2020}]{vehtari2019rank}
Vehtari, A., A.~Gelman, D.~Simpson, B.~Carpenter, and P.~B{\"u}rkner (2020).
\newblock Rank-normalization, folding, and localization: An improved r-hat for
  assessing convergence of {MCMC}.
\newblock {\em Bayesian Analysis\/}.

\bibitem[\protect\citeauthoryear{Wolpert and Macready}{Wolpert and
  Macready}{1997}]{wolpert1997no}
Wolpert, D. and W.~Macready (1997).
\newblock No free lunch theorems for optimization.
\newblock {\em IEEE transactions on evolutionary computation\/}~{\em 1\/}(1),
  67--82.

\bibitem[\protect\citeauthoryear{Yang, Cai, Li, and Lin}{Yang
  et~al.}{2006}]{yang2006stable}
Yang, K., Z.~Cai, J.~Li, and G.~Lin (2006).
\newblock A stable gene selection in microarray data analysis.
\newblock {\em BMC Bioinformatics\/}~{\em 7\/}(1), 228.

\end{thebibliography}
	
	\beginsupplement
	\section{Diagnosing convergence in joint distributions: multivariate normal models supplementary}\label{sec:multivariate_normal_supp}
	
\begin{figure}[!htb]
		\centerline{\includegraphics[width=1\textwidth]{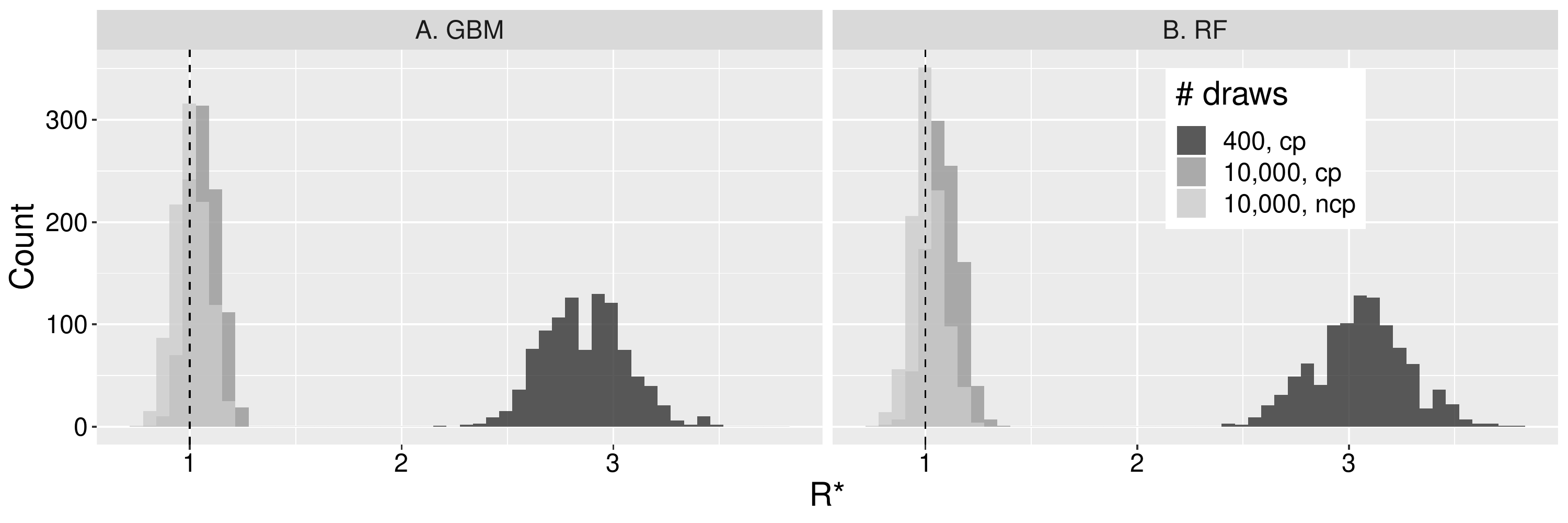}}
		\caption{\textbf{Multivariate normal example: GBM (A) versus RF (B) classifiers.} Shows $R^*$ distributions obtained for two MCMC samples (of differing numbers of draws: 400 and 10,000) from the centered parameterisation (``cp'') and one from the non-centered version (``ncp''; with 10,000 draws). Here, we show 1000 $R^*$ draws by Algorithm \ref{alg:R_star_uncertainty} for each MCMC run.}
		\label{fig:mvt_gbm_vs_rf}
	\end{figure}

	In Fig. \ref{fig:mvt_gbm_vs_rf}, we compare the performance of GBM and RF classifiers on the multivariate normal example described in \S\ref{sec:multivariate_normal_250}. The two classifiers produced similar $R^*$ distributions across the various examples: for the 400 draw example, the GBM $R^*$ distribution had a mean of 2.86 and the RF equivalent was 3.06; for the example with 10,000 samples from the centered parameterisation, the GBM classifier had a $R^*$ distribution mean of 1.07 versus 1.08 from the RF; for the example with 10,000 samples from the non-centered parameterisation, both classifiers produced a mean of 1.00.
	
	\section{Hierarchical model: Eight schools model supplementary}\label{sec:8_schools_supp}
	
\begin{figure}[!htb]
		\centerline{\includegraphics[width=1\textwidth]{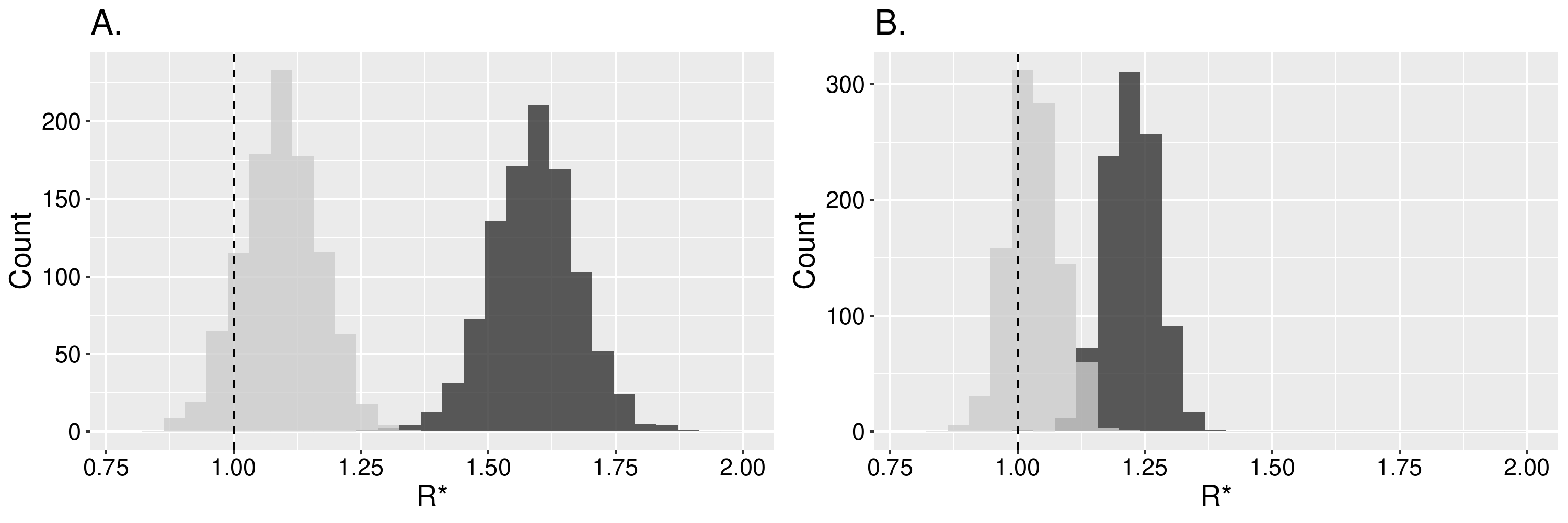}}
		\caption{\textbf{Eight schools example: $R^*$ distributions for a RF classifier.} A shows draws from the $R^*$ distribution when splitting chains in two (resulting in 8 chains); B shows the same but using the 4 original chains. The MCMC samples comprised 2000 draws in all cases, with 1000 used as post-warm-up iterations. In panels A and B, the plots show 1000 $R^*$ draws using Algorithm \ref{alg:R_star_uncertainty} for each parameterisation.}
		\label{fig:eight_schools_rf}
	\end{figure}

	Fig. \ref{fig:eight_schools_rf} shows the equivalent of Fig. \ref{fig:eight_schools} except using a RF classifier.
	
	\section{Wide datasets: multivariate normal}\label{sec:wide}
	As the number of parameter dimensions increases, it might be thought that ML algorithms will overfit the data, and, hence, testing set classification would be poor; leading to unreliable determinations of convergence. To test this hypothesis, we investigated two scenarios using a multivariate normal target.
	
	\subsection{250-dimensional model}
\begin{figure}[!htb]
		\centerline{\includegraphics[width=1\textwidth]{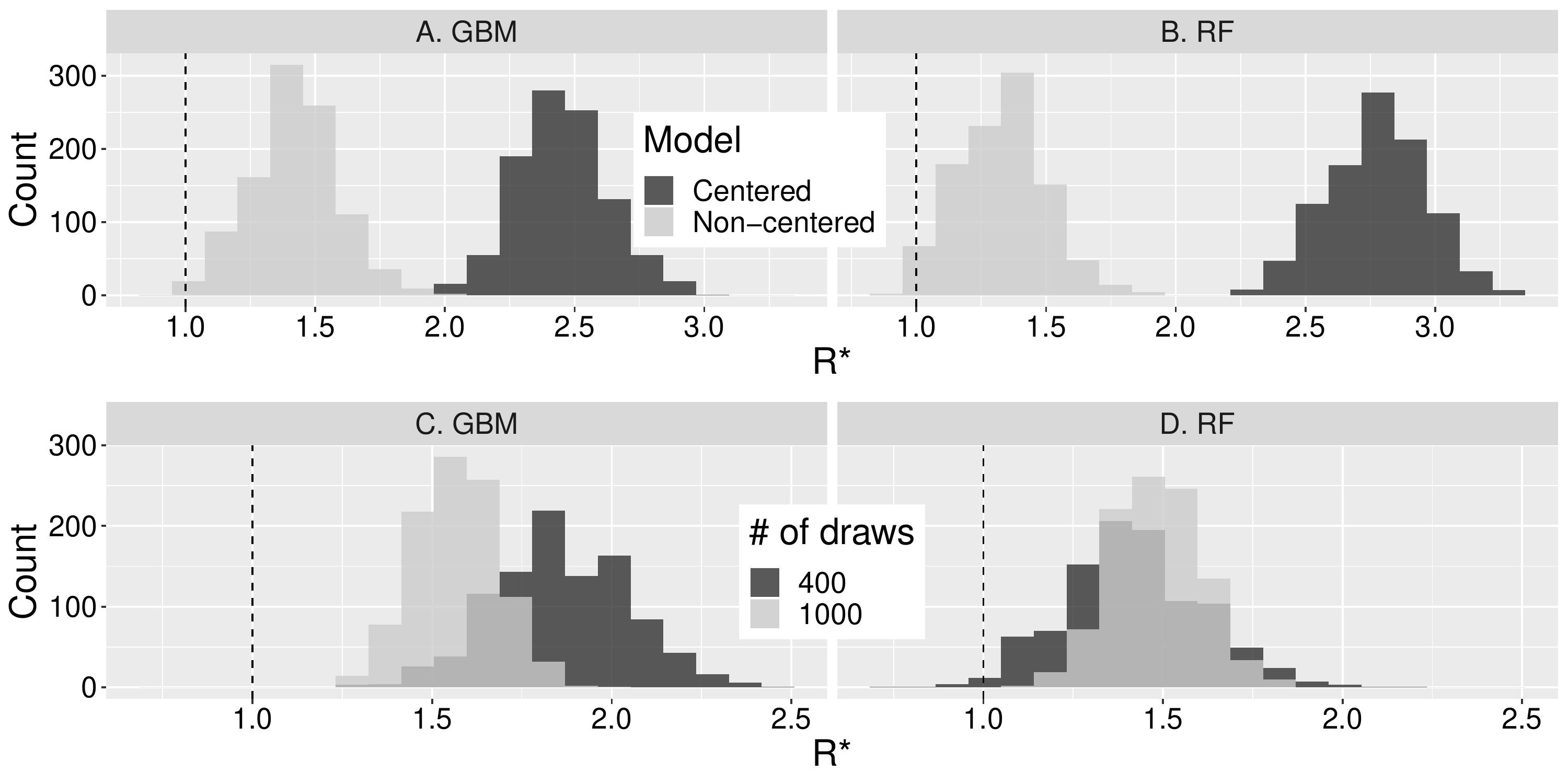}}
		\caption{\textbf{Wide data examples.} Top row shows the $R^*$ distribution for the 250-dimensional example in \S\ref{sec:wide} with 250 post-warm-up iterations per chain from Stan's NUTS algorithm across both model parameterisations; bottom row shows $R^*$ distribution for the 10,000-dimensional example with 400 and 1000 MCMC iterations per chain (although the first half of these were discarded as warm-up). In the left column, results for the GBM classifier are shown; in the right, we show the same for the RF classifier. In all cases, 1000 draws of $R^*$ are plotted as generated by Algorithm \ref{alg:R_star_uncertainty} for a single MCMC run composed of 4 chains.}
		\label{fig:mvt_wide_both}
	\end{figure}

\begin{figure}[!htb]
		\centerline{\includegraphics[width=1\textwidth]{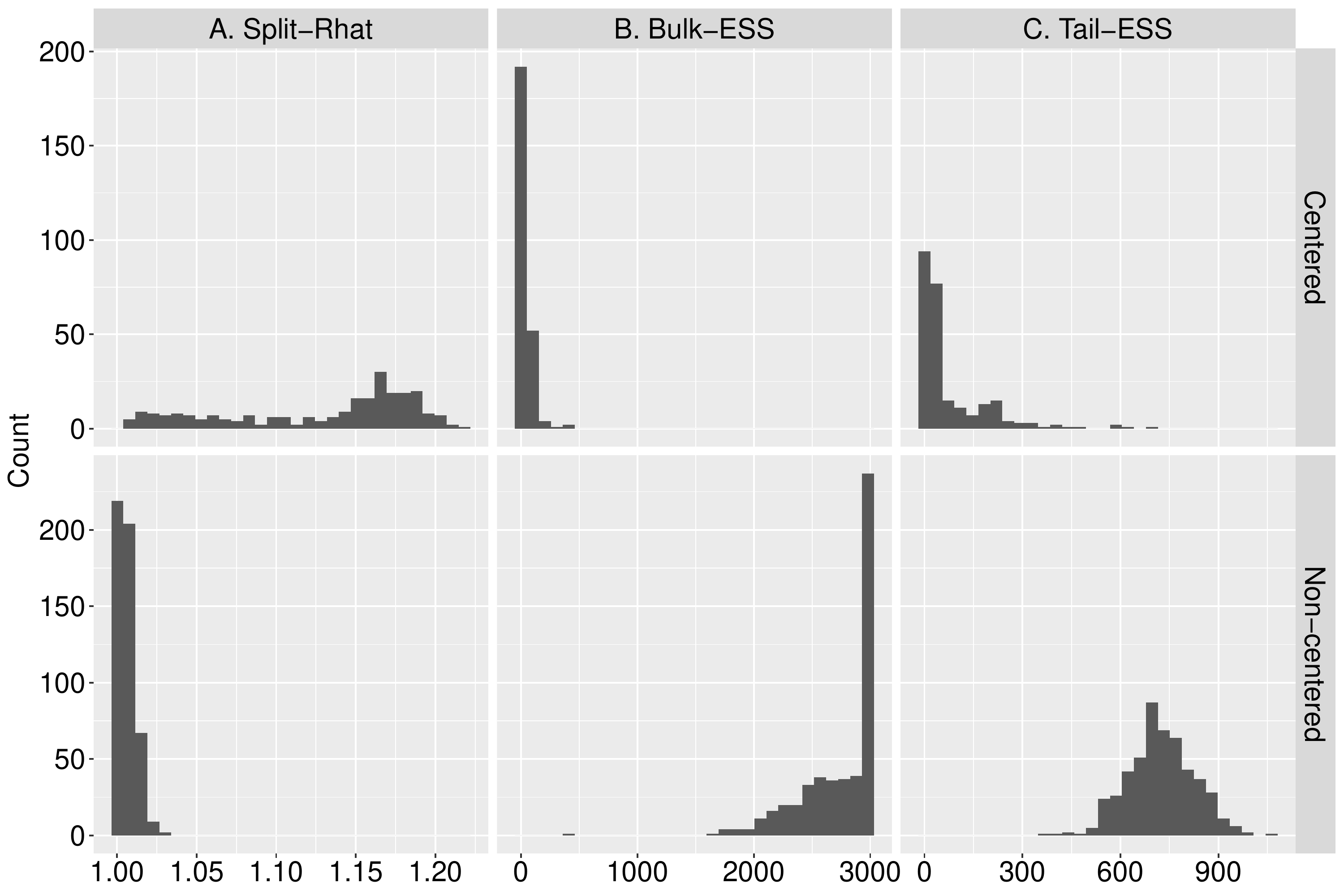}}
		\caption{\textbf{Wide data 250-dimensional example: established diagnostics.} The top row shows the results for the centered parameterisation; the bottom row for the non-centered. Column A shows split-$\widehat{R}$; columns B and C show the bulk- and tail-ESS; in each case the statistics are displayed for all model parameters and were calculated using 250 post-warm-up draws from Stan's NUTS algorithm. Note, that it is possible for the ESS to exceed the actual sample size if there is negative autocorrelation in the Markov chains' values. In both cases, the results correspond to a single MCMC run composed of 4 chains.}
		\label{fig:wide_both_diagnostics}
	\end{figure}

	In the first of these, we used the 250-dimensional multivariate normal of eq. \eqref{eq:mvt_normal_250} with 250 post-warm-up iterations (after 250 warm-up iterations) for each of 4 chains from Stan's NUTS to calculate $R^*$ distributions as in Algorithm \ref{alg:R_star_uncertainty}. Here, we considered both the centered and non-centered parameterisations, where, in both cases, the number of iterations is comparable to the number of parameters, so the training data is relatively ``wide''. We also calculated $R^*$ distributions using both the GBM and RF classifiers: the $R^*$ distribution in each case is shown in Figs. \ref{fig:mvt_wide_both}A\&B. This figure shows that, across both parameterisations, convergence has not yet been reached since all $R^*$ distributions are shifted rightwards of 1. The same conclusion is reached if rank-normalised split-$\widehat{R}$ is used instead (Fig. \ref{fig:wide_both_diagnostics}A), since, for both parameterisations, some of the parameters had $\widehat{R}>1.01$. Using bulk- or tail-ESS instead, we conclude that the non-centered parameterisation shows signs of convergence whereas the centered does not (Fig. \ref{fig:wide_both_diagnostics}B\&C). 
	
	\subsection{10,000-dimensional model}
	We next consider a more challenging example -- a target distribution with 10,000 dimensions. In this case, we assume independent standard normals for each dimension. In Figs. \ref{fig:mvt_wide_both}C\&D, we plot the $R^*$ distributions from GBM and RF classifiers for two MCMC runs targeting this distribution: one with 400 iterations, the other with 1000. In all cases, the distributions were right of $R^*=1$, indicating non-convergence. These results were also echoed by rank-normalised split-$\widehat{R}$, with 19\% of dimensions having $\widehat{R}>1.01$ for the 400 iteration case and 3\% for the 1000 iteration case.
	
	Overall, the examples in this section suggest that $R^*$ is a conservative measure of convergence: when there are not enough draws, it will tend to diagnose non-convergence. We also note that the statistic took comparable time to calculate relative to existing convergence diagnostics on a desktop computer.
	
	\section{Non-stationary marginals}\label{sec:non-stationary}
	If a Markov chain does not mix with itself, this also indicates that convergence has not occurred \citep{gelman2013bayesian}. In this section, we investigate whether $R^*$ can detect non-stationary sampling distributions. In all the examples, we present results for both GBM and RF classifiers.
	
	\subsection{Trending mean across all chains}\label{sec:non-stationary_chains}
\begin{figure}[!htb]
		\centerline{\includegraphics[width=1.0\textwidth]{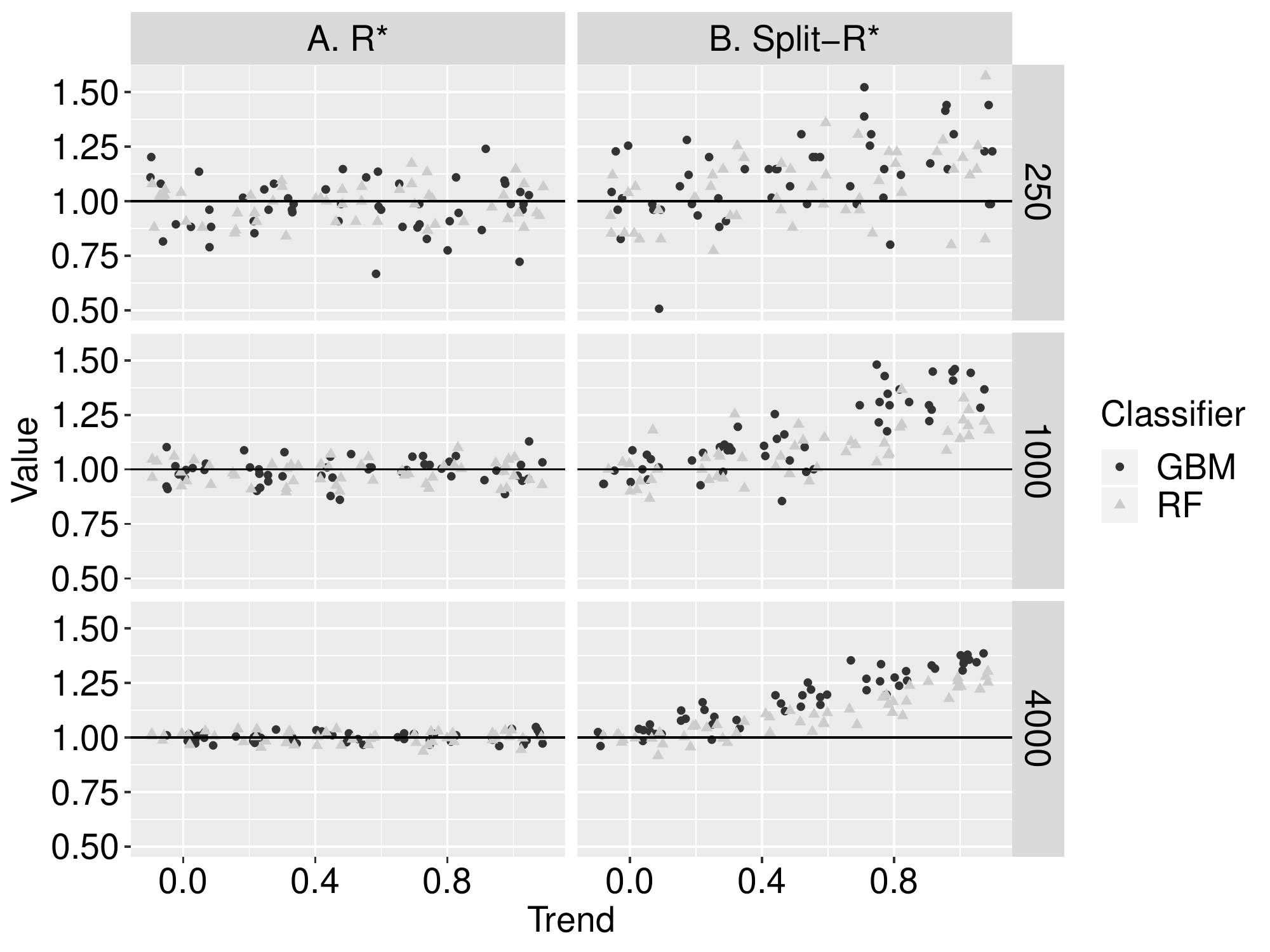}}
		\caption{\textbf{Univariate trends example.} Points and triangles represent results for the GBM and RF classifiers, respectively. Column A represents results for $R^*$ calculated using Algorithm \ref{alg:R_star} on the 4 chains; column B  shows the same calculation after each chain is split into two halves. The rows present the differing sample sizes. The horizontal axis measures (half) the change in mean across the whole sample: so a value ``1'' indicates the mean increases by 2 units from the start to end of sampling. At each parameter set, 10 replicates were run and jitter was added to the points.}
		\label{fig:trends_all_dim}
	\end{figure}

	We first recapitulate an example from appendix A in \cite{vehtari2019rank}. This example showed that split-$\widehat{R}$ could detect non-convergence caused by shifts in sampling distributions over time: in their case, they analysed chains with common linear trends in mean. Specifically, they first generated 4 chains by random sampling from a univariate normal distribution, then added a common time trend to each chain, resulting in a univariate distribution whose mean increased during sampling. We first repeat this analysis but using $R^*$ rather than $\widehat{R}$: in Fig. \ref{fig:trends_all_dim}, we show these results. In column A, we show the results for $R^*$ calculated on the 4 chains that ran; column B shows the same calculation but after the chains are split into two equal halves. The rows show the range of sample sizes investigated: 250, 1000 and 4000; the horizontal axis shows the magnitude of time trend added to each sample; for all parameter sets, we run 10 replicates. The points and triangles show $R^*$ derived from GBM and RF classifiers, respectively. This plot mirrors Fig. 4 in the supplementary materials of \cite{vehtari2019rank} and shows that, without splitting the chains, $R^*$ does not increase with trend whereas, after splitting, it does. As expected, split-$R^*$ is more reliably able to detect non-convergence as sample size increases.
	
	These results make intuitive sense: without systematic between-chain variation, it is not possible to reliably determine which of them caused a particular observation. In this case, because all chains exhibited the same secular trends over time, there would not be differences in their marginals. By splitting chains into two -- the first half being the early phase, and the second half being the later phase with higher mean -- this forces differences in the marginals. This meant it was possible to reliably pick whether an observation was caused by an early phase chain or a later one. As such, we recommend that $R^*$ always be calculated using split chains as is recommended for $\widehat{R}$ \citep{carpenter2017stan,vehtari2019rank}.
	
	\subsection{Trending mean in a single dimension}\label{sec:non-stationary_single}
\begin{figure}[!htb]
		\centerline{\includegraphics[width=1.0\textwidth]{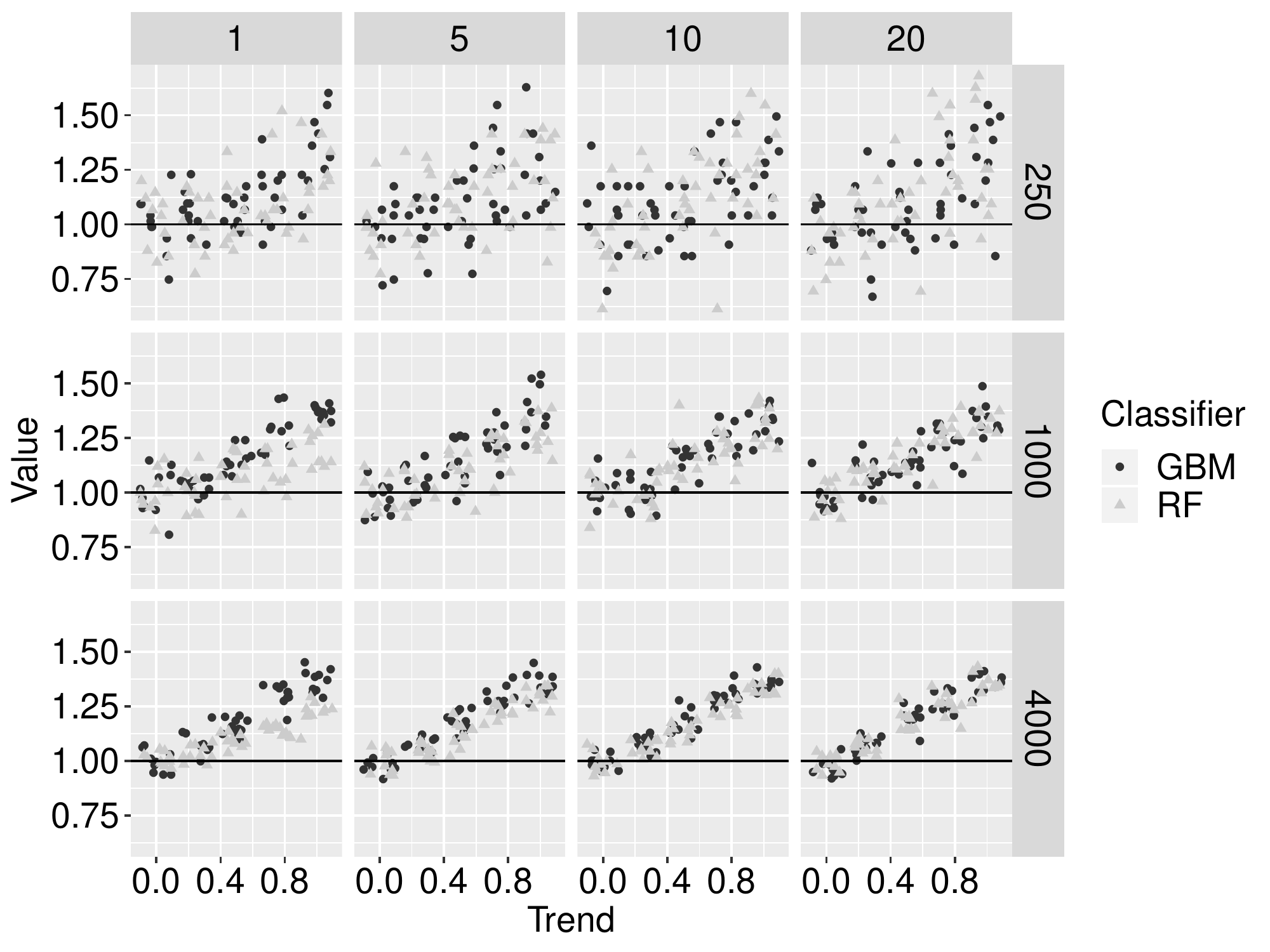}}
		\caption{\textbf{Multivariate trends example with split-$R^*$.} Points and triangles represent results for the GBM and RF classifiers, respectively. The columns present different dimensionalities of the target distribution; the rows present different sample sizes. The horizontal axis measures (half) the change in mean across the single dimension that had a trend added to it: a value ``1'' indicates its mean increases by 2 units from the start to end of sampling; all other dimensions (if dimensions exceeded 1) had stationary distributions. At each parameter set, 10 replicates were run and jitter was added to the points.}
		\label{fig:trends_one_dim}
	\end{figure}

	We next consider whether split-$R^*$ can detect non-convergence when only a single dimension trends. In Fig. \ref{fig:trends_one_dim}, we show how $R^*$ performs across a range of target dimensions. In the simulations here, all dimensions bar one are stationary; the remaining dimension has a linear trend added to it. In all cases, across both GBM and RF classifiers, split-$R^*$ increased with trend. Indeed, differences in the typical values of this metric were not apparent across the different target dimensions considered. This suggests split-$R^*$ can robustly determine chain identity if only a single dimension has a non-stationary sampling distribution.

	\subsection{Trending covariance}\label{sec:non-stationary_covariance}
\begin{figure}[!htb]
		\centerline{\includegraphics[width=1.0\textwidth]{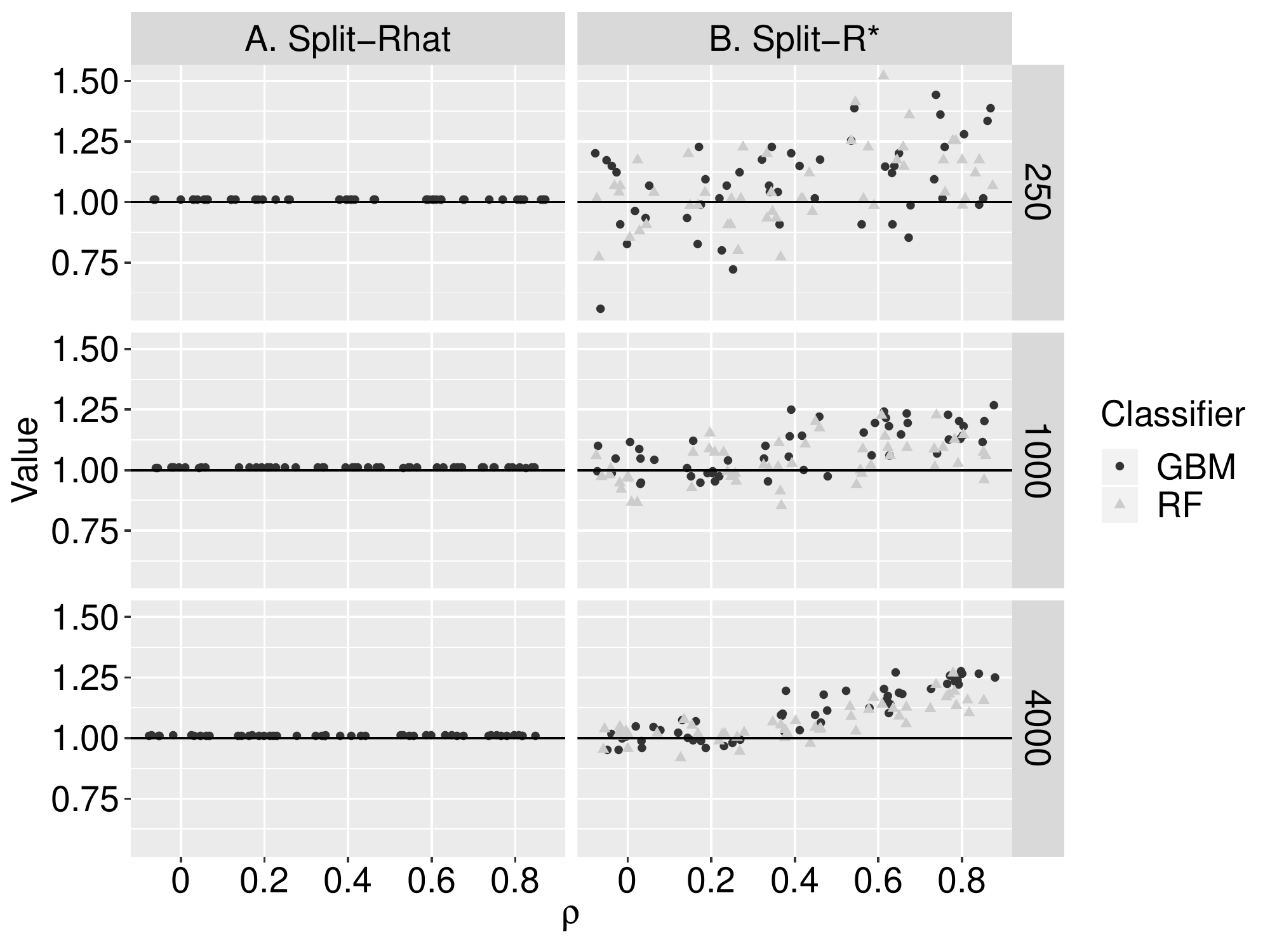}}
		\caption{\textbf{Bivariate normal with trending correlation example.} Points and triangles represent results for the GBM and RF classifiers, respectively. Column A shows the results for split-$\widehat{R}$; column B for $R^*$; the rows present the differing sample sizes. The horizontal axis measures (half) the change in correlation across the whole sample: so a value ``0.5'' indicates the correlation increases by 1 unit (from -0.5 to 0.5) from the start to end of sampling. At each parameter set, 10 replicates were run and jitter was added to the points.}
		\label{fig:trends_joint_distribution}
	\end{figure}

	Means that trend over time is one form of non-stationarity; another is a time-varying covariance. Next, we consider a bivariate normal with (constant) standard normal marginals but where the correlation between dimensions trends over time. Specifically, we allow the correlation to increase linearly from $-\rho$ and $\rho$ throughout the course of simulations and use i.i.d. draws from the process across 4 ``chains''. Again, as before, $R^*$ calculated on unsplit chains is unable to detect this form of non-stationarity, since there are no inter-chain differences in the sampling distribution. Similarly, split-$\widehat{R}$ does not detect this form of non-convergence since the marginal distribution across chains does not vary over time (Fig. \ref{fig:trends_joint_distribution}A). By contrast, split-$R^*$ can (Fig. \ref{fig:trends_joint_distribution}B), since it uses all information in the samples, including the covariance structure.
	
	\subsection{Chain persistence}\label{sec:non-stationary_persistence}
	When forming training and testing sets as part of the ML algorithm used to determine $R^*$, the testing set is effectively treated as a sort of ``independent'' hold-out dataset. Markov chains, in general, have persistence, meaning that the test set will not be truly independent and can -- according to the level of autocorrelation in the chains -- be highly related to the training set. In this section, we investigate how this autocorrelation affects the performance of $R^*$. The difficulty with this question is that higher chain autocorrelation typically means the sampling distribution is a rougher approximation of the target, so $R^*$ should be higher due to the properties of the sampling distribution. It could also be higher because the training set is less distinct from the testing set.
	
\begin{figure}[!htb]
		\centerline{\includegraphics[width=1.0\textwidth]{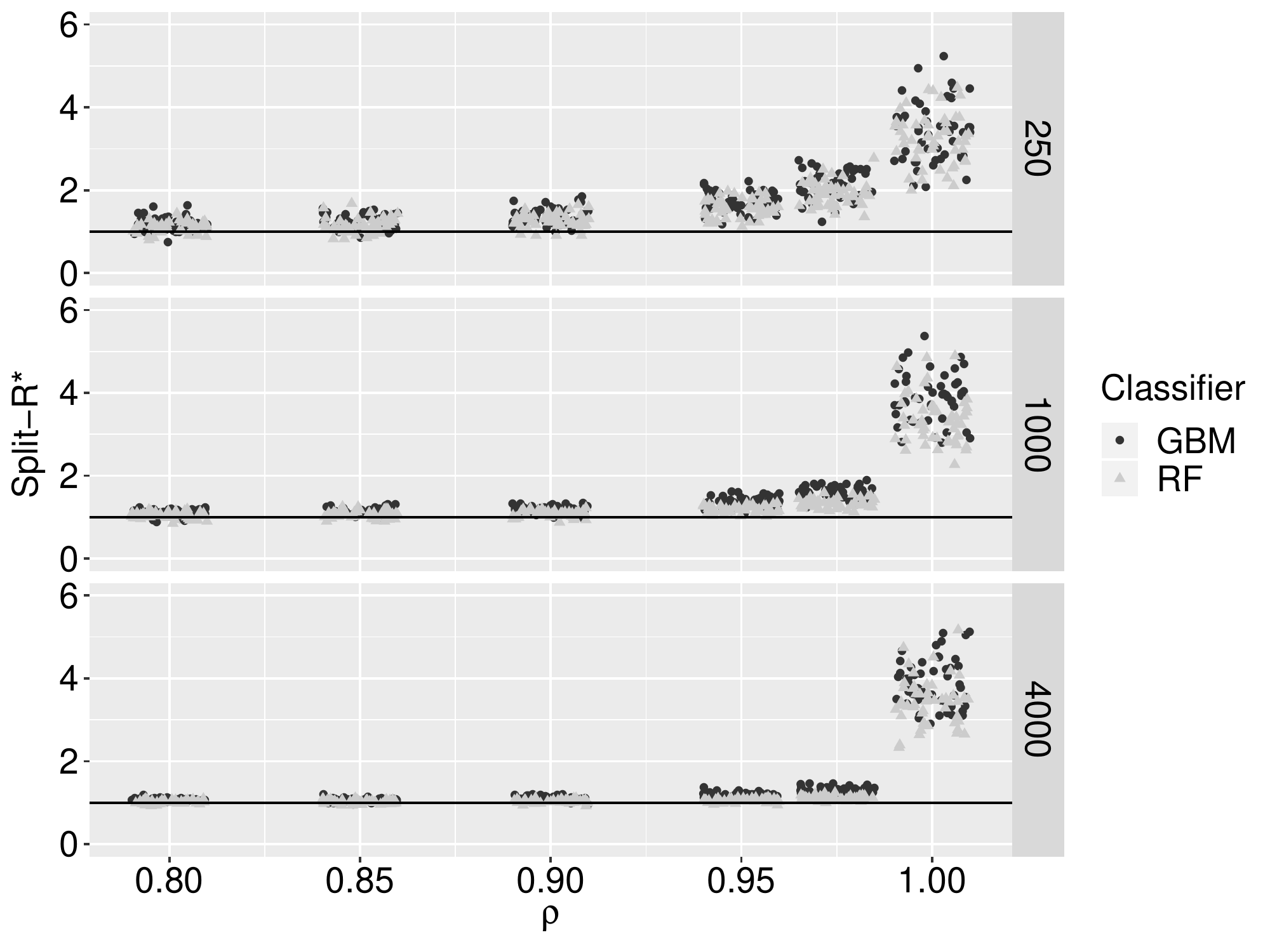}}
		\caption{\textbf{Non-stationary distribution: AR(1) example.} Points and triangles represent results for the GBM and RF classifiers, respectively. The horizontal axis measures the autocorrelation of the AR(1) processes; the vertical axis shows the value of $R^*$ calculated on chains split in half; each panel shows a different number of iterations. At each parameter set, 10 replicates were run and jitter was added to the points. The black line shows $R^*=1$.}
		\label{fig:trends_ar1}
	\end{figure}

	To investigate this, we generated AR(1) processes (as defined in eq. (\ref{eq:ar1})) with autocorrelations, $\rho$, ranging from 0.8-1 and, in each case, calculated $R^*$ via Algorithm \ref{alg:R_star}. Note, that only when $|\rho|<1$ is the marginal distribution defined by this process itself stationary; at $\rho=1$, its variance increases linearly with time, so, by definition, is not converged. In Fig. \ref{fig:trends_ar1}, we show the results of these simulations across various numbers of iterations: 250, 1000 and 4000 (different panels). In all cases, $R^*>1$ whenever $\rho=1$, indicating lack of convergence. As $\rho$ declined, so did $R^*$.
	
\begin{figure}[!htb]
		\centerline{\includegraphics[width=1.0\textwidth]{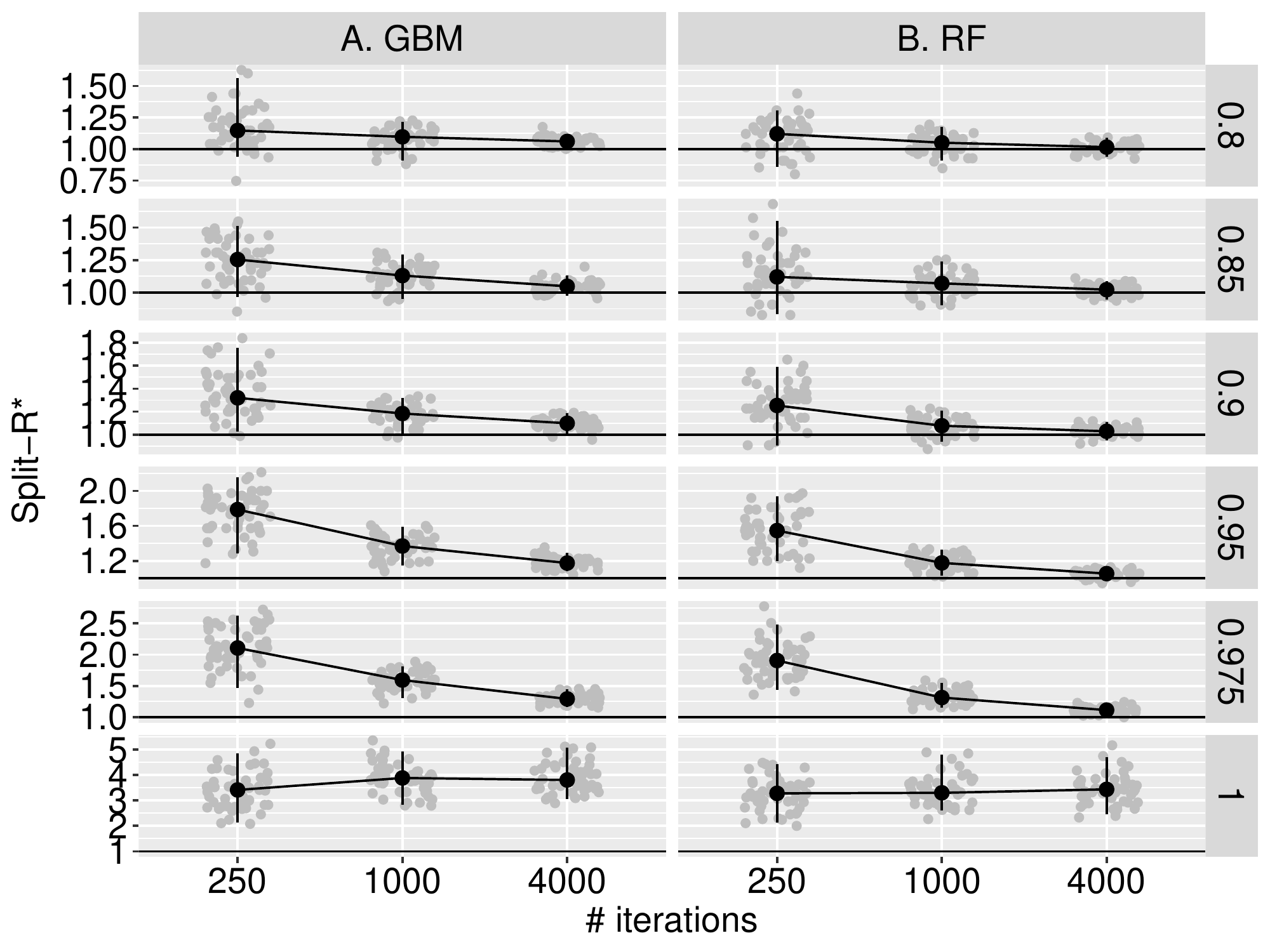}}
		\caption{\textbf{Non-stationary distribution: AR(1) example alternative view.} Column A shows the results for a GBM classifier; column B for a RF classifier. The horizontal axis measures the number of iterations; the vertical axis shows the value of $R^*$ calculated on chains split in half; each panel shows a different value of autocorrelation. At each parameter set, 10 replicates were run and jitter was added to the points. The black dots show the median $R^*$ values at each parameter set; upper and lower whiskers show 2.5\% and 97.5\% quantiles. The black horizontal line shows $R^*=1$. }
		\label{fig:trends_ar1_transposed}
	\end{figure}

	Notably, as the number of iterations increased, the values of $R^*$ for $\rho<1$ declined, whereas $R^*$ for $\rho=1$ actually increased: this is most easily seen by examining Fig. \ref{fig:trends_ar1_transposed} (which shows the same data as in Fig. \ref{fig:trends_ar1} but in a different way). This characteristic is exactly as desired: larger sample sizes should yield a sampling distribution closer to convergence for $\rho<1$; the same for $\rho=1$ produces distributions that are no closer to convergence, and more samples allows better determination of this non-convergence.
	
	Overall, it seems that $R^*$ is a conservative measure and with greater chain autocorrelation it suggests more draws are necessary for convergence.
	
	\section{Many parameter models: ovarian and prostate analysis}\label{sec:prostate}
	In this section, we analyse two Bayesian models both fit to real data. In both examples, we illustrate $R^*$ distributions derived only from GBM classifiers. The first is a logistic regression model fit to microarray ovarian cancer data with 54 data points and 1536 predictor variables; overall, this model has 4719 parameters. Since there are relatively few data points relative to the number of predictors, regularised horseshoe priors are specified on the regression coefficients \citep{piironen2017sparsity} since most are expected to be zero. This dataset has been used for benchmarking in the past (see, for example, \cite{schummer1999comparative,hernandez2010expectation,paananen2019implicitly}) and is known to result in a multimodal posterior.
	
	The other dataset we use is of similar form but for prostate cancer. The original form of the dataset is described here: \cite{singh2002gene}. We use a filtered version of the dataset as detailed here: \cite{yang2006stable}, which we also analysed using logistic regression with regularised horseshoe priors. It has 18,105 parameters in total. 
	
	For each model, we consider two MCMC runs: one with 4 chains run with 800 thinned post-warm-up iterations (9000 total iterations with 1000 discarded as warm-up; 8000 post-warm-up iterations thinned by a factor of 10); another with 16 chains run with 1000 post-warm-up iterations each (500 warm-up iterations discarded and no thinning).
	
\begin{figure}[!htb]
		\centerline{\includegraphics[width=1.0\textwidth]{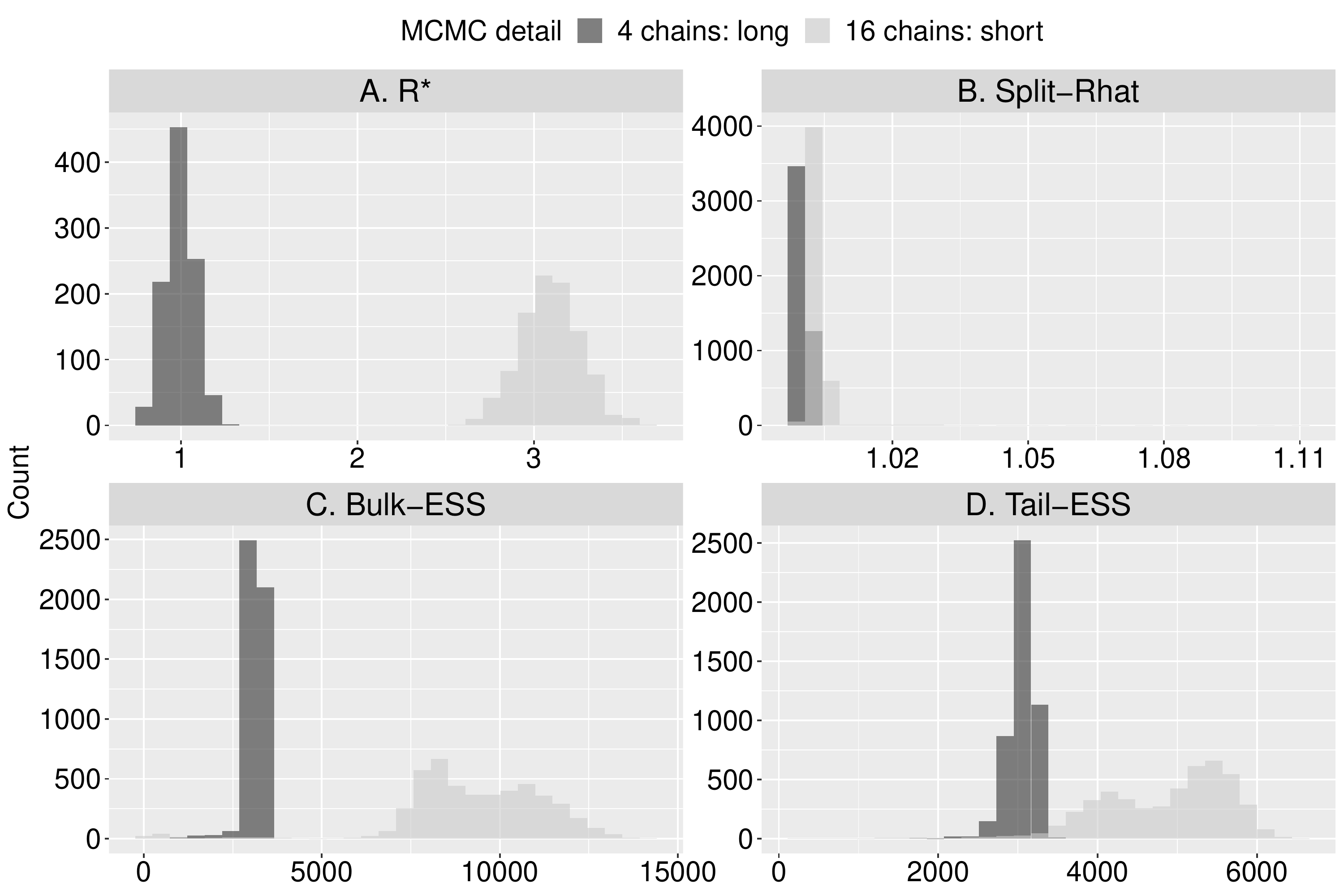}}
		\caption{\textbf{Ovarian example.} In all plots, we show the results from two model runs: one with 4 chains run with 800 thinned post-warm-up iterations (9000 total iterations with 1000 discarded as warm-up; 8000 post-warm-up iterations thinned by a factor of 10); another with 16 chains run with 1000 post-warm-up iterations each (500 warm-up iterations discarded and no thinning). In A, $R^*$ distributions (with 1000 draws each using Algorithm \ref{alg:R_star_uncertainty}) using the GBM classifier are shown; B shows rank-normalised split-$\widehat{R}$ values across all parameters; C shows bulk-ESS across all parameters; and D shows tail-ESS across parameters.}
		\label{fig:ovarian}
	\end{figure}

	For the ovarian model, we show the results in Fig. \ref{fig:ovarian}. In Fig. \ref{fig:ovarian}A, we show the $R^*$ distributions for each model run, which show that, whereas the ``long'' model run has converged, the ``short'' one has yet to do so. Whilst it is harder to discern, this pattern is mirrored in Fig. \ref{fig:ovarian}B, since the long model has $\widehat{R}<1.01$ for all parameters, whereas the short model had 83 parameters where this was not the case. Similarly, in Figs. \ref{fig:ovarian}C and \ref{fig:ovarian}D, which show the bulk-ESS and tail-ESS respectively, it is evident that, for the short model, there remain a few parameters with low effective sample sizes, whereas the long model has more consistent values for this metric.
	
\begin{figure}[!htb]
		\centerline{\includegraphics[width=1.0\textwidth]{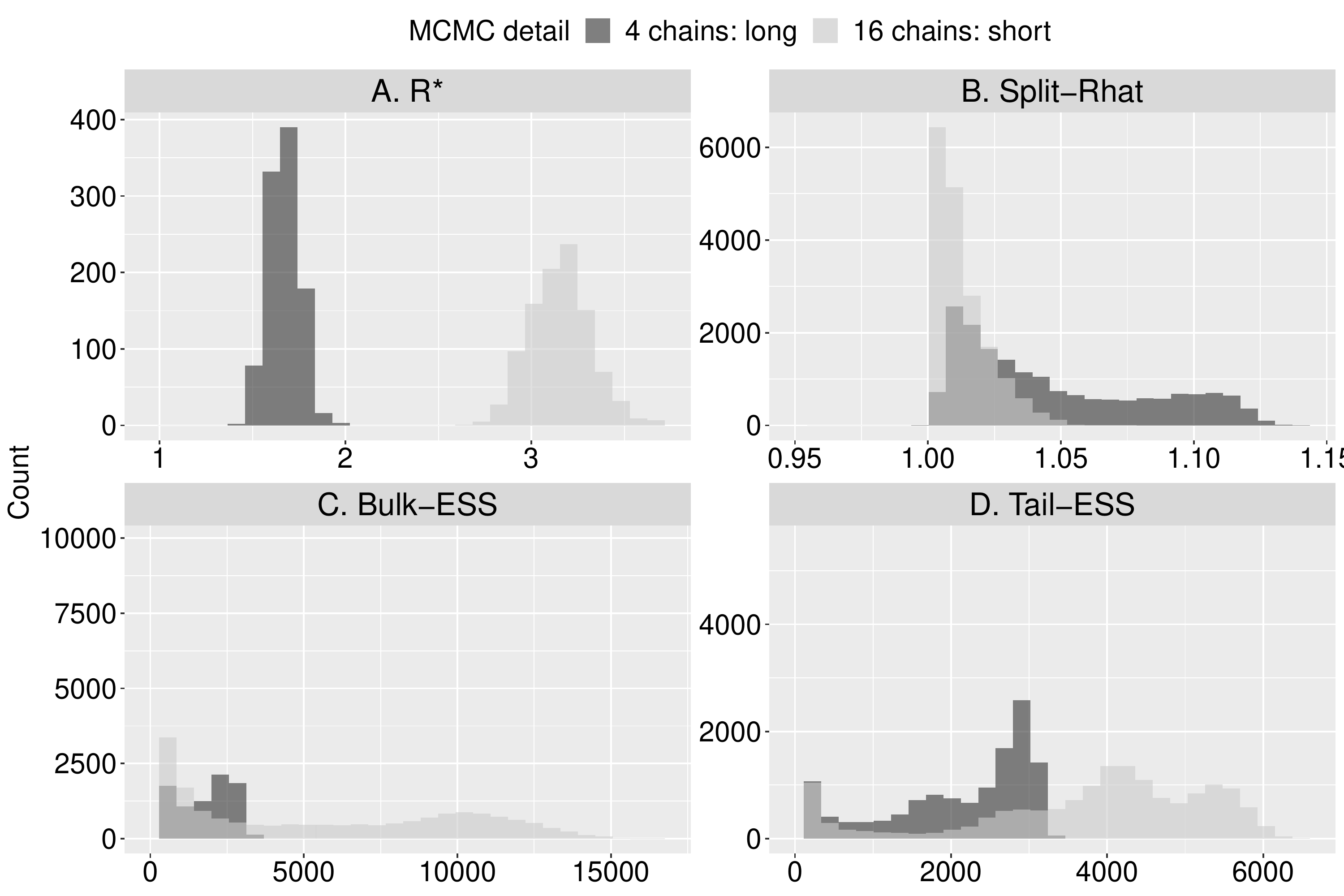}}
		\caption{\textbf{Prostate example.} In all plots, we show the results from two model runs: one with 4 chains run with 800 thinned post-warm-up iterations (9000 total iterations with 1000 discarded as warm-up; 8000 post-warm-up iterations thinned by a factor of 10); another with 16 chains run with 1000 post-warm-up iterations each (500 warm-up iterations discarded and no thinning). In A, $R^*$ distributions (with 1000 draws each using Algorithm \ref{alg:R_star_uncertainty}) using the GBM classifier are shown -- for the long run, these were calculated after thinning the parameters by a factor of 5; B shows rank-normalised split-$\widehat{R}$ values across all parameters; C shows bulk-ESS across all parameters; and D shows tail-ESS across parameters.}
		\label{fig:prostate}
	\end{figure}

	For the prostate model, we show the results in Fig. \ref{fig:prostate}. Since this model has nearly four times as many parameters as the ovarian model, it was more computationally expensive to estimate $R^*$ for it. To handle this, we thinned down the parameters by a factor of 5 for the long model, recognising that, of course, this measure will make it more likely that we diagnose convergence. Despite this, both $R^*$ measures indicated that the MCMC runs had yet to converge (Fig. \ref{fig:prostate}A), which was mirrored by the other metrics considered (Figs. \ref{fig:prostate}B,C\&D).
	
	\section{Discrete distributions}\label{sec:discrete}
	In this section, we compare the performance of $R^*$ and $\widehat{R}$ on a univariate discrete target distribution. To do so, we generate draws from four discrete Markov chains, following a 1st order Markov process with transition probabilities,
	\begin{equation}
	p_{ij} = \text{Pr}(X_{n+1}=j|X_n=i),
	\end{equation}
	where $p_{ij}$ is the probability of transitioning from state $i$ at time $n$ to state $j$ at time $n+1$, and $X(0)=1$. The target distribution is, hence, the stationary distribution of this process.
	
	\subsection{Small state-space}\label{sec:discrete_small}
	We first consider a univariate discrete distribution with four states. In three of the four chains, we generate draws using a transition probability matrix:
	\begin{gather}
	\boldsymbol{P}
	=
	\begin{bmatrix}
	0 & 1/2 & 1/2 & 0\\
	1/2 & 0 & 1/3 & 1/6\\
	1/4 & 1/4 & 1/4 & 1/4\\
	0 & 1 & 0 & 0
	\end{bmatrix},
	\end{gather}
	which implies a stationary target distribution with the following probability simplex $\pi_1 = (11/46, 15/46, 14/46, 6/46)$.
	
\begin{figure}[!htb]
		\centerline{\includegraphics[width=1.0\textwidth]{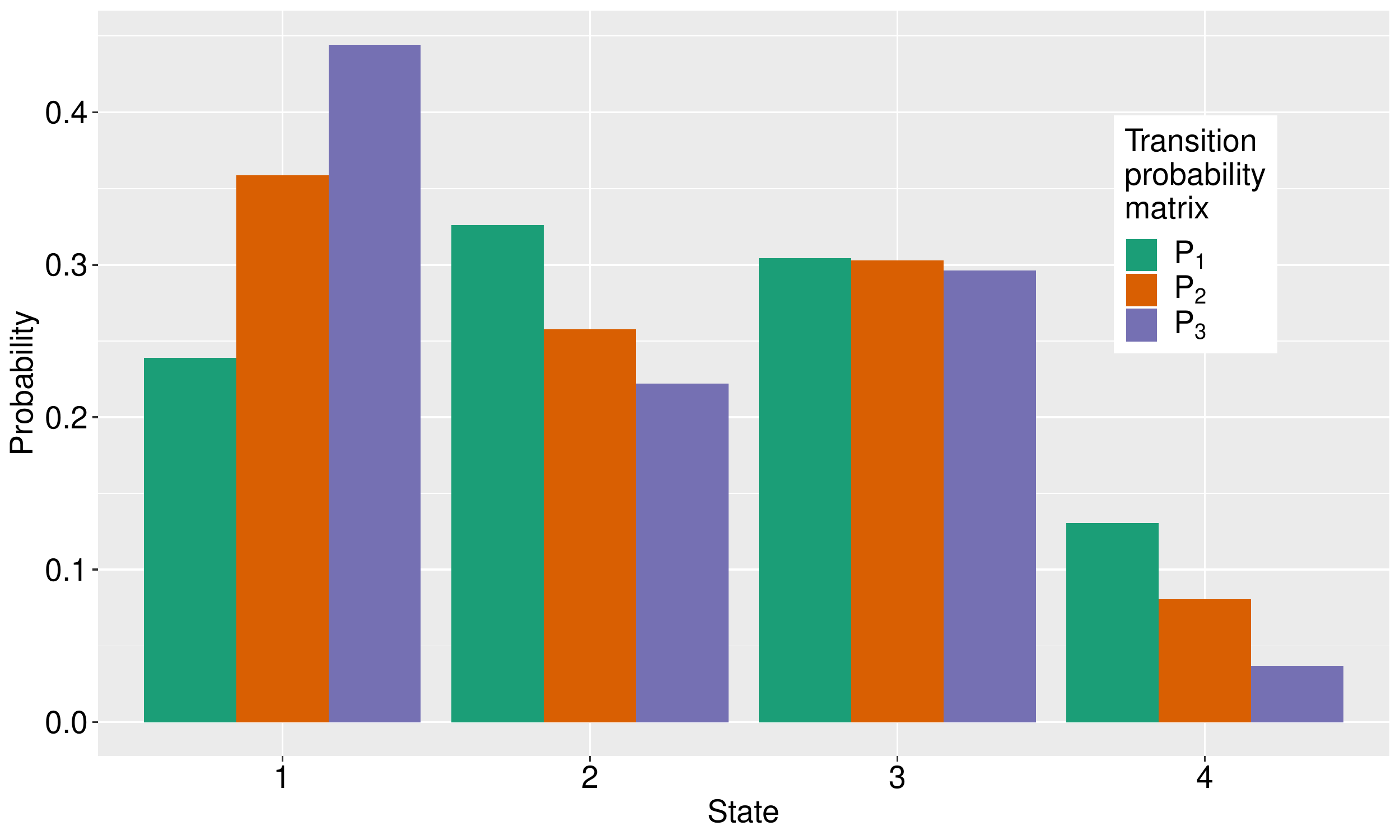}}
		\caption{\textbf{Discrete distribution example: stationary distributions.} The three colours illustrate the stationary distributions of the Markov chains corresponding to the matrices $P_1$, $P_2$ and $P_3$ defined in \S\ref{sec:discrete}.}
		\label{fig:discrete_stationary}
	\end{figure}

	For the fourth chain, we specify differing transition probability matrices in each of three sets of results. In the first, we assume $\boldsymbol{P}_1=\boldsymbol{P}$ (i.e. the same as used for the other three chains); in the second and third cases, we assume this transition matrix is given by:
	\begin{gather}
	\boldsymbol{P}_2
	=
	\begin{bmatrix}
	0 & 1/2 & 1/2 & 0\\
	1/2 & 0 & 1/3 & 1/6\\
	5/8 & 1/8 & 1/8 & 1/8\\
	1/2 & 1/2 & 0 & 0
	\end{bmatrix},\quad
	\boldsymbol{P}_3
	=
	\begin{bmatrix}
	0 & 1/2 & 1/2 & 0\\
	1/2 & 0 & 1/3 & 1/6\\
	1 & 0 & 0 & 0\\
	1 & 0 & 0 & 0
	\end{bmatrix}.
	\end{gather}
	These correspond to stationary probability simplices:
	\begin{equation}
	\pi_2 = (71/198, 17/66, 10/33, 8/99), \quad \pi_3 = (4/9, 2/9, 8/27, 1/27).
	\end{equation}
	The three stationary distributions corresponding to each transition probability matrix are shown in Fig. \ref{fig:discrete_stationary}. This illustrates that there is an ordering among the stationary distributions $\pi_1\sim\pi_2\sim\pi_3$, meaning $\pi_1$ is most similar to $\pi_2$, and $\pi_3$ is most similar to $\pi_2$. Put another way, $\pi_3$ is most different from $\pi_1$.
	
\begin{figure}[!htb]
		\centerline{\includegraphics[width=1.0\textwidth]{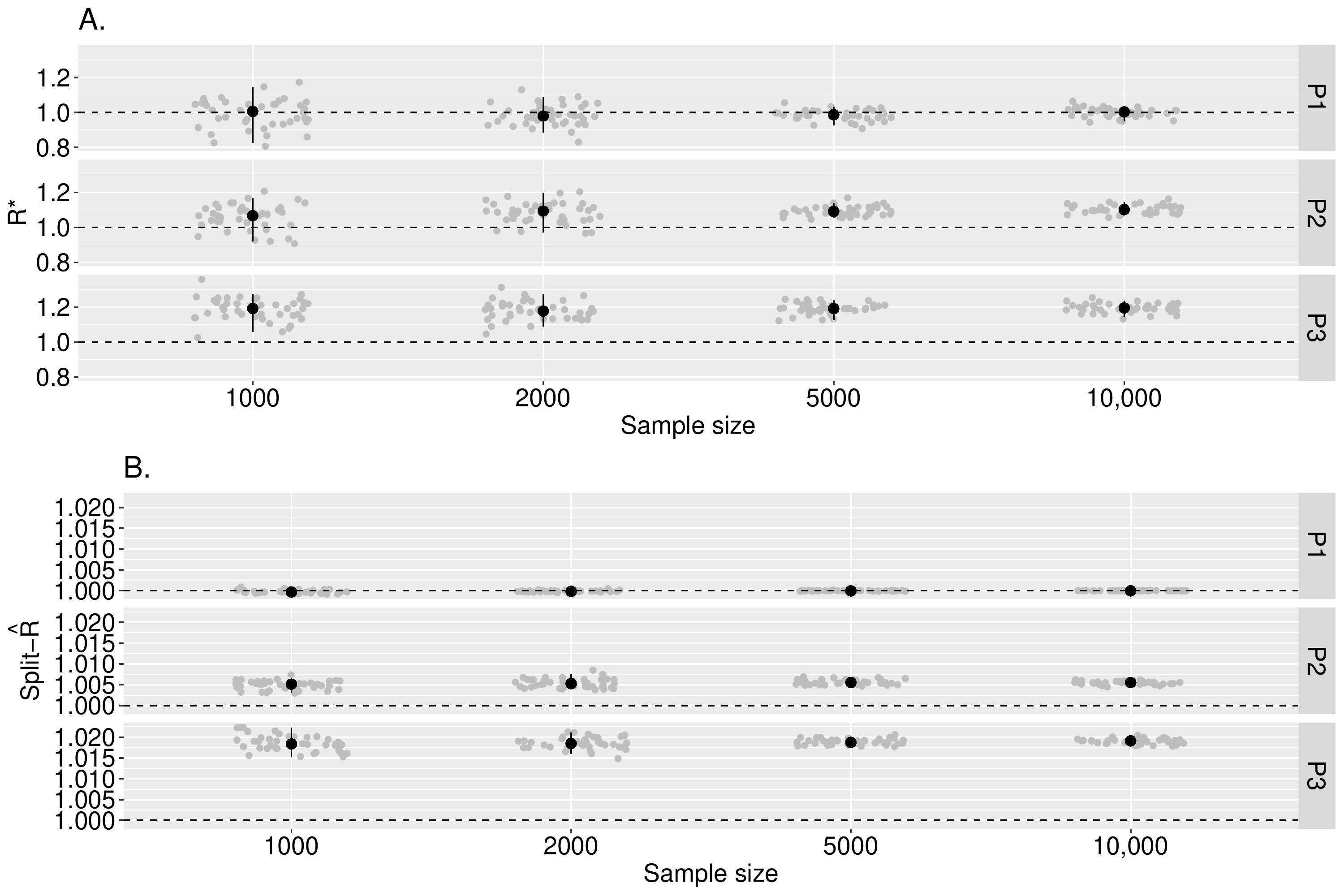}}
		\caption{\textbf{Discrete distribution example: convergence diagnostics.} On the horizontal axis, we show the sample size of each set of simulations. The panels within A. and B. indicate the transition probability matrix used for the fourth chain (the other three chains always used $\boldsymbol{P}=\boldsymbol{P}_1$). In panel A, grey points indicate $R^*$ from each replicate calculated using Algorithm \ref{alg:R_star}) , and the black point-ranges indicate 2.5\% , 50\% and 97.5\% quantiles across the 40 replicates. In panel B, we show the same, but for split-$\widehat{R}$. The dashed horizontal lines at 1 indicate the convergence threshold in both cases.}
		\label{fig:discrete}
	\end{figure}

	For each of these three cases, we consider a range of sample sizes across each of the four chains: 1000, 2000, 5000 and 10,000. At each combination of transition probability matrix (for the fourth chain) and sample size, we perform 40 replicates. In each replicate, we calculate both $R^*$ using a GBM classifier and rank-normalised split-$\widehat{R}$. The results of these experiments are shown in Fig. \ref{fig:discrete}. In the horizontal axes of both panels, we show the sample size. The vertical axis indicates the value of $R^*$ and split-$\widehat{R}$ in panels A and B, respectively. The stacked sub-panels within A and B indicate the transition matrix used for the fourth chain. Since $\boldsymbol{P}_1=\boldsymbol{P}$, the stationary distribution for the other three chains, both diagnostics should tend towards indicating convergence in these cases, and, for both statistics, this was typically the case. For $R^*$, the replicates were centred on $R^*=1$, although the stochasticity of replicates meant that, in some cases, $R^*>1$. Whereas, $\widehat{R}$ was similarly close to 1 in all cases. Considering the simulations using $\boldsymbol{P}_2$ for the fourth chain transition probability matrix, $R^*>1$ across a majority of replicates, and this majority increased with sample size. In this case, $1.01>\widehat{R}>1$ for all replicates across the various sample sizes. For simulations using $\boldsymbol{P}_3$ for the fourth chain transition probability matrix, $R^*>1$ across all replicates and the magnitude of median $R^*$ increased with sample size; in this case, $\widehat{R}>1.01$ in all replicates.
	
	This example shows that $R^*$ can detect poor convergence for discrete parameter spaces, like $\widehat{R}$.

	\subsection{Large state-space}\label{sec:discrete_large}
	We next consider a larger discrete parameter space: a univariate distribution with 20 states. As in \S\ref{sec:discrete_small}, we generate draws from three discrete Markov chains that share a common transition probability matrix, $\boldsymbol{P}$. For the fourth (and final) chain, we consider two cases: one where it shares the same matrix, $\boldsymbol{P}_1=\boldsymbol{P}$; the other, when its Markov process has a different matrix, $\boldsymbol{P}_2$. Rather than write down a given $\boldsymbol{P}$ and $\boldsymbol{P}_2$, we generated these stochastically: for each of the rows, we randomly sampled from from a Dirichlet process with shape vectors consisting of ones (of length 20).
	
\begin{figure}[!htb]
		\centerline{\includegraphics[width=1.0\textwidth]{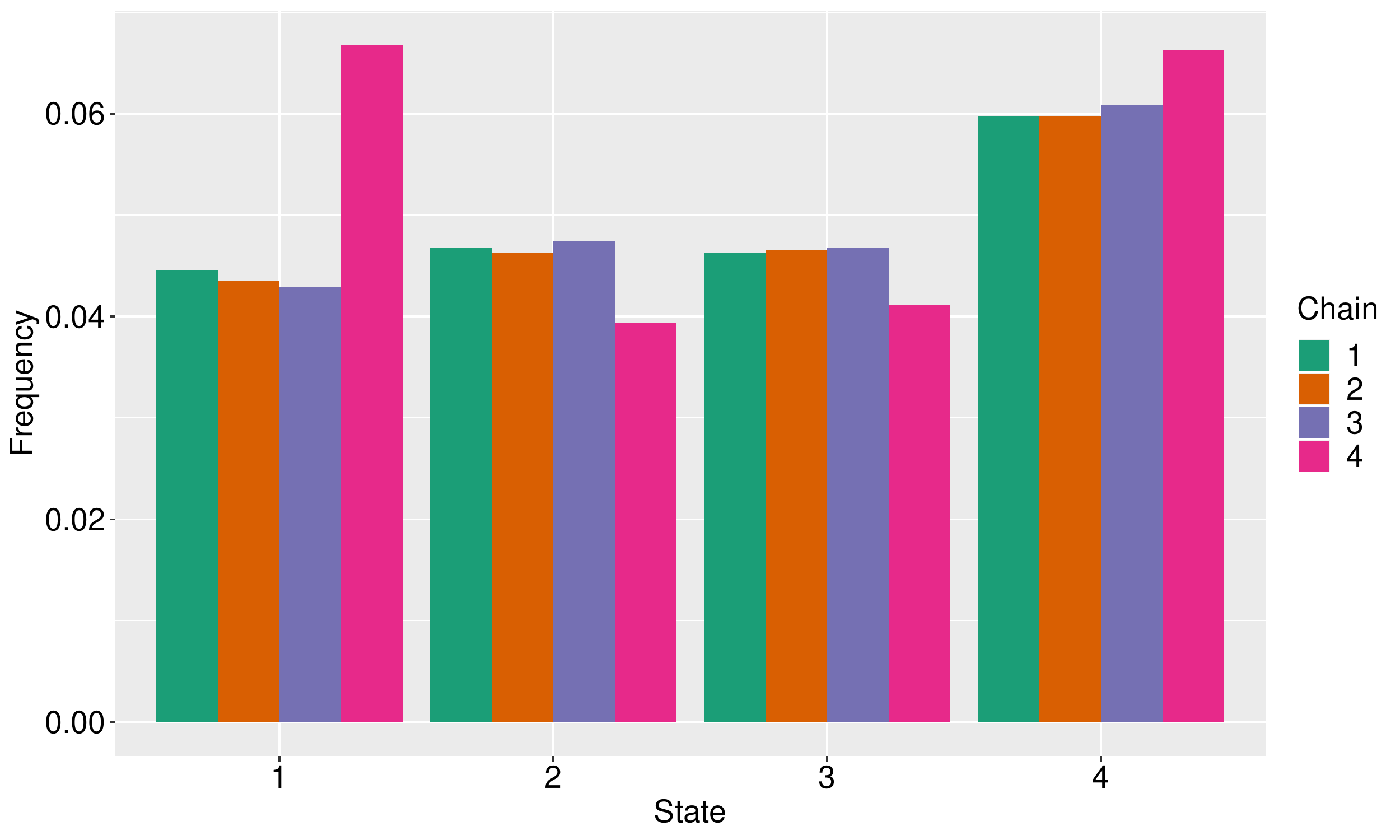}}
		\caption{\textbf{Large state-space discrete distribution example: approximate stationary distribution.} The distribution over the first four states of a 20-state discrete model achieved after drawing 100,000 samples from each of four Markov chains. Here, the first three chains shared the same transition probability matrix; the last chain used a different matrix.}
		\label{fig:discrete_stationary_higherd}
	\end{figure}

	To illustrate the differences between the two transition probability matrices, $\boldsymbol{P}$ and $\boldsymbol{P}_2$, we simulate 100,000 draws from each Markov process. In Fig. \ref{fig:discrete_stationary_higherd}, we show the marginal distribution obtained across all four chains: where the first three used $\boldsymbol{P}$, and the last used $\boldsymbol{P}_2$. In this plot, we show only the first four states of the 20-state model. The differences in marginal distributions between the first three chains and the last indicate that there are substantial differences in the target distributions.
	
\begin{figure}[!htb]
		\centerline{\includegraphics[width=1.0\textwidth]{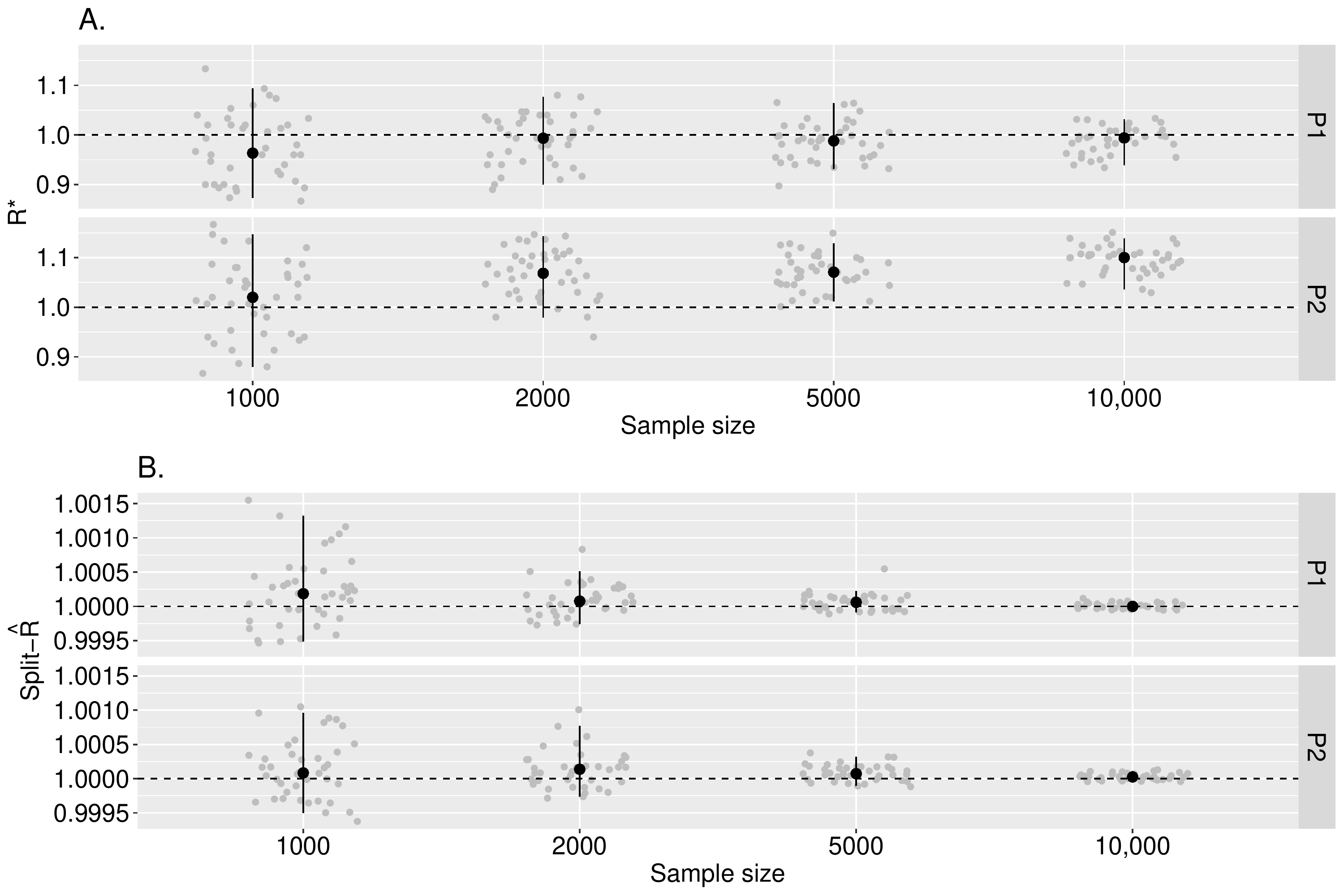}}
		\caption{\textbf{Large state-space discrete distribution example: convergence diagnostics.} On the horizontal axis, we show the sample size of each set of simulations. The panels within A. and B. indicate the transition probability matrix used for the fourth chain (the other three chains always used $\boldsymbol{P}=\boldsymbol{P}_1$). In panel A, grey points indicate $R^*$ from each replicate calculated using Algorithm \ref{alg:R_star}) , and the black point-ranges indicate 2.5\% , 50\% and 97.5\% quantiles across the 40 replicates. In panel B, we show the same, but for split-$\widehat{R}$. The dashed horizontal lines at 1 indicate the convergence threshold in both cases.}
		\label{fig:discrete_higherd}
	\end{figure}

	To probe the ability of $R^*$ and $\widehat{R}$ to detect poor convergence, we repeat the same exercise as in \S\ref{sec:discrete_small} although, here, we consider the new $\boldsymbol{P}_1$ and $\boldsymbol{P}_2$ cases. In Fig. \ref{fig:discrete_higherd}, we show the results of this analysis. This shows that $R^*\sim 1$ for the $\boldsymbol{P}_1$ case, correctly indicating no convergence issues; for the $\boldsymbol{P}_2$ case, $R^*>1$ across all replicates at sample sizes of 5000 and above. By contrast, $\widehat{R}$ struggles to differentiate between the $\boldsymbol{P}_1$ and $\boldsymbol{P}_2$ cases: in both instances, it signifies convergence.
	
	Collectively, the results of \S\ref{sec:discrete_small} and \S\ref{sec:discrete_large} show that $R^*$ is able to detect convergence issues in discrete target distributions. The results of \S\ref{sec:discrete_large} hint that $R^*$ may be superior than $\widehat{R}$ in larger discrete spaces, although further empirical investigation is needed.
	
	\section{ML sensitivity}\label{sec:ML_sensitivity}
	In this section, we investigate how two decisions about classifiers — which classifier to use (in \S\ref{sec:ml_model}) and what hyperparameters to use for it (in \S\ref{sec:hyperparameters}) — affect calculation of $R^*$. The test cases used for empirical evaluation were selected from our pool of examples introduced in the text. Specifically, they were chosen to represent cases where $R^*$ was nearer 1, since, for these, small changes in calculated $R^*$ could lead to different decisions about convergence being reached. The four examples we used were:
	
	\begin{itemize}
		\item The AR(1) example described in \S\ref{sec:heterogeneity}, except using 1000 draws per chain.
		\item The 250-dimensional non-centered multivariate normal example introduced in \S\ref{sec:multivariate_normal_250}. In each replicate, we generated 500 post-warm-up draws from the posterior (using 500 warm-up steps) using Stan's NUTS algorithm across each of 4 chains.
		\item The alternative parameterisation of the Cauchy model described in \S\ref{sec:cauchy}. For each replicate, we generated samples from the model using Stan's NUTS algorithm with 1000 post-warm-up draws (and 1000 warm-up draws) across each of 4 chains.
		\item The non-centered parameterisation of the eight schools model described in \S\ref{sec:eight_shools}. In each replicate, we used Stan's NUTS algorithm with adapt delta set to 0.95 to generate 1000 post-warm-up draws (and 1000 warm-up draws) across each of 4 chains; these were then used to calculate $R^*$.
	\end{itemize}
	
	\subsection{ML classifier comparison}\label{sec:ml_model}
	In this section, we investigate how classification accuracy depends on choice of ML classifier. We restricted our analysis to popular classifiers with relatively few hyperparameters to simplify comparison amongst them. Notably, we do not consider neural network models, since the structure of nets effectively defines a high dimensional hyperparameter space.
	
	The ML classifiers we compared were GBMs, RFs, k-nearest neighbours (KNNs), SVMs with linear kernels and a generalised linear model approach (GLM). All these models were called through the \textbf{\textsf{R}}'s Caret package \citep{kuhn2008building}: the GBM is implemented using the \textit{gbm} model in Caret which uses the ``gbm'' package \citep{greenwell2019package}; RFs are implemented using the \textit{rf} model in Caret which uses the ``randomForest'' \citep{liaw2002classification} package; KNN is natively implemented in Caret and called using the \textit{knn} model; SVMs are implemented in Caret using the \textit{svmLinear} model which uses ``kernlab'' package \citep{karatzoglou2004kernlab}; and the GLM is implemented through the \textit{multinom} model in the ``nnet'' package \citep{ripley2016package}.
	
	Each of these classifiers has hyperparameters, and, due to the differences between the examples we consider (defined in \S\ref{sec:ML_sensitivity}), the optimal hyperparameters were likely to differ between examples. For each classifier-example combination, we first performed a single replicate of each experiment and choose those hyperparameters which maximised classification accuracy. The set of all hyperparameters we selected between were chosen so that training time for each classifier remained reasonable. These sets were the same across all examples:
	
	\begin{itemize}
		\item GBM: Cartesian product of $\text{int.depth}=(3, 7)$; $\text{\# trees} = (50, 100)$;
		\item RF: $m_{\text{try}} = (1, 2)$. (Note, in \S\ref{sec:hyperparameters}, we investigate a dimension-specific approach to $m_{\text{try}}$ but, here, for ease of comparison with the other methods, we keep this static.)
		\item KNN: $K=(5,10,15, 20,40)$;
		\item SVM: $C=(0.25, 0.5, 0.75)$;
		\item GLM: $\text{decay}=(0.1, 0.2, 0.5, 1)$.
	\end{itemize}
	
	These hyperparameters were then used in subsequent replicates. We also used a different number of replicates at each unique combination of classifier and hyperparameters across the experiments due to the differing demands of each example: for the AR(1) example, we used 100 replicates; for the multivariate normal example, we used 100; for the Cauchy model, 50 replicates; and for the eight schools model, 30 replicates.
	
\begin{figure}[!htb]
		\centerline{\includegraphics[width=1.0\textwidth]{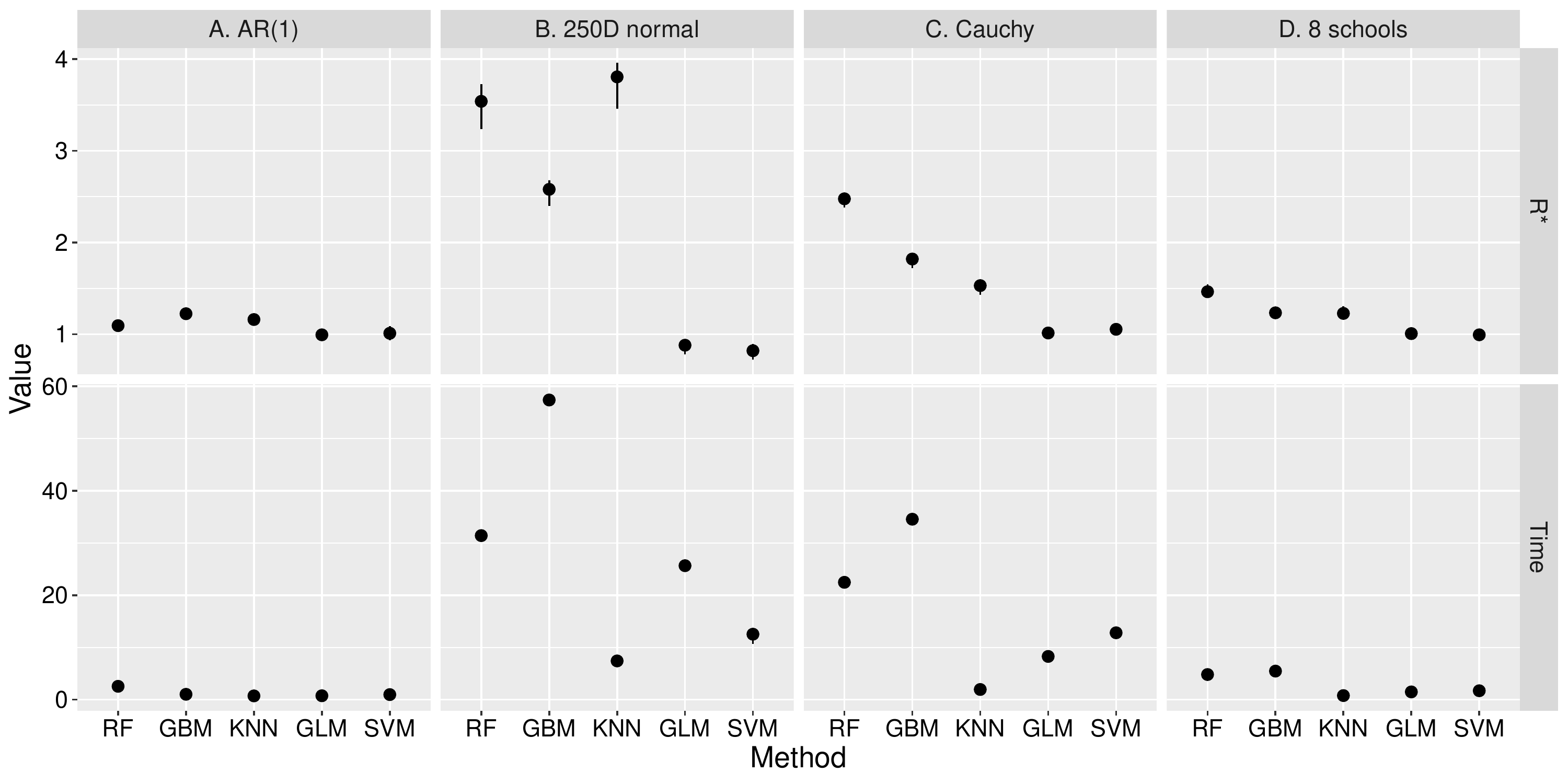}}
		\caption{\textbf{ML classifier comparison.} Top row shows $R^*$ values calculated using Algorithm \ref{alg:R_star}; bottom row show the time taken to run each method on a desktop computer. The different columns correspond to the examples described in \S\ref{sec:ML_sensitivity}. Black point-ranges indicate 25\% , 50\% and 75\% quantiles across all replicates; note, that the number of replicates varies across examples as given in \S\ref{sec:ml_model}.}
		\label{fig:ml_comparison_all}
	\end{figure}

	The results of these experiments are shown in Fig. \ref{fig:ml_comparison_all}. The top row shows $R^*$ as calculated by each classifier (in each case, chains split into two halves); the bottom row shows the time (in seconds) taken for $R^*$ calculation. The panels correspond to the examples described in \S\ref{sec:ML_sensitivity}. Across the examples, the classification accuracy of RFs and GBMs were, generally, highest; followed by KNNs. An exception was for the 250-dimensional normal where KNN performed best. In the AR(1) example, GBM outperformed RF. In higher dimensional cases, RF outperformed GBM. In all cases, RF and GBM classifiers took the greatest time to train; in higher dimensional examples, RF-based $R^*$ were cheaper to calculate than those from GBMs.

	\subsection{Hyperparameter sensitivity}\label{sec:hyperparameters}
	In this section, we considered only the two classifiers — GBMs and RFs — that performed best on our comparison between ML methods in \S\ref{sec:ml_model}. The performance of GBMs and RFs, like most ML methods, depends on their hyperparameters. In GBMs, a regression function is approximated by additive components, where each of those corresponds to a tree. The number of such regression trees is thus a hyperparameter of the model. Each regression tree can split amongst any integer number of variables (the \textit{interaction depth}). For GBMs, we analyse how their performance depends on these two hyperparameters. For RFs, there is a single hyperpameter: $m_{\text{try}}$, the number of members of the subset of all features over which to search for an optimal split when forming a decision tree. Note, that the maximum allowed value of $m_{\text{try}}=K$ — the dimensionality of the target distribution. 
	
	In this section, we investigate the sensitivity of $R^*$ calculated by GBMs and RFs to each of these sets of hyperparameters across a range of examples (defined in \S\ref{sec:ML_sensitivity}). Since each example is of different dimensionalities, we compare different sets of hyperparameters in each, although include amongst the sets the default values we suggest for each classifier in \S\ref{sec:method} (these are shown as triangles on the plots). Note, that for the GBM model, we did not investigate a full range of hyperparameters needed to optimise $R^*$ for many of the examples due to the extensive training time needed to do so. For RFs, runtime was less restrictive and we were better able to survey the sensitivity across hyperparameter space.
	
	We also used a different number of replicates at each unique combination of classifier and hyperparameters across the experiments due to the differing demands of each example: for the AR(1) example, we used 200 replicates; for the multivariate normal example, we used 20; for the Cauchy model, 20 replicates; and for the eight schools model, 50 replicates.

	\subsubsection{Autoregressive example}\label{sec:hyperparameters_ar1}
\begin{figure}[!htb]
		\centerline{\includegraphics[width=1.0\textwidth]{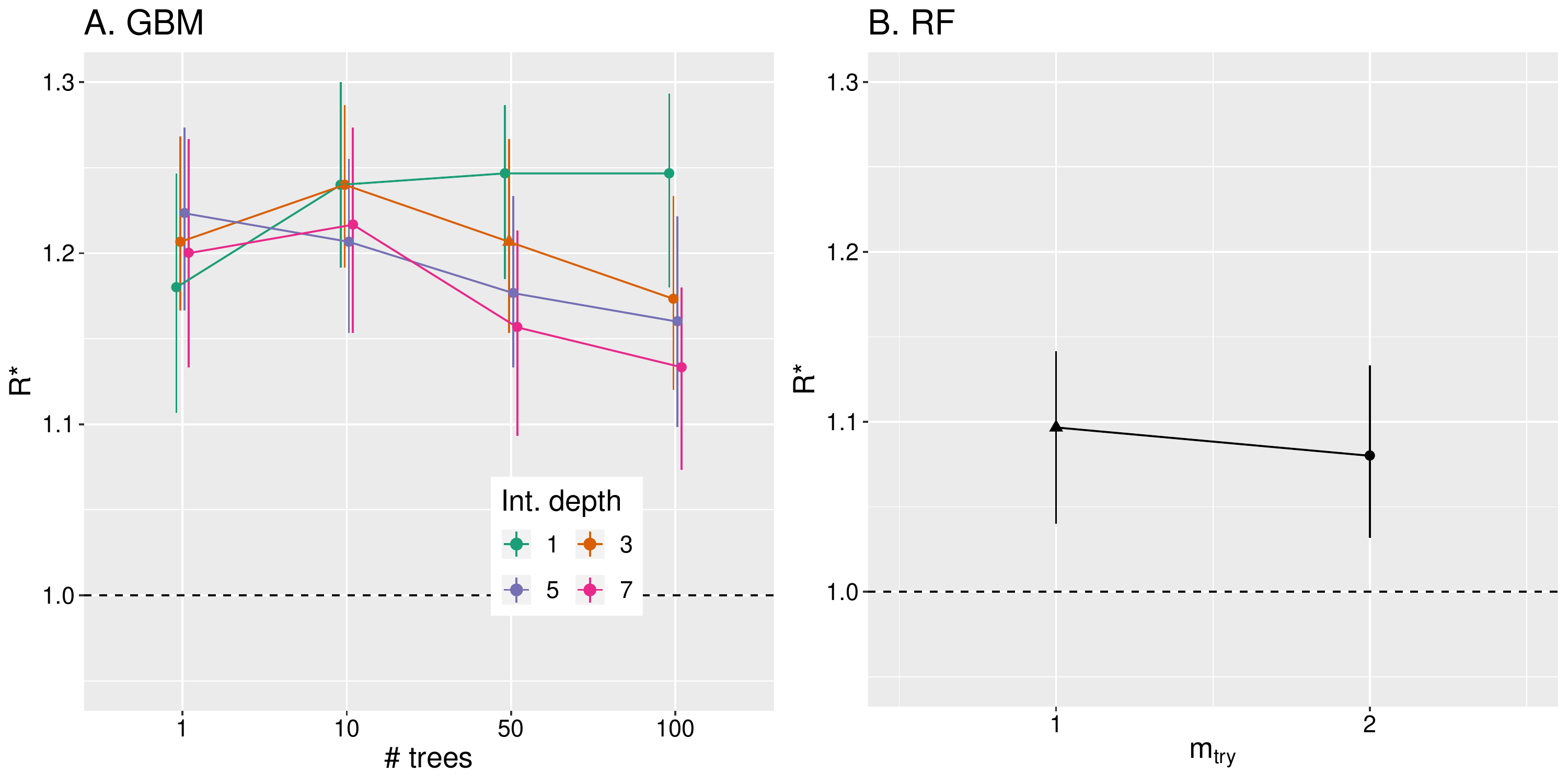}}
		\caption{\textbf{Hyperparameter sensitivity: AR(1) example.} Panel A shows the sensitivity of GBM-derived $R^*$ to hyperparameter variation for the AR(1) example described in \S\ref{sec:hyperparameters_ar1}; Panel B shows the same but for a RF classifier. Black point-ranges indicate 25\% , 50\% and 75\% quantiles across all replicates. In Panel A, the coloured lines shows the different interaction depths investigated. In both panels, the results for the parameterisation we suggest as defaults is shown as a triangle.}
		\label{fig:hypers_ar1}
	\end{figure}

	In Fig. \ref{fig:hypers_ar1}, we show the results of the AR(1) analysis. In panel A, we show the sensitivity of $R^*$ as calculated by GBMs to variation in hyperparameters. In panel B, we show the same but for RFs. In both panels, the results for the parameterisation we suggest as defaults is shown as a triangle. For the GBM model, $R^*$ varied according to hyperparameters: of the set investigated, an interaction depth of 1 appears optimal with 10+ trees. In panel B, the values of $R^*$ are generally lower than those achieved by GBMs. As shown in \S\ref{sec:comparison_gbm_rf}, RFs appear less well-suited to lower dimensional problems. Across the two possible hyperparameter values for RFs, there was not a substantial difference in the values of $R^*$ calculated.
	
	\subsubsection{Multivariate normal: 250-dimensional model}\label{sec:hyperparameters_normal}
\begin{figure}[!htb]
		\centerline{\includegraphics[width=1.0\textwidth]{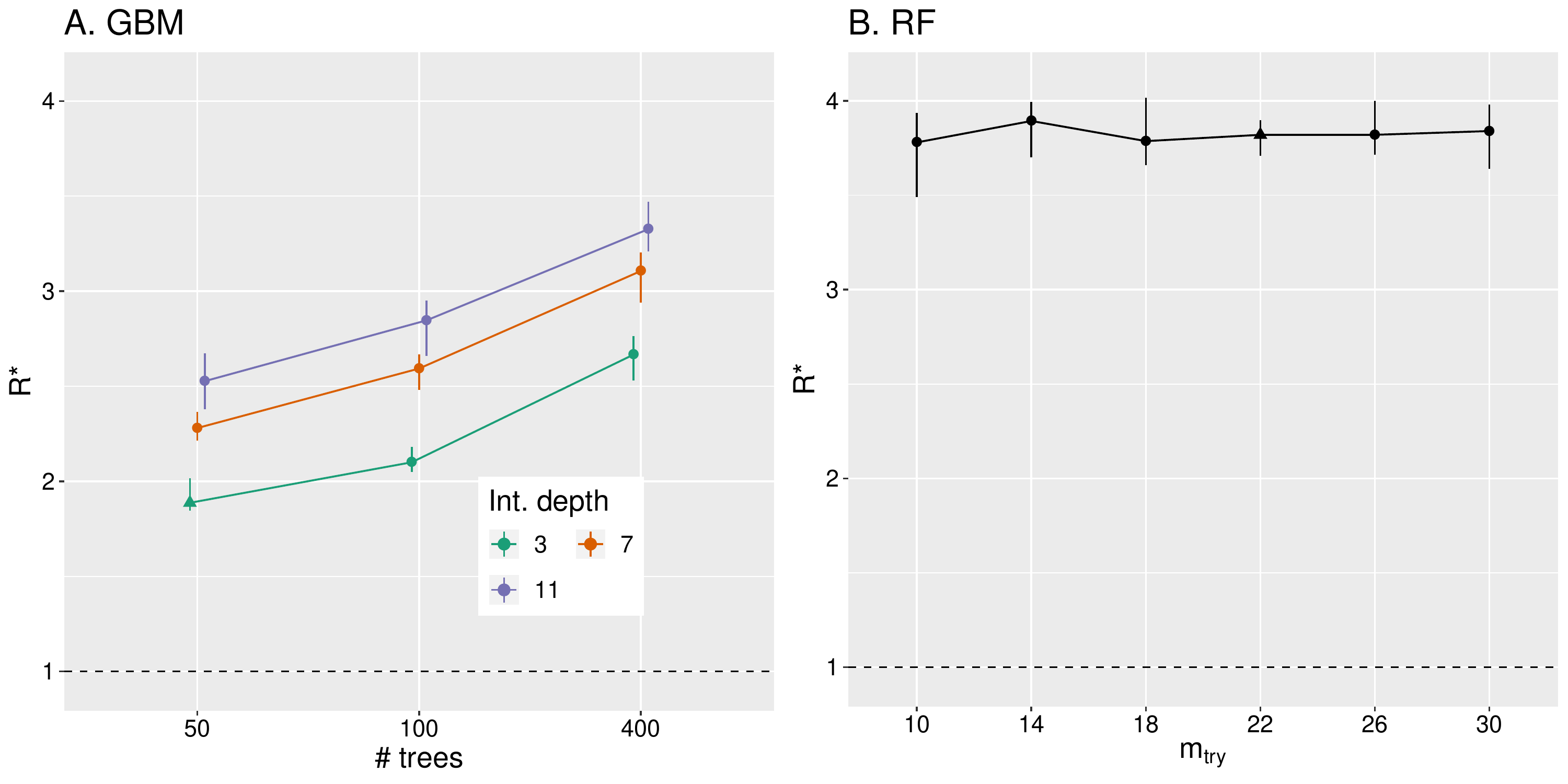}}
		\caption{\textbf{Hyperparameter sensitivity: 250-dimensional normal example.} Panel A shows the sensitivity of GBM-derived $R^*$ to hyperparameter variation for the Cauchy example described in \S\ref{sec:hyperparameters_normal}; Panel B shows the same but for a RF classifier. Black point-ranges indicate 25\% , 50\% and 75\% quantiles across all replicates. In Panel A, the coloured lines shows the different interaction depths investigated. In both panels, the results for the parameterisation we suggest as defaults is shown as a triangle.}
		\label{fig:hypers_normal}
	\end{figure}

	In Fig. \ref{fig:hypers_normal}, we show the results of the analysis for the 250-dimensional normal example. In panel A, we show the sensitivity of $R^*$ as calculated by GBMs to variation in hyperparameters. In panel B, we show the same but for RFs. This shows that $R^*$ was higher as calculated by RFs and that performance of GBMs depended strongly on hyperparameters.
	
	\subsubsection{Cauchy model: alternative parameterisation}\label{sec:hyperparameters_cauchy}
\begin{figure}[!htb]
		\centerline{\includegraphics[width=1.0\textwidth]{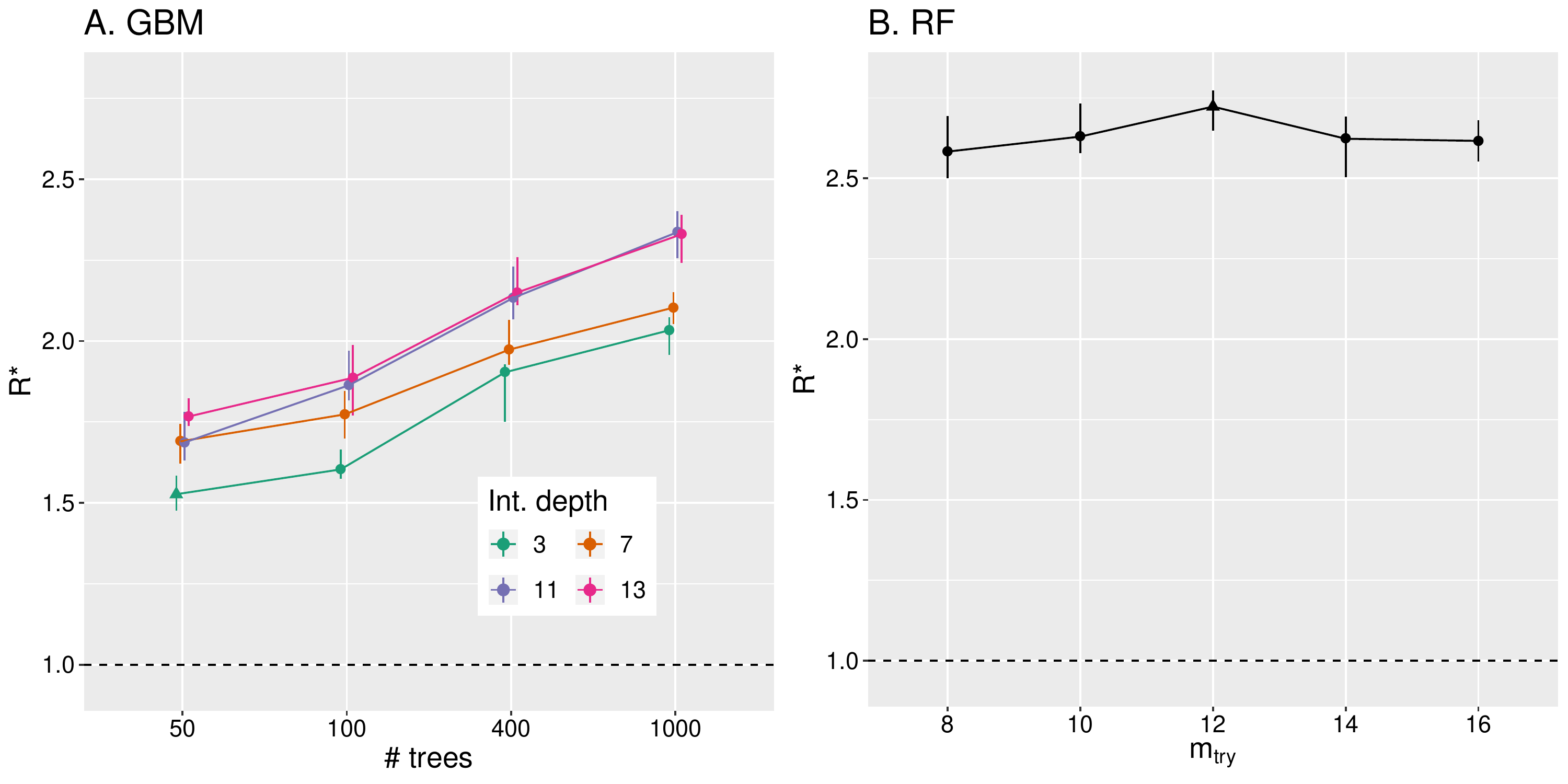}}
		\caption{\textbf{Hyperparameter sensitivity: Cauchy example.} Panel A shows the sensitivity of GBM-derived $R^*$ to hyperparameter variation for the Cauchy example described in \S\ref{sec:hyperparameters_cauchy}; Panel B shows the same but for a RF classifier. Black point-ranges indicate 25\% , 50\% and 75\% quantiles across all replicates. In Panel A, the coloured lines shows the different interaction depths investigated. In both panels, the results for the parameterisation we suggest as defaults is shown as a triangle.}
		\label{fig:hypers_cauchy}
	\end{figure}

	In Fig. \ref{fig:hypers_cauchy}, we show the results of the Cauchy model analysis. In panel A, we show $R^*$ as calculated by a GBM classifier; in panel B, the same but using a RF classifier. In both panels, the results for the parameterisation we suggest as defaults is shown as a triangle. Across the range of hyperparameters investigated, the RF classifier outperformed the GBM and was less sensitive to variation in its hyperparameters.
	
	\subsubsection{Eight schools: non-centered parameterisation}\label{sec:hyperparameters_8_schools}
\begin{figure}[!htb]
		\centerline{\includegraphics[width=1.0\textwidth]{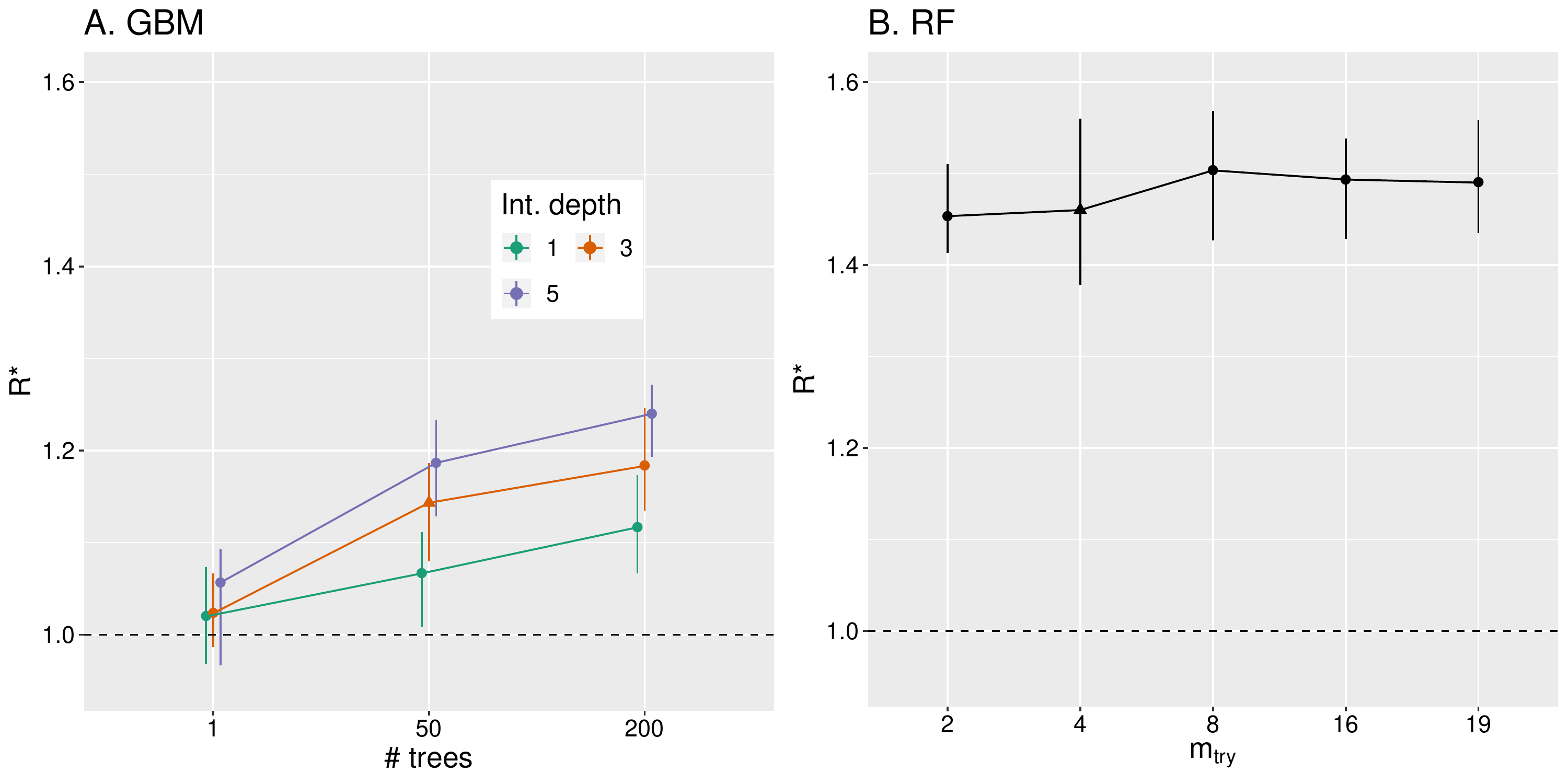}}
		\caption{\textbf{Hyperparameter sensitivity: eight schools example.} Panel A shows the sensitivity of GBM-derived $R^*$ to hyperparameter variation for the eight schools example described in \S\ref{sec:hyperparameters_8_schools}; Panel B shows the same but for a RF classifier. Black point-ranges indicate 25\% , 50\% and 75\% quantiles across all replicates. In Panel A, the coloured lines shows the different interaction depths investigated. In both panels, the results for the parameterisation we suggest as defaults is shown as a triangle.}
		\label{fig:hypers_8_schools}
	\end{figure}

	In Fig. \ref{fig:hypers_8_schools}, we show the results of the experiments on the eight schools model. Panel A shows how $R^*$ calculated using a GBM varies with its two hyperparameters; panel B shows the same but for RFs. This shows that, in this case, the value of $R^*$ calculated by RFs was higher than that for GBMs. Further, whereas GBMs were relatively sensitive to their hyperparameter values, RFs were less so.
	
	\section{Comparing GBMs and RFs}\label{sec:comparison_gbm_rf}
	In \S\ref{sec:ML_sensitivity}, we present a series of experiments using popular ML methods and find that, among these methods, GBMs and RFs perform most consistently across them. In this section, we compare $R^*$ derived from using a GBM classifier with that from a RF classifier. In these experiments, we fix the hyperparameters of each as given in \S\ref{sec:method}. In \S\ref{sec:joint_distribution}, we compare how both methods are able to diagnose lack of convergence in a joint distribution. In \S\ref{sec:tail_fatness}, we compare the ability of both approaches to diagnose differences in the tails of the marginal distributions between chains. Since we use independent sampling to generate draws, we are also able to calculate the \textit{Bayes optimal} classification accuracy, which is the classifier that minimises the probability of misclassification error \citep{devroye2013probabilistic}. This classifier assigns chain identities according to the one which has the maximum posterior probability of having generated a draw. Here, we assume that \textit{a priori} all chains are equally likely to have generated the data, so the classifier categorises draws to the chain which maximises the likelihood of that observation.
	
	\subsection{Joint distribution}\label{sec:joint_distribution}
	Here, we consider a multivariate normal target distribution with dimensionality ranging from 1 to 32. For three `chains', we generated 2000 draws by independently sampling from a multivariate normal with zero mean and given covariance matrix; for the fourth chain, the same number of draws are generated in the same way except with a different covariance matrix. To generate the covariance matrices, we randomly sampled two 32-dimensional covariance matrices from the LKJ distribution with degrees of freedom 1 \citep{lewandowski2009generating}: note, that this results in matrices with unit diagonal values, meaning the marginal distributions of each dimension are standard normals. To obtain covariance matrices for the targets with dimensions, $d<32$, submatrices, $A^{\{d\}}$, were constructed from the 32-dimensional covariance matrices by taking leading blocks from them: $A^{\{d\}}:=A^{\{32\}}_{1:d, 1:d}$. For each target, we performed 20 replicates, where in each case, $R^*$ was calculated using GBM and RF classifiers. In addition, we determined an $R^*$ using the Bayes optimal classifier: to do so, we generated 10,000 draws from the four-chain process and assign chain identities to the maximum likelihood class.  
	
\begin{figure}[!htb]
		\centerline{\includegraphics[width=1\textwidth]{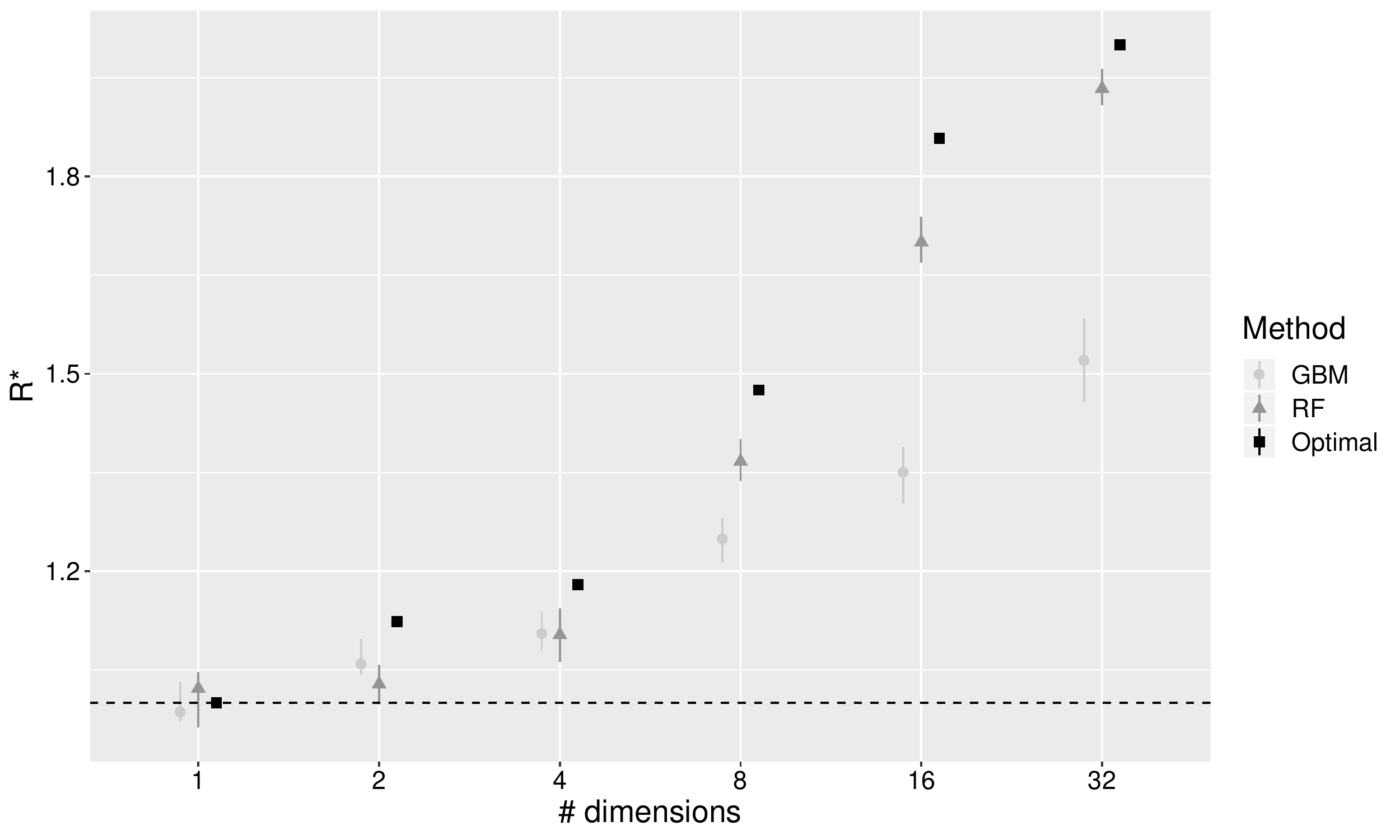}}
		\caption{\textbf{GBM versus RF: multivariate normal.} The horizontal axis gives the dimensions of the target distribution; the vertical axis gives the value of $R^*$ as calculated by Algorithm \ref{alg:R_star} using a GBM (light grey points), a RF (dark grey triangles) and the Bayes optimal classifier (black squares). The upper and lower point ranges show the 75\% and 25\% quantiles and the filled shapes show the medians. For the Bayes optimal classifier, there is only a point estimate since the optimal accuracy was estimated by Monte Carlo sampling using 10,000 draws from the four chain process. The dashed line shows $R^*=1$. Here, $R^*$ was calculated using chains split into two halves.}
		\label{fig:gbm_vs_rf_normal}
	\end{figure}

	The results of these experiments are shown in Fig. \ref{fig:gbm_vs_rf_normal}. In this plot, the bottom axis shows the dimensionality of the target distribution; the vertical axis shows $R^*$, and the point colour and shadings give the method used to calculate this statistic. When the target is unidimensional, the four chains have the same target distribution and the optimal $R^*=1$. In this case, both the GBM and RF methods produce classification rates that overlap with this value, indicating convergence. As the number of dimensions increases, it becomes easier to differentiate between samples from the two processes and the optimal $R^*$ grows. $R^*$ as calculated by GBMs and RFs also increases. For a two-dimensional target, the GBM method outperforms the RF one, getting closer to optimal classification. For higher dimensional targets, the RF classifier outperforms, achieving near-optimal $R^*$ in 32 dimensions.
	
	\subsection{Fat tails}\label{sec:tail_fatness}
	We now compare how the GBM and RF methods are able to diagnose lack of convergence in the tails of a target distribution. To do so, we again consider four chains. In three of these chains, we generated 2000 draws by randomly sampling from a multivariate Student-t distribution with mean zero and 3 degrees of freedom. In the remaining chain, we generated the same number of independent draws but use a mean-zero multivariate Student-t distribution with differing degrees of freedom, $\nu$. We consider a range of target dimensions: to generate the shape matrix for the first three chains, we use the same approach as in \S\ref{sec:joint_distribution} to build up appropriate shape matrices for each target dimensions, $A^{\{d\}}$. We examine a range of values of $\nu=\{4, 8, 16, 32\}$, representing decreasing tail fatness. Across the range of $\nu$ considered, the covariance matrix for the first three chains is given by $\nu/{(\nu-2)} A^{\{d\}}$. To ensure that the fourth chain has the same covariance as the first three, we use a shape matrix ${(\nu-2)}/\nu A^{\{d\}}$. For each combination of dimensions and $\nu$, we performed 20 replicates, where in each, we again calculate $R^*$ for both the GBM and RF classifiers. Additionally, we estimate an optimal $R^*$ using the Bayes optimal classifier by calculating the predictive accuracy of 10,000 draws of the four-chain process.
	
\begin{figure}[!htb]
		\centerline{\includegraphics[width=1\textwidth]{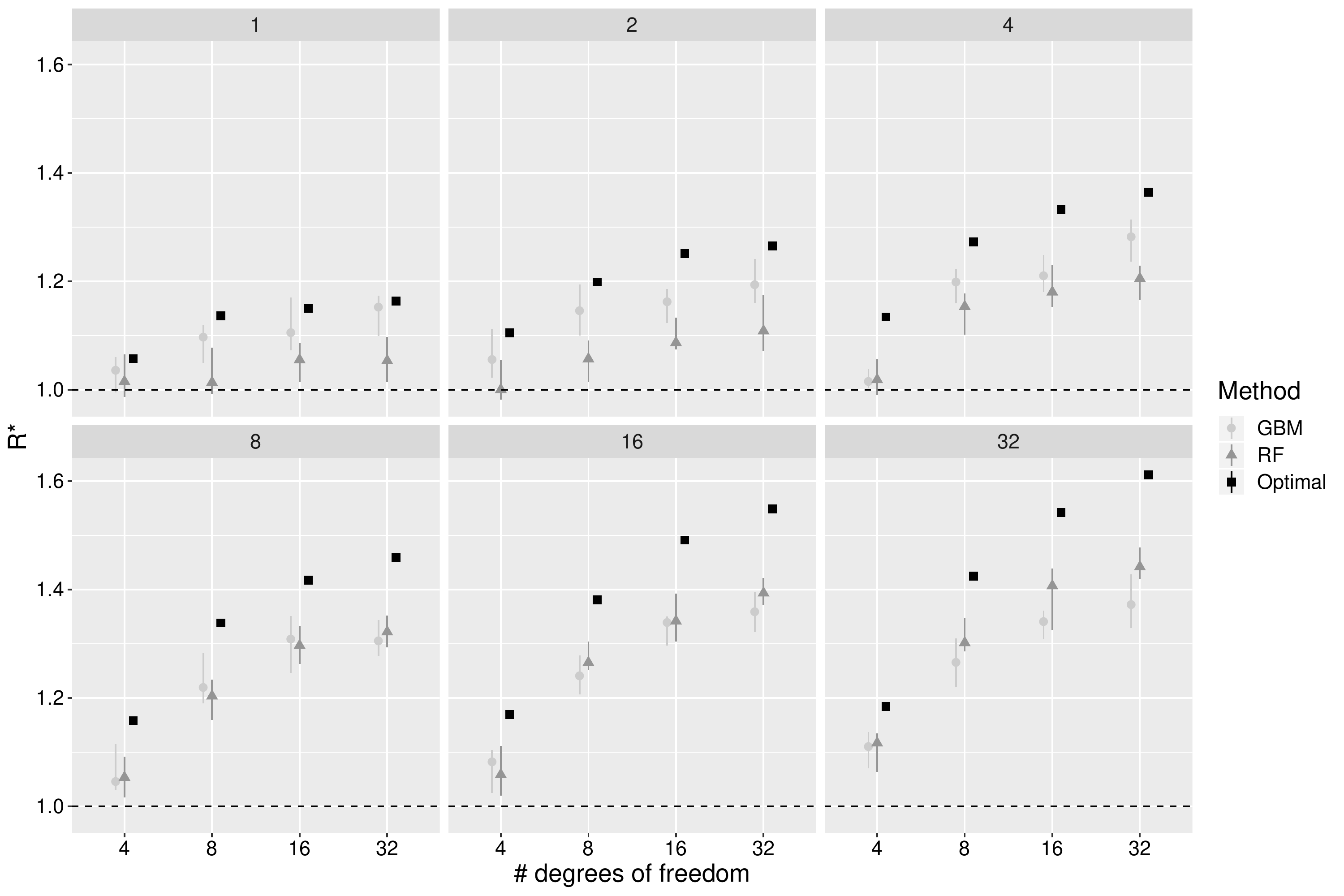}}
		\caption{\textbf{GBM versus RF: multivariate Student-t.} Different panels show the various dimensions of the target distribution. The horizontal axis gives the degrees of freedom of the multivariate Student-t distribution for the fourth chain; the vertical axis gives the value of $R^*$ as calculated by Algorithm \ref{alg:R_star} using a GBM (light grey points), a RF (dark grey triangles) and the Bayes optimal classifier (black squares). The upper and lower point ranges show the 75\% and 25\% quantiles and the filled shapes show the medians. For the Bayes optimal classifier, there is only a point estimate since the optimal accuracy was estimated by Monte Carlo sampling using 10,000 draws from the four chain process. The dashed line shows $R^*=1$. Here, $R^*$ was calculated using chains split into two halves.}
		\label{fig:gbm_vs_rf_studdentt}
	\end{figure}

	The results of these experiments are shown in Fig. \ref{fig:gbm_vs_rf_studdentt}. Here, each panel shows a target distribution of different dimensionality: 1, 2, 4, 8, 16 and 32 dimensions. Within each panel, the horizontal axis shows the degrees of freedom, $\nu$, of the fourth chain; the vertical axis gives $R^*$. The point colour and shadings give the method used to calculate $R^*$. In each panel, increases in $\nu$ make it easier to differentiate draws from the fourth chain from those of the first three: accordingly $R^*$ tends to increase for each method. As the target dimensionality increases (panels left-right along each row), it also becomes easier to contrast draws from the fourth chain, and $R^*$ increases. In all cases, the optimal $R^*$ exceeds those using GBM or RF classifiers. In 1-4 dimensions, the GBM approach outperforms the RF one, getting closer to the optimal classification rate. If the number of dimensions is 16 or greater, the order switches, and RFs often outperform GBMs.

\end{document}